\documentclass[aip,APLQuantum,reprint,showpacs,superscriptaddress]{revtex4-2}
\usepackage{amsmath,amssymb,mathrsfs,amsfonts,dsfont}
\usepackage{bm}
\usepackage{color}
\usepackage{appendix}
\usepackage{graphicx}
\usepackage{footmisc}
\usepackage{bm}
\usepackage{color}
\usepackage{CJKutf8}
\usepackage{orcidlink}
\usepackage{hyperref}
\hypersetup{colorlinks=true,
	linkcolor=red,
	citecolor=blue,
	urlcolor=blue}

\begin{document}
	
	\title{Quantum speed limits in dephasing dynamics of a qubit system coupled to thermal environments}
	
	\author{Xiangji Cai~\orcidlink{0000-0001-6655-5736}}
	\email[Authors to whom correspondence should be addressed: ]{xiangjicai@foxmail.com, renjing19@sdjzu.edu.cn, mxj@sdyu.edu.cn\\ and  aczerwin@umk.pl}
	\affiliation{School of Science, Shandong Jianzhu University, Jinan 250101, China}
	\author{Yanyan Feng}
	\affiliation{School of Science, Shandong Jianzhu University, Jinan 250101, China}
	\author{Jing Ren}
	\email[Authors to whom correspondence should be addressed: ]{xiangjicai@foxmail.com, renjing19@sdjzu.edu.cn, mxj@sdyu.edu.cn\\  and  aczerwin@umk.pl}
	\affiliation{School of Science, Shandong Jianzhu University, Jinan 250101, China}
	\author{Kang Lan~\orcidlink{0000-0001-7796-1529}}
	\email{lank@qfnu.edu.cn}
	\affiliation{Department of Physics, Qufu Normal University, Qufu 273165, China}
	\author{Shuning Sun}   
	\email{shuningsun@sdyu.edu.cn}
	\affiliation{School of Information Engineering, Shandong Youth University of Political Science, Jinan 250103, China}
	\author{Xiangjia Meng~\orcidlink{0000-0001-6742-3532}}
	\email[Authors to whom correspondence should be addressed: ]{xiangjicai@foxmail.com, renjing19@sdjzu.edu.cn, mxj@sdyu.edu.cn\\  and  aczerwin@umk.pl}
	\affiliation{School of Information Engineering, Shandong Youth University of Political Science, Jinan 250103, China}
	\author{Artur Czerwinski~\orcidlink{0000-0003-0625-8339}}
	\email[Authors to whom correspondence should be addressed: ]{xiangjicai@foxmail.com, renjing19@sdjzu.edu.cn, mxj@sdyu.edu.cn\\  and  aczerwin@umk.pl}
	\affiliation{Institute of Physics, Faculty of Physics, Astronomy and Informatics, Nicolaus Copernicus University in Torun, ul. Grudziadzka 5, 87-100 Torun, Poland}
	\begin{abstract}
	We theoretically study the quantum speed limits (QSLs) of a qubit system coupled to a thermal dephasing environment with an Ohmic-like spectral density. Based on the geometric QSLs time bound, which is derived by employing the trace distance to quantify the geodesic between two distinguishable states in dynamical evolution, we study the influences of the temperature and spectral density of the environment on the QSLs time of the dephasing qubit. We also investigate the interplay between the QSLs time, the environmental temperature, and the spectral density of the environment. It has been demonstrated that the QSLs time closely depends on the transition frequency and the dynamical behavior (e.g., coherence trapping) of the dephasing qubit. For a fixed Ohmicity parameter of the environmental spectral density, the increase in environmental temperature can enhance the QSLs time bound. In addition, when the environmental temperature remains constant, the increase in the Ohmicity parameter initially leads to a reduction in the QSLs time bound, which is then followed by an increase in the time bound of QSLs. Our results can help to better understand the QSLs in the dynamics of open quantum systems and have potential application in the modulation of QSLs time in the dephasing qubit by engineering the spectral density of the environment.
	\end{abstract}
	
	\maketitle
	
	\section{Introduction}
	\label{sec:intr}
	Quantum speed limits (QSLs) determine the maximum speed of the dynamical evolution of quantum systems~\cite{JPhysA50.453001}.
	QSLs play a crucial role in various aspects of quantum physics, ranging from many fundamental questions to potential applications in quantum information science.
	For instance, in quantum computing, QSLs can guide algorithm design to minimize computational time~\cite{PhysRevLett.81.5442,Nature406.1047,Nature532.210,PhysRevRes.2.032016}.
	In quantum communication, QSLs can be used to assess the efficiency of information transfer~\cite{NewJPhys.24.065003,PhysRevA82.022318,PhysRevA95.042314}. 
	Moreover, QSLs provide deep insights into the fundamental limitations in quantum optimal control, quantum metrology, quantum sensing, as well as in the processes of optimizing and enhancing the performance of quantum devices~\cite{PhysRevLett.103.240501,PhysRevLett.111.260501,PhysRevLett.113.010502,PhysRevLett.127.110506,PhysRevLett.130.170801,PhysRevLett.133.210802,PhysRevE97.062116,RevModPhys.92.021001,PhysRevRes.5.043194}. 
	
	For a closed quantum system, which is governed by a time-independent Hamiltonian $H$ and evolves unitarily between pure orthogonal states, there are two independent types of QSLs time bound related to the variance and average of the system energy established by Mandelstam and Tamm (MT)~\cite{JPhys.9.249} and Margolus and Levitin (ML)~\cite{PhysicaD120.188}, respectively. 
	Therefore, the QSLs time bound for a closed quantum system between orthogonal states under unitary dynamics is unified as 
	\begin{equation}
		\label{eq:uniQSLs}
		\tau_{\mathrm{QSL}}=\max\left\{\frac{\pi\hbar}{2\Delta E},\frac{\pi\hbar}{2(E-E_{0})}\right\},
	\end{equation}
	where $(\Delta E)^{2}=\langle\psi(0)|H^{2}|\psi(0)\rangle-E^{2}$ denotes the variance of energy, $E=\langle\psi(0)|H|\psi(0)\rangle$ is the average energy with respect to initial state $|\psi(0)\rangle$ and $E_{0}$ is the ground state energy.
	The theoretical studies of the ultimate speed limits in dynamical evolution have also been expanded to various cases, such as nonorthogonal mixed states~\cite{PhysRevA67.052109,PhysRevA82.022107}, time-dependent Hamiltonian~\cite{JPhysA46.335302,JPhysA47.215301,PhysRevA99.042116}, quantum phase space~\cite{NewJPhys.19.103018,PhysRevA101.042107}, non-Hermitian quantum systems~\cite{PhysRevLett.123.180403,PhysRevLett.127.100404,PhysRevA104.052620} and classical systems~\cite{PhysRevLett.120.070401,PhysRevLett.120.070402,PhysRevLett.121.070601,PRXQuantum2.040349}.
	
	Due to unavoidable environmental effects, any quantum system loses coherence in its dynamical evolution~\cite{RevModPhys.59.1,Breuerbook,Schlosshauerbook,PhysRep.831.1}. 
	The studies of decoherence dynamics of open quantum systems have increasingly become an important research field~\cite{PhysRevB78.235311,PhysRevA97.012104,PhysRevLett.100.180402,PhysRevLett.103.210401,PhysRevLett.105.050403,PhysRevLett.109.170402,RepProgPhys.77.094001,RevModPhys.88.021002,RevModPhys.89.015001,PhysRevLett.112.120404,PhysRevLett.112.210402,PhysRevLett.114.190502,PhysRevA100.062112,PhysRevA103.013714}.
	Concurrently, it has recently drawn extensive attention to study QSLs in the dynamical evolution of open quantum systems.
	These studies are not merely limited to finding different bounds, but more importantly, they focus on the exploration of the underlying physical mechanisms~\cite{PhysRevLett.110.050402,PhysRevLett.110.050403,PhysRevLett.111.010402,PhysRevA91.022102,PhysRevA91.032112,PhysRevA93.020105,PhysRevA94.052125,PhysRevA96.012105,PhysRevA100.052305,PhysRevX6.021031,PhysRevA95.052104,NewJPhys.21.123041,PhysRevLett.115.210402,SciRep.4.4890,PhysRevA98.022114,PhysRevA98.042132,PhysRevA100.022118,NewJPhys.21.013006,PhysRevRes.2.023299,PhysRevA103.022210,PhysRevA104.052424,PhysRevLett.124.110601,PhysRevLett.132.230404,PhysRevA110.052433,PhysRevLett.120.060409,Quantum3.168,PhysRevA108.052207,PhysRevA106.012403,PhysRevX12.011038,PhysRevX13.011013,Photonics9.875,PhysRevD108.126011,PhysRevA111.022445}.
	In recent years, the contribution of the time unitary evolution, which is governed by the Hamiltonian of the system, on QSLs of open quantum systems has attracted significant attention~\cite{PhysRevA93.052331,PhysRevA95.052104,NewJPhys.24.055003,PhysRevA108.012204,ResultsPhys.57.107315,PhysRevA110.042425}.
	In a closed quantum system undergoing a time unitary evolution between pure states, the geodesic connecting two states can be uniquely quantified by the quantum Fisher information (QFI) metric~\cite{PhysRevA103.022210}.
	By contrast, in the quantum evolution of an open system, an infinite number of geometric metrics can be used to quantify the geodesic between two distinguishable states~\cite{PhysRevX6.021031}.
	In general, each geometric metric corresponds to a distinct QSLs time bound.
	This implies that in the dynamical evolution of an open quantum system, there exists an infinite family of time bounds of QSLs\cite{PhysRevX6.021031}.
	It is extremely meaningful, by comparing different QSLs time bounds, to obtain the tightest one derived from the metric which quantifies the geodesic closest to the total distance between the initial and final states~\cite{PhysRevLett.120.060409,Quantum3.168,PhysRevA103.022210,PhysRevA108.052207}.
	In recent years, the observations of QSLs in dynamical evolution of open quantum systems have been increasingly realized theoretically and  experimentally~\cite{JChemPhys.156.134113,PhysRevLett.114.233602,PhysRevLett.126.180603,NewJPhys.26.013043,CommunPhys.7.142,PhysRevA110.042215}.

	For open quantum systems, the loss of coherence results mainly from the decoherence and dissipation processes induced by the environments.
	The characteristic time of the former process is much smaller than that of the latter one.
	In certain specific quantum platforms, the time scales of numerous useful information-processing tasks are generally either smaller than or commensurate with the characteristic time of decoherence. This implies that the quantum systems engaged in handling these tasks exhibit a higher degree of sensitivity to decoherence compared to dissipation~\cite{ProcRSocA452.567,PhysRevA65.032326}.
	The study of pure decoherence (dephasing) dynamics of a qubit coupled to an environment, 
	which is the physical model originally proposed to study the impact of decoherence in quantum computers, has recently emerged as a crucial area of research, due to the facts that the environmental spectral density can be effectively engineered in experiments and that the dephasing qubit occurs coherence trapping in specific dynamical regions~\cite{PhysRevA84.031602,PhysRevA87.010103,PhysRevA89.024101,PhysRevA90.052103,SciRep.5.13359,JChemPhys.149.094107,PhysRevA97.012126,PhysRevA101.032112,Entropy25.634}.
	The QSLs time bounds of a dephasing qubit for the case of the environment at zero temperature have been studied based on different approaches~\cite{SciRep.4.4890,PhysRevA110.042425}.
	Nevertheless, it is a fact that in practical scenarios, the environment around the qubit system does not maintain a zero-temperature condition.
	Some important questions arise naturally: How does the environmental temperature affect the QSLs time of the dephasing qubit? 
	What is the interplay between the QSLs time, the environmental temperature and the spectral density of the environment?
	
	In this paper, we theoretically study the QSLs in the dephasing dynamics of a qubit system coupled to a thermal environment with an Ohmic-like spectral density. 
	By employing the trace distance to quantify the geodesic that connects two evolving states, we derive the geometric QSLs time bound in dynamical evolution of open quantum systems.
	Based on the derived time bound of QSLs, we study how the temperature of the environment affects the QSLs time bound in the weak and strong coupling regions and study the influence of the Ohmicity parameter of the environmental spectral density on the QSLs time bound for fixed environmental temperatures in detail. 
	We investigate the interplay between the time bound of QSLs, environmental temperature and spectral density of the environment and explore the latent potentiality of the enhancement of QSLs by means of the engineering of the environmental spectral density.
	
	This paper is organized as follows. In Sec.~\ref{sec:theo1} we derive the geometric QSLs time bound of open quantum systems by employing the trace distance to quantify the geodesic distance between two evolving states.
	In Sec.~\ref{sec:theo2} we study the QSLs time of a qubit system interacting with a thermal dephasing environment with an Ohmic-like spectral density and investigate the interplay between the time bound of QSLs, the environmental temperature and the environmental spectral density.
	In Sec.~\ref{sec:conc} we give the conclusions drawn from the present study.
	
	\section{QSLs in open system dynamics}
	\label{sec:theo1}
	We consider the quantum process of an open system evolving from an initial state $\rho(0)$ to the state $\rho(t)$.
	Correspondingly, the dynamical evolution of the open quantum system is governed by a generalized master equation as~\cite{Breuerbook,PhysRevA70.012106}
	\begin{equation}
		\label{eq:masequ}
		\frac{d}{dt}\rho(t)=\mathcal{L}\rho(t)\equiv\mathcal{L}_{c}\rho(t)+\mathcal{L}_{r}\rho(t),
	\end{equation}
	where the two parts of the dynamical generator $\mathcal{L}\rho(t)$ can be generally expressed as the time-local structures
	\begin{equation}
		\label{eq:dyngen}
		\begin{split}
			\mathcal{L}_{c}\rho(t)&=-\frac{i}{\hbar}[H(t),\rho(t)],\\
			\mathcal{L}_{r}\rho(t)&=\sum_{k}\gamma_{k}(t)\Big[L_{k}(t)\rho(t)L_{k}^{\dag}(t)-\frac{1}{2}\{L_{k}^{\dag}(t)L_{k}(t),\rho(t)\}\Big].
		\end{split}
	\end{equation}
	The first part $\mathcal{L}_{c}\rho(t)$ is associated with the unitary contribution for coherent evolution generated by the Hamiltonian $H(t)=H_{0}+H_{LS}(t)$ where $H_{0}$ 
	is the unperturbed Hamiltonian of the quantum system and $H_{LS}(t)$ denotes a time-dependent Lamb shift Hamiltonian which describes the renormalization of the unperturbed energy levels of the system induced by the coupling to the environment.
	The second part $\mathcal{L}_{r}\rho(t)$ is related to the nonunitary contribution for relaxation arising from the environmental effects, which is determined by the Lindblad operators $L_{k}(t)$ and decay rates $\gamma_{k}(t)$.
	Although the quantum master equation~\eqref{eq:masequ} is expressed in terms of time-local structures in the Lindblad form, it remains highly effective for describing non-Markovian dynamics~\cite{PhysRevA81.062115}. 
	In the description of a quantum system under non-Markovian dynamical evolution, at least one of the decay rates $\gamma_{k}(t)$ in the master equation exhibits negative values during certain time intervals.

	In investigations of environmental effects on the QSLs time of an open quantum system evolving from an initial state $\rho(0)$, the final state $\rho(\tau)$ cannot be set arbitrarily because arbitrary final states may not lie on the physically allowed path determined by the system and the environment.
	The final state $\rho(\tau)$ must emerge from this specific evolution of the initial state $\rho(0)$ under the quantum master equation~\eqref{eq:masequ}.
	The QSLs time for an open quantum system is not merely a geometric property of initial and final states but reflects the interplay between the intrinsic characteristic of the system and the environmental effects. 
	Therefore, it must be computed from the actual evolution under the master equation, taking these constraints into account.
	
	To derive the QSLs time required for a quantum system to evolve from an initial state $\rho(0)$ to a final target state $\rho(\tau)$, a widely used treatment is to parameterize the quantum state as $\rho\bm(\lambda(t)\bm)$ using the parameterization $t\in[0, \tau]\rightarrow\lambda(t)$ such that $\lambda_{I}=\lambda(0)$ and $\lambda_{F}=\lambda(\tau)$ with $\tau$ being the evolution time.
	The geodesic distance $\mathcal{G}\bm(\rho(0),\rho(\tau)\bm)$ is the lower bound to the length of the actual evolution path that connect the initial state $\rho(0)$ and the target state $\rho(\tau)$~\cite{PhysRevX6.021031,PhysRevA103.022210}
	\begin{equation}
		\label{eq:geopat}
		\mathcal{G}\bm(\rho(0),\rho(\tau)\bm)\leq\ell\bm(\rho(0),\rho(\tau)\bm),
	\end{equation}
with 
	\begin{equation}
		\label{eq:lenpat}
		\ell\bm(\rho(0),\rho(\tau)\bm)=\int_{0}^{\tau}\left(\frac{ds}{dt}\right)dt=\int_{0}^{\tau}\sqrt{\sum_{j,k=1}^{r}g_{jk}\frac{d\lambda_{j}}{dt}\frac{d\lambda_{k}}{dt}}dt,
\end{equation}
where $g_{jk}$ denotes a metric tensor and $r$ is the number of elements within the set $\lambda(t)=\{\lambda_{1},\lambda_{2},\cdots,\lambda_{r}\}$.
By means of the definition of the average speed along the evolution path
	\begin{equation}
	\label{eq:lenpat}
	\bar{v}(\tau)=\frac{\ell\bm(\rho(0),\rho(\tau)\bm)}{\tau}=\frac{1}{\tau}\int_{0}^{\tau}\left(\frac{ds}{dt}\right)dt,
\end{equation}
it is straightforward for us to derive the time bound of QSLs~\cite{PhysRevA103.022210}
	\begin{equation}
	\label{eq:geoQSLpar}
	\tau\geq\tau_{\rm{QSL}}=\frac{\mathcal{G}\bm(\rho(0),\rho(\tau)\bm)}{\bar{v}(\tau)}.
\end{equation}
In this treatment, the evolution time $\tau$ parameterizes the quantum state $\rho(t)$ to study the QSLs between the initial state $\rho(0)$ and the target state $\rho(\tau)$.

	We employ here the trace distance to quantify the geodesic for an open quantum system evolving from the state $\rho(t)$ to the state $\rho(\tau)$ as
	\begin{equation}
		\label{eq:tradis}
		\mathcal{D}\bm(\rho(t),\rho(\tau)\bm)=\frac{1}{2}\|\rho(\tau)-\rho(t)\|_{\mathrm{tr}},
	\end{equation}
	where $\|A\|_{\mathrm{tr}}=\mathrm{tr}|A|=\mathrm{tr}\sqrt{A^{\dag}A}$ denotes the trace norm of an operator $A$.
	The trace distance between the state $\rho(t)$ and $\rho(\tau)$ is bounded by $0\leq\mathcal{D}\bm(\rho(t),\rho(\tau)\bm)\leq1$ and it satisfies the properties of a geometric metric, namely, (i) non-negativity $\mathcal{D}\bm(\rho(t),\rho(\tau)\bm)\geq0$, where the equality holds if and only if $\rho(t)=\rho(\tau)$, (ii) symmetry $\mathcal{D}\bm(\rho(t),\rho(\tau)\bm)=\mathcal{D}\bm(\rho(\tau),\rho(t)\bm)$, and (iii) the triangle inequality $\mathcal{D}\bm(\rho(t),\rho(\tau)\bm)\leq \mathcal{D}\bm(\rho(t),\rho(t')\bm)+\mathcal{D}\bm(\rho(t'),\rho(\tau)\bm)$.  
	
	It is worth mentioning that the trace distance for two distinguishable states in dynamical evolution can be preserved under time-independent unitary transformations, which suggests that the trace distance in Eq.~\eqref{eq:tradis} is invariant for the quantum state expressed in terms of different time-independent basis.
	However, the trace distance in Eq.~\eqref{eq:tradis} cannot be preserved under time-dependent unitary transformations since the states $\rho(t)$ and $\rho(\tau)$ are related to two different times in dynamical evolution. This means that the trace distance for two evolving states will be changed if we transform the dynamical equation in Eq.~\eqref {eq:masequ} into the interaction picture with respect to the Hamiltonian of the quantum system.
	
	We consider the time evolution of the state $\rho(t)$ of the open quantum system in the time interval $t\in[0, \tau]$.
	The geodesic distance quantified by the trace distance metric $\mathcal{D}\bm(\rho(0),\rho(\tau)\bm)$ is a straight line, namely, the shortest path, that connects the initial state $\rho(0)$ and the final state $\rho(\tau)$, i.e., $\rho(t)=[1-p(t)]\rho(0)+p(t)\rho(\tau)$, where $p(t)$ is any monotonically function of time $t$ obeying $p(0)=0$ and $p(\tau)=1$~\cite{PhysRevA103.022210}.
	The geodesic paths connecting the initial and final states characterized by some other metrics, e.g., QFI and Wigner-Yanase information (WYI) metrics, have also been derived
	by Uhlmann~\cite{RepMathPhys.36.461} and Gibilisco and Isola~\cite{JMathPhys.44.3752}, respectively.
	O' Conner \textit{et al}. made a comparison of three QSLs time bounds derived respectively in terms of the trace distance, QFI and WYI metrics for a qubit system undergoing a generalized amplitude damping channel~\cite{PhysRevA103.022210}. 
	Their comprehensive analysis revealed that, under any fixed evolution path, there is no definite magnitude relation among the QSLs time bounds derived from different metrics and one bound consistently emerged as the tighter one than the others which is closely dependent on the parameter of the initial state.
	
	Different from the parameterization treatment, we here divide 
	the time interval $[0, \tau]$, in which the quantum system evolves from an initial state $\rho(0)$ to the final state $\rho(\tau)$, into $n$  small sub-intervals $\Delta t=\tau/n$.
	The total distance in the dynamical evolution of the quantum system that connects the initial state $\rho(0)$ and final state $\rho(\tau)$ can be expressed as the sum of the distances in each infinitesimal time interval~\cite{PhysRevLett.110.050402,NewJPhys.24.055003}
	\begin{equation}
		\label{eq:burfis}
		\ell\bm(\rho(0),\rho(\tau)\bm)=\lim_{n \rightarrow\infty}\sum_{k=1}^{n}\mathcal{D}\bm(\rho(t_{k-1}),\rho(t_{k})\bm),
  \end{equation}
where $t_{k}=k\Delta t$.
Based on the fact that the geodesic distance provides a lower bound of the total distance in dynamical evolution, i.e.,  $\mathcal{D}\bm(\rho(0),\rho(\tau)\bm)\leq\ell\bm(\rho(0),\rho(\tau)\bm)$, we can obtain the geometric QSLs time bound for the quantum system evolving from an initial state $\rho(0)$ to the final state $\rho(\tau)$ as
	\begin{equation}
		\label{eq:geoQSL}
		\tau\geq\tau_{\rm{QSL}}=\frac{\mathcal{D}\bm(\rho(0),\rho(\tau)\bm)}{(1/\tau)\int_{0}^{\tau}v(t)dt},
	\end{equation}
	where $v(t)$ is the instantaneous speed of dynamical evolution defined as
	\begin{equation}
		\label{eq:insspe}
		v(t)=\frac{d}{dt}\ell(t)=\frac{1}{2}\|\mathcal{L}\rho(t)\|_{\mathrm{tr}}.
	\end{equation}
	The quantum speed $v(t)$ is completely determined by the dynamical generator containing both unitary and nonunitary contributions.
	In the case of unitary dynamics $\mathcal{L}_{r}\rho(t)=0$, the evolution speed $v(t)$ is a measure of asymmetry relative to time translation and quantifies the coherence of the system associated with the Hamiltonian $H(t)$ since it vanishes for all incoherent states~\cite{NatCommun.5.3821,PhysRevA93.052331}.
	The geometric QSLs time bound of an open quantum system in Eq.~\eqref{eq:geoQSL} closely depends on the dynamical evolution, such as, the dynamical generator $\mathcal{L}\rho(t)$ and the evolution time $\tau$.
	It is saturated if the system always evolves along the geodesic that connects the initial state $\rho(0)$ and final state $\rho(\tau)$.
	
	\section{QSLs in dephasing dynamics in thermal environments}
	\label{sec:theo2}
	In the following, we will study the QSLs of a qubit system interacting with a thermal dephasing environment with a spectral density that can be effectively engineered.
	
	\subsection{Dephasing dynamics of a qubit system coupled to a bosonic thermal environment}
	
	We first consider the dephasing dynamics of a qubit system coupled to a thermal environment with an infinite bosonic field modes.
	This physical model was initially proposed with the aim of investigating the impact of decoherence in quantum computers~\cite{PhysRevA51.1015}. Subsequently, it has been widely employed to explore the decoherence dynamics of open quantum systems.
	The total system, which is composed of the qubit system and the bosonic environment, is described by the Hamiltonian as follows~\cite{Breuerbook,Schlosshauerbook}
	\begin{equation}
		\label{eq:sinintHam}
		\mathcal{H}=\frac{\hbar}{2}\omega_{0}\sigma_{z}+\sum_{k}\hbar\omega_{k}a_{k}^{\dag}a_{k}+\sigma_{z}\sum_{k}\left(g_{k}a_{k}^{\dag}+g_{k}^{*}a_{k}\right),
	\end{equation}
	where $\omega_{0}$ is the transition frequency between the qubit states $|1\rangle$ and $|0\rangle$, $\omega_{k}$ is the frequency of the environmental modes, $a_{k} (a_{k}^{\dag})$ are the creation (annihilation) operators and $g_{k}$ are the coupling constants between environmental modes and the qubit system through the Pauli matrix $\sigma_{z}$.
	
The qubit system undergoes dephasing dynamics since the interaction Hamiltonian commutes with the system Hamiltonian. 
The dynamical evolution of the dephasing qubit is governed by the quantum master equation
	\begin{equation}
		\label{eq:depmasequ}
		\frac{d}{dt}\rho(t)=-\frac{i}{2}[\omega_{0}\sigma_{z},\rho(t)]+\frac{\gamma(t)}{2}\Big[\sigma_{z}\rho(t)\sigma_{z}-\frac{1}{2}\{\sigma_{z}\sigma_{z},\rho(t)\}\Big].
	\end{equation}
where $\gamma(t)$ denotes the dephasing rate.
By assuming that the states of the qubit system and the environment are  initially independent of each other, namely, $\rho_{\rm tot}(0)=\rho(0)\otimes\rho_{E}$, and performing the continuum limit of the environmental modes  $\sum_{k}|g_{k}|^{2}\rightarrow\int d\omega J(\omega)\delta(\omega_{k}-\omega)$ with $J(\omega)$ denoting the spectral density of the environmental coupling, 
	the dynamics of the qubit system can be exactly solved.
	The diagonal elements of the reduced density matrix do not evolve with time while its off-diagonal elements are time dependent 
	\begin{equation}
		\label{eq:elesinden}
		\begin{split}
			\rho_{11}(t)&=\rho_{11}(0),\\
			\rho_{00}(t)&=1-\rho_{11}(t),\\
			\rho_{10}(t)&=\rho_{01}^{*}(t)=e^{-i\omega_{0}t}F(t)\rho_{10}(0),
		\end{split}
	\end{equation}
	where $F(t)=e^{-D(t)}$ is the dephasing factor with the decay function $D(t)$ expressed as
	\begin{equation}
		\label{eq:decfun0}
		D(t)=\int_{0}^{\infty}J(\omega)\frac{1-\cos(\omega t)}{\hbar^{2}\omega^2}\coth\frac{\hbar\omega}{2k_{B}T}d\omega,
	\end{equation}
with $T$ being the environmental temperature.
By taking the time derivative of the decay function $D(t)$, we can obtain the dephasing rate $\gamma(t)$ in Eq.~\eqref{eq:depmasequ} as
\begin{equation}
	\label{eq:deprat0}
	\gamma(t)=\int_{0}^{\infty} J(\omega)\frac{\sin(\omega t)}{\hbar^{2}\omega}\coth\frac{\hbar\omega}{2k_{B}T}d\omega.
\end{equation}
	
	The dynamics of the qubit system is closely associated with the spectral density of the environment $J(\omega)$.
	In the physical realizations of different dephasing processes, the environmental spectral density can assume various forms~\cite{Weissbook}. Moreover, the structure of the environmental spectral density can be effectively modified by means of environment engineering techniques~\cite{NewJPhys.11.103055,PhysRevA87.012127}.
	We consider here a common class of environmental spectral density characterized by the family of Ohmic-like spectra~\cite{PhysRevA87.010103,PhysRevA89.024101,PhysRevA90.052103,PhysRevA97.012126,PhysRevA101.032112}
	\begin{equation}
		\label{eq:speden}
		J(\omega)=\frac{\omega^s}{\omega_{c}^{s-1}}e^{-\omega/\omega_{c}},
	\end{equation}
	where $\omega_{c}$ is the characteristic cutoff frequency of the environmental modes and $s$ represents the Ohmicity parameter.
	By changing the Ohmicity parameter $s$, the environmental coupling can be classified into  sub-Ohmic $(0<s<1)$, Ohmic $(s=1)$ and super-Ohmic $(s>1)$ spectra, respectively.
	The Ohmicity parameter $s$ and the characteristic cutoff frequency $\omega_{c}$ determine the shape of the spectral density $J(\omega)$.
	For instance, in the case $s=1$, $J(\omega)$ increases approximately linearly in the frequency range $\omega<\omega_{c}$ and decreases in the frequency range $\omega>\omega_{c}$.
	It is feasible to engineer the Ohmicity parameter of the environmental spectral density during the simulation of dephasing dynamics in trapped ultracold atoms~\cite{PhysRevA84.031602}.

	For the case of the environment in its vacuum state initially (i.e., the environmental temperature $T=0$), the decay function $D(t)$ can be analytically written as
	\begin{equation}
		\label{eq:decfunT0}
		D(t)=\frac{1}{\hbar^{2}}\left\{
		\begin{aligned}
			&\frac{1}{2}\ln(1+\omega_{c}^{2}t^{2}),\\
			&\left\{1-\frac{\cos[(s-1)\arctan(\omega_{c}t)]}{(1+\omega_{c}^{2}t^{2})^{(s-1)/2}}\right\}\frac{\Gamma(s)}{s-1},
		\end{aligned}\right.
	\end{equation}
	respectively for the cases $s=1$ and $s\neq1$, where $\Gamma(s)=\int_{0}^{\infty}e^{-x}x^{s-1}dx$ denotes the Gamma function.
	Correspondingly, the dephasing rate $\gamma(t)$ can be analytically expressed as
	\begin{equation}
		\label{eq:depratT0}
		\gamma(t)=\frac{\omega_{c}\sin[s \arctan(\omega_{c}t)]}{\hbar^{2}(1+\omega_{c}^{2}t^{2})^{s/2}}\Gamma(s).
	\end{equation}
In the high environmental temperature limit $\omega\ll\omega_{T}$ with $\omega_{T}=k_{B}T/\hbar$ being the characteristic thermal frequency, 
the decay function can be analytically written as
$D(t)=2\omega_{T}\alpha(t)/(\hbar^{2}\omega_{c})$ with 
\begin{align}
	\label{eq:alphat}
	\alpha(t)=\left\{
	\begin{aligned}
		&\frac{1}{2}\left[\pi t-\frac{1}{\omega_{c}}\ln(1+\omega_{c}^{2}t^{2})-2t\arctan\frac{1}{\omega_{c}t}\right],\\
		&\frac{1}{2}\ln(1+\omega_{c}^{2}t^{2}),\\
		&\left\{1-\frac{\cos[(s-2)\arctan(\omega_{c}t)]}{(1+\omega_{c}^{2}t^{2})^{(s-2)/2}}\right\}\frac{\Gamma(s)}{(s-2)(s-1)},\\
	\end{aligned}\right.
\end{align}
for the case $s=1$, the case $s=2$ and the cases $s\neq1$ and $s\neq2$, respectively.	
The corresponding dephasing rate $\gamma(t)$ can be analytically expressed as
\begin{equation}
	\label{eq:deprathigT}
	\gamma(t)=\frac{2\omega_{T}}{\hbar^{2}}\left\{
	\begin{aligned}
		&\arctan(\omega_{c}t),\\
		&\frac{\sin[(s-1)\arctan(\omega_{c}t)]}{(1+\omega_{c}^{2}t^{2})^{(s-1)/2}}\frac{\Gamma(s)}{s-1},
	\end{aligned}\right.
\end{equation}
for the cases $s=1$ and $s\neq1$, respectively.
In the case of the environment in a thermal equilibrium state with an arbitrary finite temperature $T>0$, the decay function $D(t)$ and the dephasing rate $\gamma(t)$ can be computed numerically.
	
The dephasing rate $\gamma(t)$ can be used to distinguish the Markovian and non-Markovian regions for the dephasing dynamics of the qubit system
which stem from weak and strong couplings between the system and the environment, respectively~\cite{RepProgPhys.77.094001,PhysRevA94.042110}. 
The dynamics of the qubit system is Markovian if the dephasing rate $\gamma(t)>0$ at any time $t$ whereas it is non-Markovian if the dephasing rate $\gamma(t)<0$ in some time intervals.
Based on the expressions in Eqs.~\eqref{eq:depratT0} and~\eqref	{eq:deprathigT} and the mathematical property $0<\arctan(\omega_{c}t)<\pi/2$ for $t>0$, we can identify the critical values of the Ohmicity parameter $s$ for the dephasing rate $\gamma(t)$  being always positive and taking temporarily negative values in the cases of zero environmental temperature $T=0$ and high environmental temperature limit $T\gg\hbar\omega/k_{B}$, respectively.

In the former case, the dephasing rate $\gamma(t)$ always takes positive values for $0<s\leq s_{\mathrm{cri}}=2$ whereas it can be negative in some time intervals for $s>2$.
In the latter case, the dephasing rate $\gamma(t)$ is always positive for $0<s\leq s_{\mathrm{cri}}=3$ whereas it can be temporarily negative for $s>3$.
It is worth mentioning that in the case of the environment with a finite temperature $T>0$, the critical value of the Ohmicity parameter $s$ for clarifying Markovian and non-Markovian regions satisfy $2<s_{\mathrm{cri}}\leq3$ and it increases monotonically the environmental temperature $T$~\cite{PhysRevA87.010103}.
This indicates that in the case of weak environmental coupling $0<s\leq2$, the dephasing dynamics of the qubit system always exhibits Markovian characteristics while in the case of strong environmental coupling $s>3$, it always displays non-Markovian characteristics.
In contrast, the dynamical dephasing of the system undergoes a transition from Markovian to non-Markovian in the case of intermediate environmental coupling $2<s_{\mathrm{cri}}\leq3$
which closely depends on the environmental temperature $T$.

Numerous approaches have been proposed to quantitatively evaluate the degree of non-Markovian characteristics in the dynamics of open quantum systems (see, e.g., review papers~\onlinecite{RepProgPhys.77.094001,RevModPhys.88.021002}).
One of the most widely adopted approaches, developed by Breuer, Laine and Piilo (BLP), quantifies the non-Markovianity as the cumulative maximum backflow of information from the environment to the quantum system.
Based on the framework established by BLP, the non-Markovianity in the quantum dynamics of an open system is defined as~\cite{PhysRevLett.103.210401}
\begin{equation}
	\label{eq:nonMar}
	\mathcal{N}=\max\limits_{\rho_{1,2}(0)}\int_{\sigma>0}\sigma\textbf{(}t,\rho_{1,2}(0)\textbf{)}dt.
\end{equation}
where $\sigma\textbf{(}t,\rho_{1,2}(0)\textbf{)}=dD\textbf{(}\rho_{1}(t),\rho_{2}(t)\textbf{)}/dt$
denotes the rate of change of the trace distance with respect to time $t$, 
and the maximum is obtained by maximizing over all initial states \(\rho(0)\), with the integration extended over all time intervals in which $\sigma\textbf{(}t,\rho_{1,2}(0)\textbf{)}>0$.
By taking optimization over all pairs of initial states, the non-Markovianity in the dynamics of the dephasing qubit can be expressed as~\cite{SciRep.10.88} 
\begin{equation}
	\label{eq:non-Mar}
	\mathcal{N}=-\int_{0 \ \gamma(t)<0}^{\infty}\gamma(t)|F(t)|dt,
\end{equation}
where the optimal pair of the initial states is chosen as the maximally coherent states $|\psi_{\pm}\rangle=\left(|1\rangle\pm|0\rangle\right)/\sqrt{2}$.

On the basis of the expressions of the decay function $D(t)$ in Eqs.~\eqref{eq:decfunT0} and~\eqref{eq:alphat},  we can analyze the dephasing factor in the long time limit under different environmental temperature conditions.
When the environment is at zero temperature, in the long time limit the dephasing factor $F(\infty)$ approaches zero in the dynamical region $0<s\leq1$ while it reaches the steady value $F(\infty)=e^{-\Gamma(s-1)/\hbar^{2}}$ in the dynamical region $s>1$.
Conversely, in the high temperature limit of the environment, the steady-state dephasing factor converges to $F(\infty)=e^{-2\omega_{T}\Gamma(s-2)/(\hbar^{2}\omega_{c})}$  in the dynamical region $s>2$ while it vanishes in the long time limit in the dynamical region $0<s\leq2$.
This indicates that the qubit system exhibits coherence trapping in these dynamical regions where the dephasing factor takes nonzero steady value in the long time limit.
As the temperature of the environment increases from zero, the critical  Ohmicity parameter $s_{\mathrm{cri}}$ corresponding to the lower boundary of the dynamical region, where the dephasing factor attains a nonzero steady value in the long time, exhibits a monotonic increase from 2 to 3.
In the coherence trapping region, for a given Ohmicity parameter $s$, the steady value of the dephasing factor $F(\infty)$ decreases gradually with the increase of the environmental temperature $T$~\cite{PhysRevA89.024101}.

Based on the above analysis, the dynamical regions for non-Markovian characteristics and coherence trapping of the dephasing qubit are determined by the Ohmicity parameter $s$.
It is worth noting that the two dynamical regions are not consistent and the dynamical region for non-Markovian characteristics is smaller than that for coherence trapping.
In the dynamical region where coherence trapping does not occur, the  dynamics of the dephasing qubit is always Markovian.
By contrast, in the coherence trapping region, the dephasing dynamics of the qubit system can be Markovian or non-Markovian.
In other words, in the non-Markovian region, the dephasing qubit always exhibits coherence trapping while coherence trapping can occur or not in the Markovian region.
Addis \textit{et al}. showed that the low-frequency part of the environmental spectra $\omega/\omega_{c}\ll1$, determine the occurrences of both the 
information backflow and coherence trapping while its high-frequency band 
$\omega/\omega_{c}\gg1$, rule the maximum coherence in the steady state limit~\cite{PhysRevA89.024101}.

\subsubsection{Influence of environmental temperature on the dephasing dynamics}
	
	\begin{figure}[ht]
		\centering
		\includegraphics[width=3.6in]{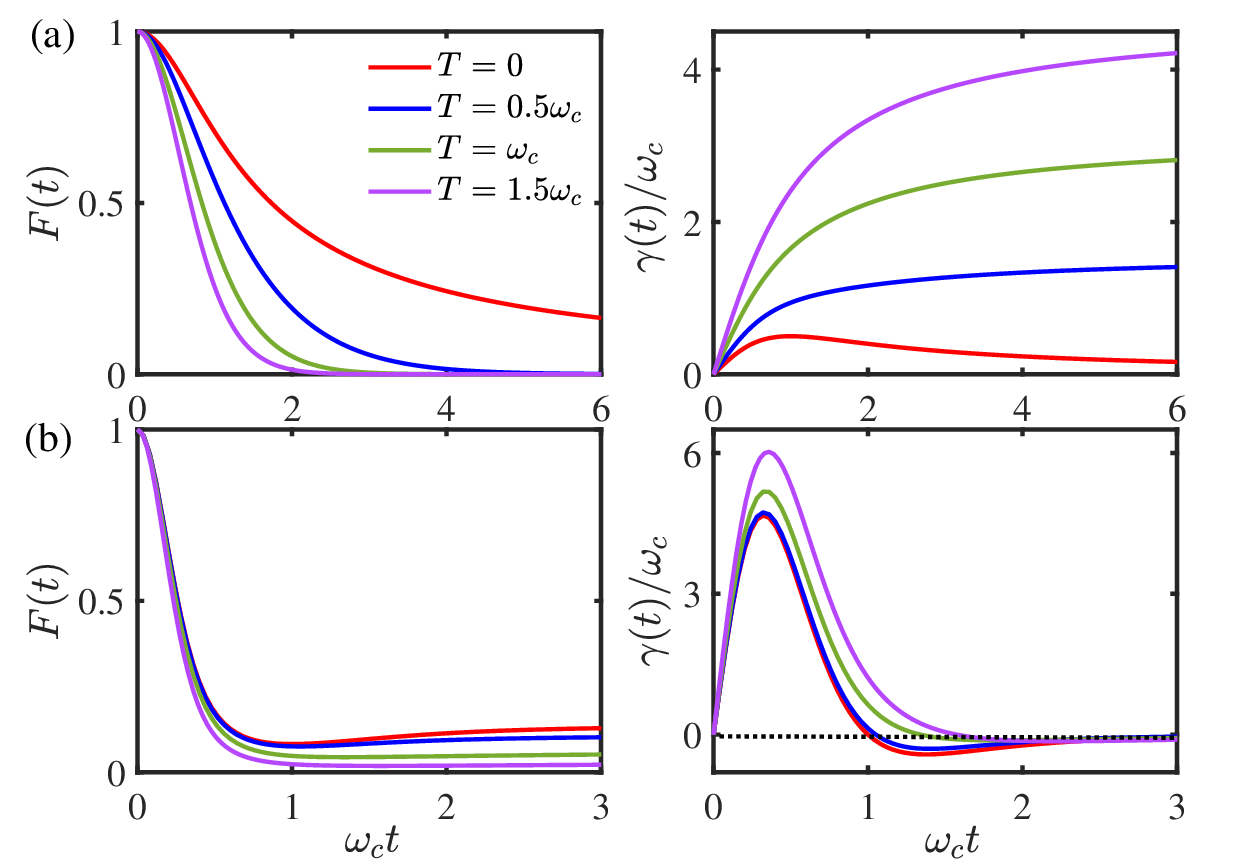}
		\caption{(Color online) Time evolution of the dephasing factor $F(t)$ (left column) and the dephasing rate $\gamma(t)$ (right column) for different values of environmental temperature $T$ in (a) weak coupling region $s=1$ and (b) strong coupling region  $s=4$. Here and in the figures plotted below, we set $\hbar=k_{B}=1$.}
		\label{fig:dephasingT}
	\end{figure}
	
	Figure~\ref{fig:dephasingT} shows the time dependence of the dephasing factor $F(t)$ and the dephasing rate $\gamma(t)$ in the weak ($s=1$) and strong ($s=4$) coupling regions for various environmental temperatures $T$. These plots provide insight into the impact of the thermal noise on the dephasing dynamics of a quantum system.
	Here and in the following, we scale the physical parameters with
	the characteristic cutoff frequency  $\omega_{c}$ to show the numerical results. For instance, we set the evolution time in units of $\omega_{c}^{-1}$ and the environmental temperature in units of $\omega_{c}$.
	
	As depicted in Fig.~\ref{fig:dephasingT} (a) in the weak coupling region, the dephasing factor $F(t)$ decreases monotonically and tends to zero in the long time limit. This indicates a continuous and irreversible loss of coherence in the system. The dephasing rate $\gamma(t)$ remains strictly positive for all evolution times $t > 0$, reflecting that the dynamics are Markovian throughout. With the increase of the environmental temperature $T$, the decay in the dephasing factor $F(t)$ becomes more remarkable and the corresponding dephasing rate $\gamma(t)$ increases. Physically, this behavior demonstrates that higher environmental temperatures enhance thermal noise, which accelerates the dynamical decoherence. This confirms that the  environmental weak coupling leads to Markovian dephasing dynamics, where the memory effects in the dynamical evolution of the qubit system do not occur.
	
	In contrast, Fig.~\ref{fig:dephasingT} (b) depicts the dephasing dynamics in the strong coupling region. The dephasing factor $F(t)$ demonstrates nonmonotonic behavior, indicative of coherence revivals. These revivals correspond to non-Markovian dynamics, where the system temporarily regains some of its coherence due to memory effects induced by the environment. The dephasing rate $\gamma(t)$ can take negative values during certain time intervals, further confirming the non-Markovian characteristics of the dynamics. 
	In the long time limit, the dephasing factor $F(t)$ approaches a steady value rather than decaying to zero. Concurrently, the dephasing rate $\gamma(t)$ converges to zero, indicating that the dephasing process stabilizes over time.
	As the environmental temperature $T$ rises, the non-monotonic behavior within the dephasing factor $F(t)$ gradually becomes increasingly less conspicuous. This indicates that the non-Markovian characteristics in the dephasing dynamics gets suppressed by the thermal noise.
	This phenomenon can be attributed to two factors: first, the critical time at which the dephasing rate $\gamma(t)$ starts to take negative values is gradually being postponed; second, the time interval during which $\gamma(t)<0$ is becoming smaller.
    In addition,  the steady value of the dephasing factor $F(t)$ in the long time decreases with the increase of the environmental temperature $T$. 
    This implies that, similar to the weak coupling region, the thermal noise can enhance the dynamical decoherence of the qubit system. 
	This is because the increase of the environmental temperature $T$ 
	not only raises the dephasing rate $\gamma(t)$ but also prolongs the duration during which it remains positive.
	
	\begin{figure}[ht]
		\centering
		\includegraphics[width=3.6in]{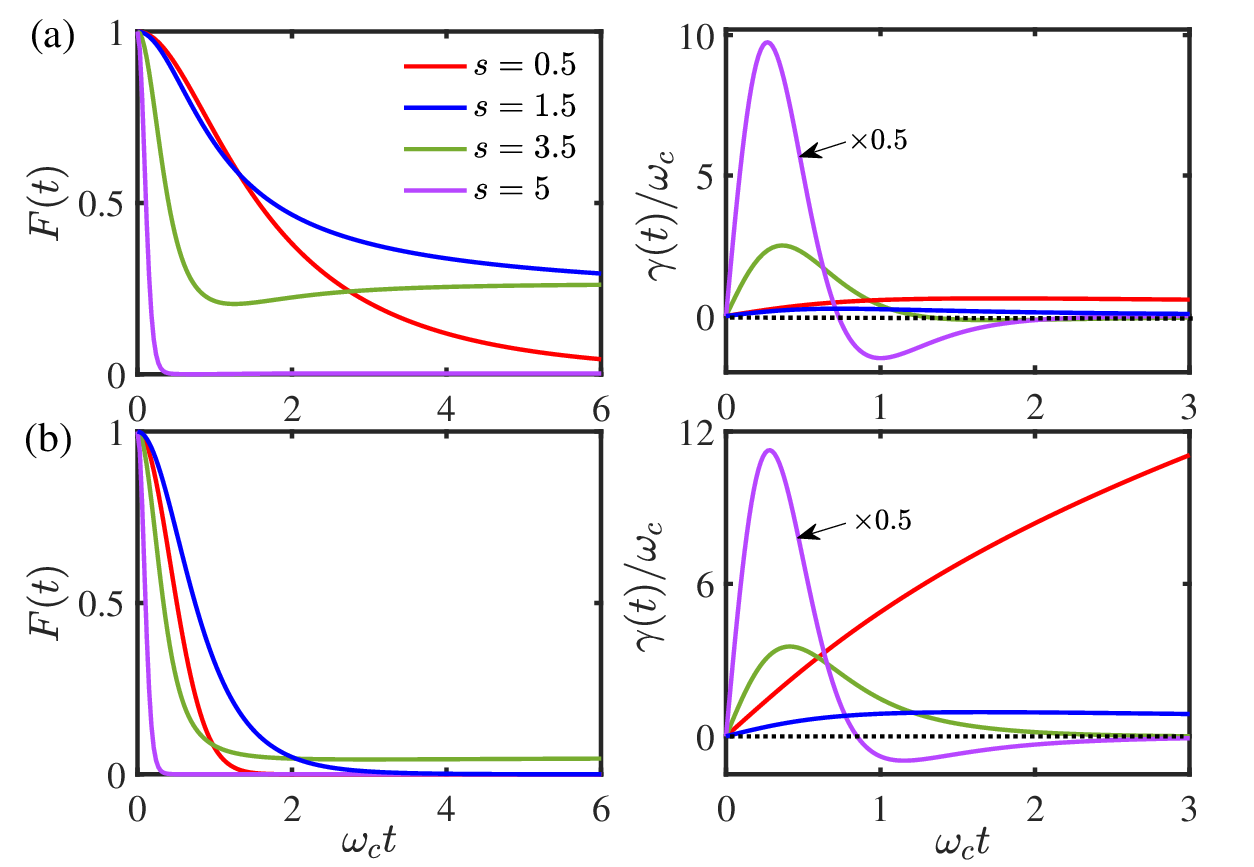}
		\caption{(Color online) Time evolution of the dephasing factor $F(t)$ (left column) and the dephasing rate $\gamma(t)$ (right column) for different values of Ohmicity parameter $s$ with the environmental temperature (a) $T=0$ and (b) $T=1.5\omega_{c}$.}
		\label{fig:dephasings}
	\end{figure}
	
	Figure~\ref{fig:dephasings} displays the time evolution of the dephasing factor $F(t)$ and the dephasing rate $\gamma(t)$ for different values of the Ohmicity parameter $s$ at environmental temperatures: $T=0$ [Fig.~\ref{fig:dephasings} (a)] and $T=1.5\omega_{c}$ [Fig.~\ref{fig:dephasings} (b)], respectively.
	
	As shown in Fig.~\ref{fig:dephasings} (a), the dephasing factor $F(t)$ exhibits monotonically decaying behavior, tending to zero in the long-time limit for $0 < s \leq s_{\mathrm{cri}}$, where the critical value of the Ohmicity parameter is $s_{\mathrm{cri}} = 2$. In this region, the dephasing rate $\gamma(t)$ remains strictly positive throughout the evolution, indicating that the dynamics are Markovian, with no memory effects influencing the coherence of the qubit system.
	In contrast, for $s>s_{\mathrm{cri}}$, the dephasing factor $F(t)$ shows nonmonotonic behavior and tends to a nonzero value in the steady state. In this region, the dephasing rate $\gamma(t)$ takes negative values during specific time intervals, clearly indicating non-Markovian dynamics. 
	Similar dynamical behavior is depicted in Fig.~\ref{fig:dephasings} (b) for the case of the environmental temperature $T=1.5\omega_{c}$. 
	However, the critical value of the Ohmicity parameter $s_{\mathrm{cri}}$ increases and $2<s_{\mathrm{cri}}<3$ compared to the zero-temperature case. This indicates that at higher temperatures $T$, stronger environmental coupling (larger $s$) is required to induce non-Markovian dynamics. Physically, the thermal noise associated with finite temperatures tends to suppress the memory effects induced by the environment, thereby delaying the transition to non-Markovian dynamics.
	
   \subsubsection{Impact of environmental temperature on the  coherence trapping and non-Markovianity of the dephasing qubit}
	
	\begin{figure}[ht]
		\centering
		\includegraphics[width=3.5in]{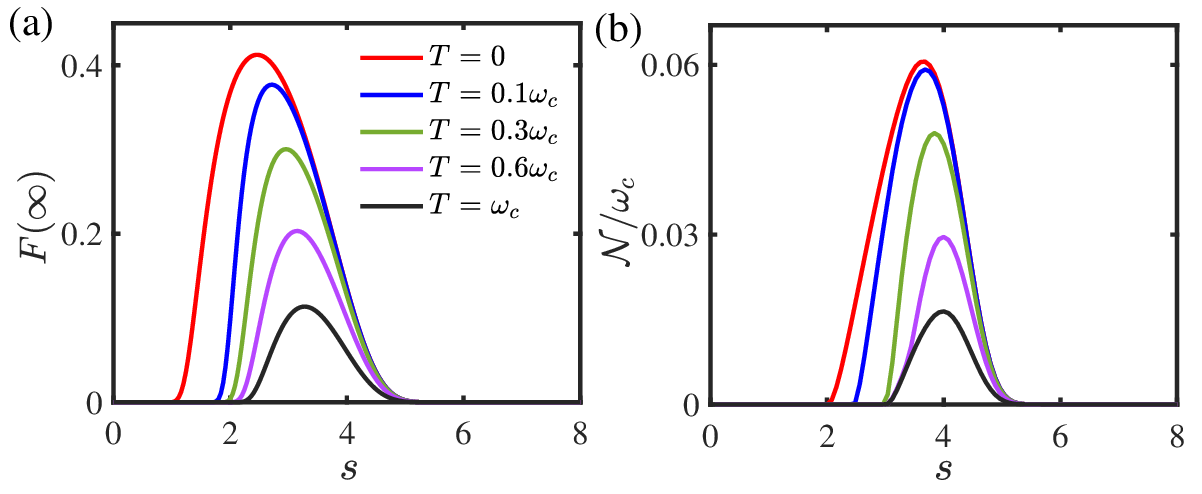}
		\caption{(Color online) (a) The dephasing factor in the steady state limit $F(\infty)$ and (b) the non-Markovianity $\mathcal{N}$ of dephasing dynamics of the qubit system as a function of the Ohmicity parameter $s$ for different environmental temperatures $T$.}
		\label{fig:cohnonmar}
	\end{figure}
	
	We plot Fig.~\ref{fig:cohnonmar} to illustrate the influence of the environmental temperature $T$ on the steady value of the dephasing factor $F(\infty)$ and the non-Markovianity $\mathcal{N}$ of the dephasing qubit. 
	In specific dynamical regions, the steady dephasing factor $F(\infty)$ [Fig.~\ref{fig:cohnonmar} (a)] and the non-Markovianity $\mathcal{N}$ [Fig.~\ref{fig:cohnonmar} (b)] exhibit nonmonotonic dependence on the Ohmicity parameter $s$, initially increasing and then decreasing as $s$ increases, for various environmental temperatures $T$.
	Additionally, for any environmental temperature $T$, both quantities asymptotically vanish in the dynamical region $s>5$.
	For the environment at zero temperature $T=0$, the dephasing factor approaches a nonzero steady value in the dynamical region $s>1$ while the depahsing qubit exhibits non-Markovian characteristics in the dynamical region $s>2$.
	The critical Ohmicity parameters $s_{\mathrm{cri}}$ where coherence trapping and non-Markovian characteristics of the dephasing qubit start to occur increase with the temperature of the environment $T$.
	As the environmental temperature $T$ increases from zero to the high-temperature limit, the critical Ohmicity parameter $s_{\mathrm{cri}}$ gradually increases from 1 to 2  in the dynamical region for the dephasing qubit exhibiting coherence trapping as shown in Fig.~\ref{fig:cohnonmar} (a) while it increases from 2 to 3 in the dynamical region for the dephasing qubit displaying non-Markovian characteristics as depicted in Fig.~\ref{fig:cohnonmar} (b).
	In these specific dynamical regions, for a given Ohmicity parameter $s$, both the steady dephasing factor $F(\infty)$ and the non-Markovianity $\mathcal{N}$ decrease gradually as the increase of the environmental temperature $T$.
	Furthermore, the Ohmicity parameter $s$ for the maximum steady value of the dephasing factor $F_{\mathrm{max}}(\infty)$ and non-Markovianity $\mathcal{N}_{\mathrm{max}}$ increases with the environmental temperature $T$.
	These results indicate that the increase in environmental temperature 
	can suppress  coherence trapping and non-Markovian characteristics in the dephasing dynamics and can reduce the dynamical regions of the dephasing qubit exhibiting these two 
	phenomena.
	
	\subsection{Quantum speed limits of the qubit system under dynamical dephasing}
	
	Based on the dephasing dynamics in terms of Eqs.~\eqref{eq:elesinden} and~\eqref{eq:depmasequ}, we can obtain the geodesic distance for the qubit system evolving from an initial state $\rho(0)$ to the final state $\rho(\tau)$ as
	\begin{equation}
		\label{eq:geodisqubsyst}
		\begin{split}
			\mathcal{D}\bm(\rho(0),\rho(\tau)\bm)
			&=|\rho_{10}(0)||e^{-i\omega_{0}\tau}F(\tau)-1|\\
			&=\frac{1}{2}\mathcal{C}(0)\sqrt{[F(\tau)-\cos(\omega_{0}\tau)]^{2}+\sin^{2}(\omega_{0}\tau)},
		\end{split}
	\end{equation}
	and the instantaneous speed of dynamical evolution
	\begin{equation}
		\label{eq:insspequbsyst}
		\begin{split}
			v(t)&=|\rho_{10}(0)|\left|\frac{d}{dt}\left[e^{-i\omega_{0}t}F(t)\right]\right|\\
			&=\frac{1}{2}\mathcal{C}(0)\sqrt{\omega_{0}^{2}+\gamma^{2}(t)}F(t),
		\end{split}
	\end{equation}
	where $\mathcal{C}(0)$ is the coherence of the quantum system initially in terms of $l_{1}$ norm.
	Obviously, for the dephasing qubit with initial coherence $\mathcal{C}(0)\neq0$, the geodesic distance between the initial and final states and the evolution  speed of the qubit are closely associated with both the Hamiltonian of the system and the dephasing dynamics induced by the environment.
	
	It is worth noting that the transition frequency $\omega_{0}$ plays essential contributions to both the geodesic distance in Eq.~\eqref{eq:geodisqubsyst} and the instantaneous speed in Eq.~\eqref{eq:insspequbsyst}.
	In the dynamical region where the coherence of the dephasing qubit is lost completely in the long time limit, i.e., $F(\infty)\rightarrow0$, the dynamics of the qubit system is Markovian and the dephasing factor $F(t)$ decays monotonically as time passes.
	The geodesic distance $\mathcal{D}\bm(\rho(0),\rho(t)\bm)$ first increases from zero ($t=0$), then displays periodic oscillations with a damped amplitude and the period $\mathcal{T}=2\pi/\omega_{0}$, and converges to $1/2\mathcal{C}(0)$ in the long time limit.
	In this dynamical region, the dephasing rate $\gamma(t)$ is always positive and as time evolves it first increases and then approaches to a steady value as shown in Fig.~\ref{fig:dephasingT}.
	The instantaneous speed $v(t)$ first increases from $1/2\omega_{0}\mathcal{C}(0)$ at $t=0$, then decreases as time passes and vanishes in the long time limit.
	In the coherence trapping region, namely, $F(\infty)\neq0$, the geodesic distance $\mathcal{D}\bm(\rho(0),\rho(t)\bm)$ first increases from zero  at $t=0$, and then displays periodic oscillations between the middle-line $1/2\mathcal{C}(0)\sqrt{1+F^{2}(\infty)}$ with the amplitude $\mathcal{A}=1/2\mathcal{C}(0)F(\infty)$ and the period $\mathcal{T}$.
	In this dynamical region, the instantaneous speed $v(t)$ first increases, then decreases as time evolves and approaches to a nonzero steady value in the long time limit.
	
	For the qubit system with initial coherence $\mathcal{C}(0)\neq0$ and  based on the geometric bound in Eq.~\eqref{eq:geoQSL}, the QSLs time for the dephasing qubit evolving from an initial state $\rho(0)$ to the final state $\rho(\tau)$ can be expressed as
	\begin{equation}
		\label{eq:geoQSL1qubsys}
		\tau\geq\tau_{\rm{QSL}}	=\frac{\sqrt{[F(\tau)-\cos(\omega_{0}\tau)]^{2}+\sin^{2}(\omega_{0}\tau)}}{(1/\tau)\int_{0}^{\tau}\sqrt{\omega_{0}^{2}+\gamma^{2}(t)}F(t)dt}.
	\end{equation}
	The QSLs time for the qubit system under dynamical dephasing is independent of the initial state due to the fact that the geodesic distance in Eq.~\eqref {eq:geodisqubsyst} and the evolution speed in Eq.~\eqref{eq:insspequbsyst} are both directly proportional to the initial coherence of the system.
	
	There is a competition between contributions arising from the Hamiltonian of the qubit system and the dephasing dynamics induced by the environment that saturates the QSLs time bound in Eq.~\eqref{eq:geoQSL1qubsys}.
	The geometric QSLs for the dephasing qubit arising from the bound between the geodesic and total distances connecting the initial and final states, namely,  $\mathcal{D}\bm(\rho(0),\rho(\tau)\bm)\leq\ell\bm(\rho(0),\rho(\tau)\bm)=\int_{0}^{\tau}v(t)dt$ is equivalent to the mathematical inequality
	\begin{equation}
		\label{eq:geoQSLssat}
		\begin{split}
			|e^{-i\omega_{0}\tau}F(\tau)-1|&=\left|\int_{0}^{\tau}\frac{d}{dt}\left[e^{-i\omega_{0}t}F(t)\right]dt\right|\\
			&\leq\int_{0}^{\tau}\left|\frac{d}{dt}\left[e^{-i\omega_{0}t}F(t)\right]\right|dt.
		\end{split}
	\end{equation}
	The inequality in the second line of  Eq.~\eqref{eq:geoQSLssat} can be saturated only in the limit case $\omega_{0}\rightarrow0$ and under the condition that the dephasing factor $F(t)$ decays monotonically, namely, the dephasing rate $\gamma(t)>0$ in the time interval $(0, \tau)$.
	Therefore, in the Markovian region, the contribution of the transition frequency $\omega_{0}$ makes the QSLs time bound unsaturated while it is the 
	combined contributions of the transition frequency $\omega_{0}$ and the information backflow from the environment that give rise to the unsaturation of the QSLs time bound in non-Markvoian dynamical region.
	
	For simplicity and without loss of generality, in the following we will display the influences of the environmental temperature and spectral density on QSLs time by setting $\omega_{0}=\omega_{c}$.
	Evidently, according to Eq.~\eqref{eq:geoQSL1qubsys}, the QSLs time is strongly dependent on the transition frequency $\omega_{0}$ of the dephasing qubit.
	However, it is worth mentioning that, the trends of the QSLs time in the case $\omega_{0}\neq\omega_{c}$ are consistent with the case $\omega_{0}=\omega_{c}$.
	The relation between the transition frequency $\omega_{0}$ and the cutoff frequency $\omega_{c}$ of the environmental spectral density merely affects the value of $\tau_{\mathrm{QSL}}$ and the duration required for $\tau_{\mathrm{QSL}}$ to attain a specific value.

	\subsubsection{Impact of environmental temperature on QSLs time}
	
	\begin{figure}[ht]
		\centering
		\includegraphics[width=3.5in]{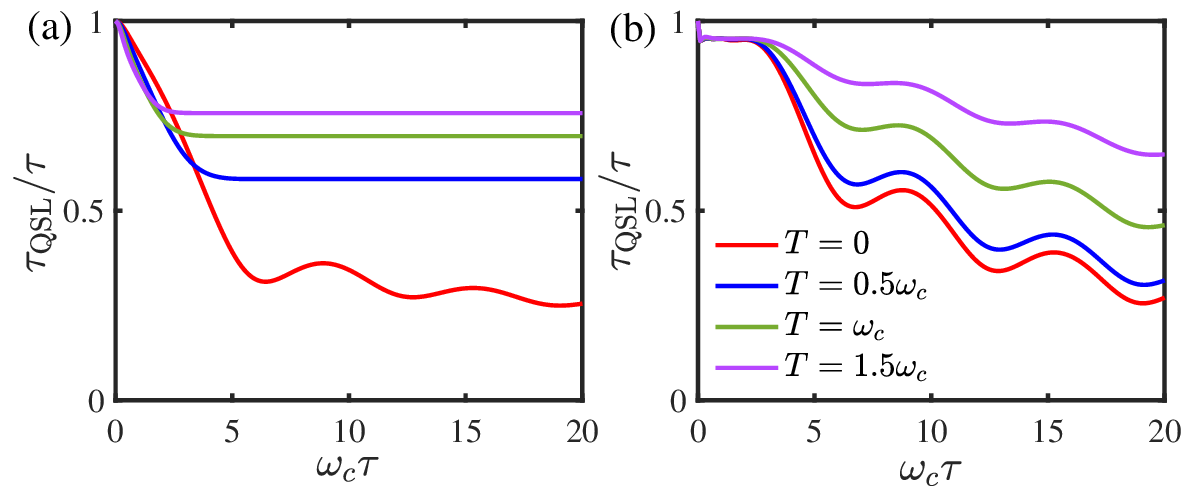}
		\caption{(Color online) The ratio $\tau_{\mathrm{QSL}}/\tau$ as a function of the evolution time $\tau$ for different values of environmental temperature $T$ in (a) weak coupling region $s=1$ and (b) strong coupling region $s=4$. The transition frequency of the qubit system is chosen as $\omega_{0}=\omega_{c}$.}
		\label{fig:QSLT}
	\end{figure}
	
	Figure~\ref{fig:QSLT} illustrates the ratio $\tau_{\mathrm{QSL}}/\tau$ for different values of environmental temperature $T$ as a function of the evolution time $\tau$.
	In the weak coupling region ($s=1$) as shown in Fig.~\ref{fig:QSLT} (a), as time evolves, the ratio $\tau_{\mathrm{QSL}}/\tau$ first undergoes a sharp decrease, followed by damped oscillatory decay, and ultimately approaches a positive steady value in the long time limit. 
	As the environmental temperature $T$ increases, the amplitude in the damped decay decreases and the steady value of the ratio $\tau_{\mathrm{QSL}}/\tau$ increases.
	This is mainly due to the fact that in this dynamical region, the coherence of the dephasing qubit will be lost completely in the long time limit. 
	In the strong coupling region ($s=4$) as depicted in Fig.~\ref{fig:QSLT} (b), as time passes, the ratio $\tau_{\mathrm{QSL}}/\tau$ first decreases sharply and then displays damped periodic oscillations. 
	It is obvious that the ratio $\tau_{\mathrm{QSL}}/\tau$ is unsaturated, i.e., $\tau_{\mathrm{QSL}}/\tau<1$ for any evolution time $\tau>0$, in both the weak and strong coupling dynamical regions. 
	In addition, with the increase of the environmental temperature $T$, the ratio $\tau_{\mathrm{QSL}}/\tau$ increases at relatively in both the weak and strong coupling regions.
	Furthermore, as the environmental temperature $T$ increases, the oscillations in $\tau_{\mathrm{QSL}}/\tau$ become less and less obvious in strong coupling region.
	The results imply that the increase of the environmental temperature can extremely enhance the QSLs time.
	
	\begin{figure}[ht]
		\centering
		\includegraphics[width=3.5in]{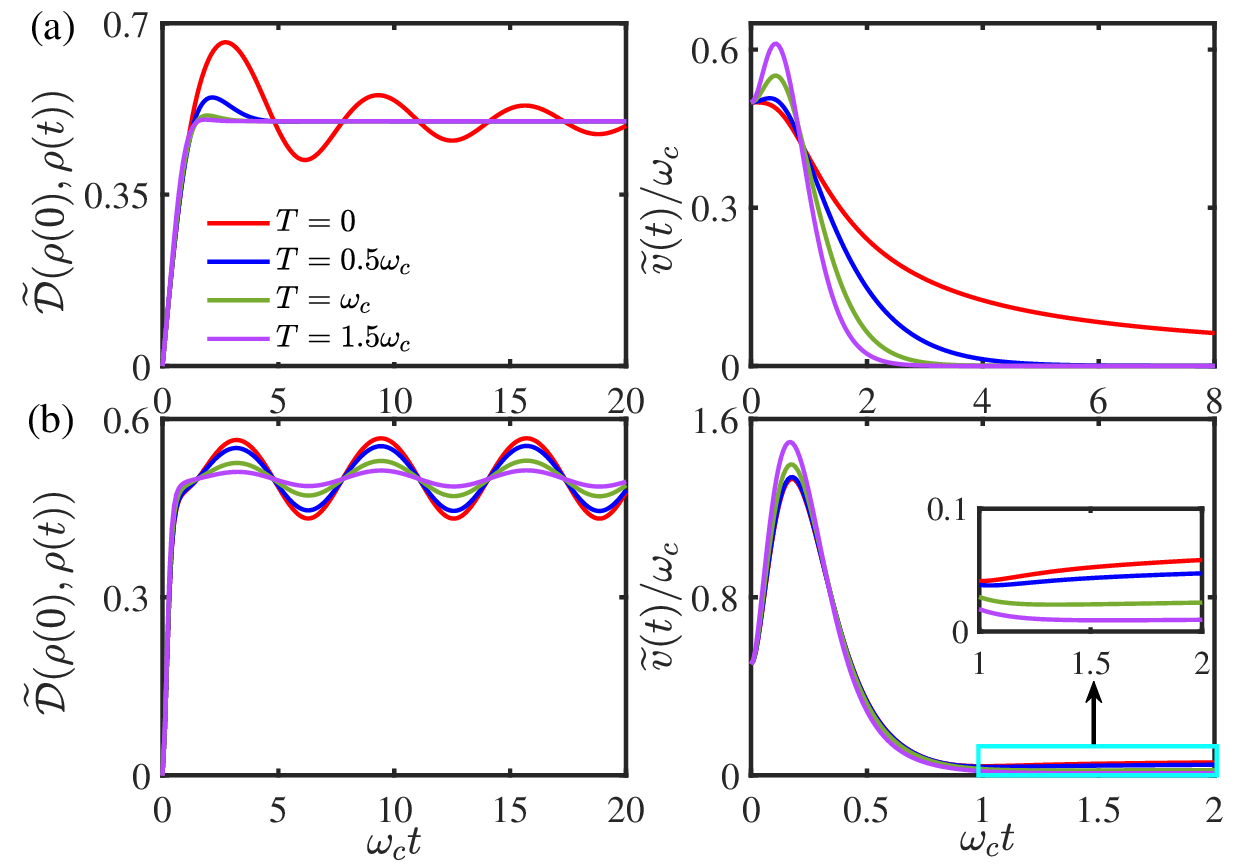}
		\caption{(Color online) Time evolution of the scaled geodesic distance $\widetilde{\mathcal{D}}\bm(\rho(0),\rho(t)\bm)=\mathcal{D}\bm(\rho(0),\rho(t)\bm)/\mathcal{C}(0)$ (left column) and the scaled instantaneous evolution speed $\widetilde{v}(t)=v(t)/\mathcal{C}(0)$ (right column) for different values of environmental temperature $T$ as a function of the evolution time $t$ in (a) weak coupling region $s=1$ and (b) strong coupling region $s=4$. The transition frequency of the qubit system is set as $\omega_{0}=\omega_{c}$.}
		\label{fig:geospe}
	\end{figure}
	
	To explain the phenomena in the QSLs time arising from the environmental temperature as shown in Fig.~\ref{fig:QSLT}, we plot the time evolution of the geodesic distance $\mathcal{D}\bm(\rho(0),\rho(t)\bm)$ and the instantaneous speed of dynamical evolution $v(t)$ for different values of environmental temperature $T$ in Fig.~\ref{fig:geospe}.
	In terms of the dephasing factor depicted in Fig.~\ref{fig:dephasingT} and the expressions in Eqs.~\eqref{eq:geodisqubsyst} and~\eqref{eq:insspequbsyst}, we further analyze the mechanism for the unsaturation of the QSLs time bound.
	In the weak coupling region ($s=1$) as depicted in Fig.~\ref{fig:geospe} (a), the geodesic distance $\mathcal{D}\bm(\rho(0),\rho(t)\bm)$ shows an oscillating increase and asymptotically tends to a same steady value $1/2\mathcal{C}(0)$ for different environmental temperature $T$. The instantaneous evolution speed $v(t)$ first increases, then decreases and approaches zero which means that the qubit system stops evolving in the long time limit. 
	As the environmental temperature $T$ increases, the instantaneous speed $v(t)$ approaches zero more quickly. This reflects that the higher environmental temperature makes the qubit system stop evolving earlier.
	This is primarily attributed to the fact that the dephasing factor $F(t)$ tends to approach zero under the steady state limit as shown in Fig.~\ref{fig:dephasingT} (a).
	Therefore, the ratio $\tau_{\mathrm{QSL}}/\tau$ decreases in short evolution time and then tends to a steady nonzero value for long evolution time in the weak coupling region as depicted in Fig.~\ref{fig:QSLT} (a).
	In the strong coupling region ($s=4$) as displayed in Fig.~\ref{fig:geospe} (b), the geodesic distance $\mathcal{D}\bm(\rho(0),\rho(t)\bm)$ first increases and then becomes a periodically oscillating function of time.
	The instantaneous evolution speed $v(t)$ first increases, then decreases and tends to a nonzero steady value in the long time limit. 
	This is mainly due to the fact that the dephasing factor $F(t)$ approaches a nonzero value in the steady state limit as displayed in Fig.~\ref{fig:dephasingT} (b).
	As the environmental temperature $T$ increases, the steady value of the instantaneous speed $v(t)$ decreases. 
	Consequently, in the strong coupling  region, the ratio $\tau_{\mathrm{QSL}}/\tau$ first decreases sharply, then shows damped oscillations and vanishes in the long time limit as illustrated in Fig.~\ref{fig:QSLT} (b).
	
	\subsubsection{Influence of environmental spectral density on QSLs time}
	
	\begin{figure}[ht]
		\centering
		\includegraphics[width=3.5in]{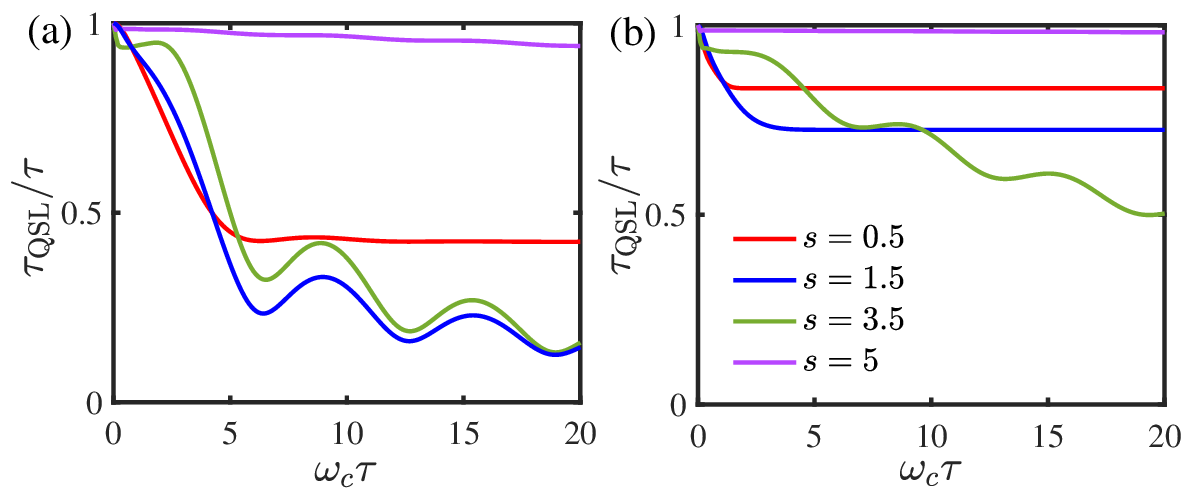}
		\caption{(Color online) The ratio $\tau_{\mathrm{QSL}}/\tau$ for different values of Ohmicity parameter $s$ as a function of the evolution time $\tau$ with the environmental temperatures (a) $T=0$ and (b) $T=1.5\omega_{c}$. The transition frequency of the qubit system is chosen as $\omega_{0}=\omega_{c}$.}
		\label{fig:QSLTs}
	\end{figure}
	
	We show the ratio $\tau_{\mathrm{QSL}}/\tau$ as a function of the evolution time $\tau$ for different values of Ohmicity parameter $s$ with the environmental temperatures $T=0$ and $T=1.5\omega_{c}$ in Fig.~\ref{fig:QSLTs} (a) and  Fig~\ref{fig:QSLTs}. (b), respectively.
	In the dynamical region where coherence trapping does not occur, for $s=0.5$ in Fig.~\ref{fig:QSLTs} (a) and for $s=0.5$ and $s=1.5$ in Fig.~\ref{fig:QSLTs} (b), as time evolves, the ratio $\tau_{\mathrm{QSL}}/\tau$ decays rapidly in a short time, then displays slightly oscillating decay and reaches a steady value in the long time limit.
	By contrast, in the coherence trapping region, for $s=1.5$, $s=3.5$ and $s=5$ in Fig.~\ref{fig:QSLTs} (b) and for $s=3.5$ and $s=5$ in Fig.~\ref{fig:QSLTs} (a), as time passes, the ratio $\tau_{\mathrm{QSL}}/\tau$ decays rapidly in a short time, then exhibits oscillating damped decay and approaches zero in the long time limit.
	In this dynamical region, the amplitude in the damped decay first increases and then decreases as the Ohmicity parameter $s$ increases. 
	This is mainly due to the fact that the steady dephasing factor $F(\infty)$ first increases and then decreases with the increase of the Ohmicity parameter $s$ as shown in Fig.~\ref{fig:cohnonmar}.

	\begin{figure}[ht]
		\centering
		\includegraphics[width=3.5in]{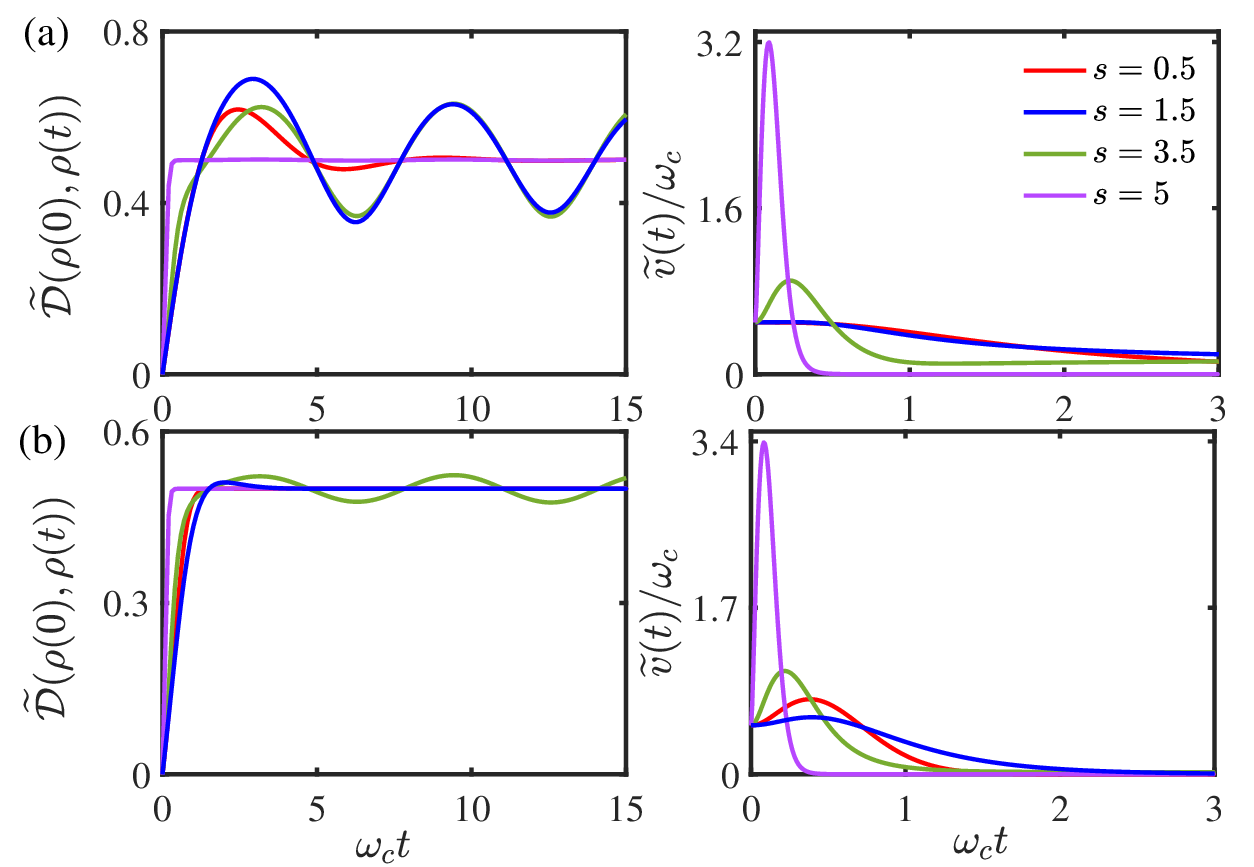}
		\caption{(Color online) Time evolution of the scaled geodesic distance $\widetilde{\mathcal{D}}\bm(\rho(0),\rho(t)\bm)=\mathcal{D}\bm(\rho(0),\rho(t)\bm)/\mathcal{C}(0)$ (left column) and the scaled instantaneous evolution speed $\widetilde{v}(t)=v(t)/\mathcal{C}(0)$ (right column) as a function of the evolution time $t$ for different values of Ohmicity parameter $s$ with the environmental temperature (a) $T=0$ and (b) $T=1.5\omega_{c}$. The transition frequency of the qubit system is set as $\omega_{0}=\omega_{c}$.}
		\label{fig:geospes}
	\end{figure}

	Figure~\ref{fig:geospes} illustrates the time evolution of the geodesic distance $\mathcal{D}\bm(\rho(0),\rho(t)\bm)$ and the instantaneous evolution speed $v(t)$.
	As displayed in the left column, the geodesic distance $\mathcal{D}\bm(\rho(0),\rho(t)\bm)$ show an increasing trend over time. For small Ohmicity parameter $s$, it tends to a steady value with the passage of time while it displays periodic oscillations for large Ohmicity parameter $s$ for both the environmental temperatures $T=0$ and $T=1.5\omega_{c}$.
	In addition, as the Ohmicity parameter $s$ increases, the amplitude in the geodesic distance $\mathcal{D}\bm(\rho(0),\rho(t)\bm)$ at long timescales first increases and then decreases.
	This is due to the fact that the dephasing dynamics of the qubit system 
	undergoes a crossover from the dynamical region of complete loss of coherence to the coherence tapping region.
	As depicted in the right column, for both values of the environmental temperature, i.e. $T=0$ and $T=1.5\omega_{c}$, the instantaneous evolution speed $v(t)$ first increases, then decreases and approaches zero in the long time limit for small Ohmicity parameter $s$.
	The long-time behavior of the geodesic distance provides insights into the residual coherence of the dephasing qubit. 
	The steady value of the geodesic distance $\mathcal{D}\bm(\rho(0),\rho(t)\bm)$ means the qubit system stops evolving corresponding to the instantaneous evolution speed $v(t)\rightarrow 0$ as shown in the right column. 
	By contrast, for large values of $s$, the instantaneous evolution speed $v(t)$ first increases, then decreases and gradually tends to a non-zero steady value. 
	This means that for small Ohmicity parameter $s$, the qubit system stops evolving with the steady-state speed of dynamical evolution $v(t\rightarrow\infty)=0$ whereas it keeps evolving for large Ohmicity parameter $s$ since the evolution speed $v(t\rightarrow\infty)>0$ in the long time limit.

	\subsubsection{Interplay between QSLs time, environmental temperature and environmental spectral density}

	By examining the results presented in Fig.~\ref{fig:QSLT} and Fig.~\ref{fig:QSLTs}, we can see that both the environmental temperature $T$ and the Ohmicity parameter $s$ significantly influence the QSLs time for the dephasing qubit.
		In the dynamical region where the dephasing qubit loses its coherence completely in the long time limit, the $\tau_{\mathrm{QSL}}/\tau$ reaches a steady value after a period of time and the steady value does not depend on the evolution time.
		By contrast, the $\tau_{\mathrm{QSL}}/\tau$ displays damped oscillating decay with the evolution time in the coherence trapping region.
	To further explore the interplay between the QSLs time, the environmental temperature and the environmental spectral density, we plot the ratio $\tau_{\mathrm{QSL}}/\tau$ for different values of environmental temperature $T$ and Ohmicity parameter $s$ for a fixed evolution time $\tau$.
		In order to comprehensively and systematically study the effects of the  Hamiltonian of the qubit system and the dephasing dynamics on the QSLs time bound, we choose a fixed evolution time $\tau=10/\omega_{c}>\mathcal{T}=2\pi/\omega_{0}$. 
		On this time scale, the contribution of the transition frequency and the trends of the dynamical evolution of the dephasing qubit can be clearly identified.
	
	\begin{figure}[ht]
		\centering
		\includegraphics[width=3.2in]{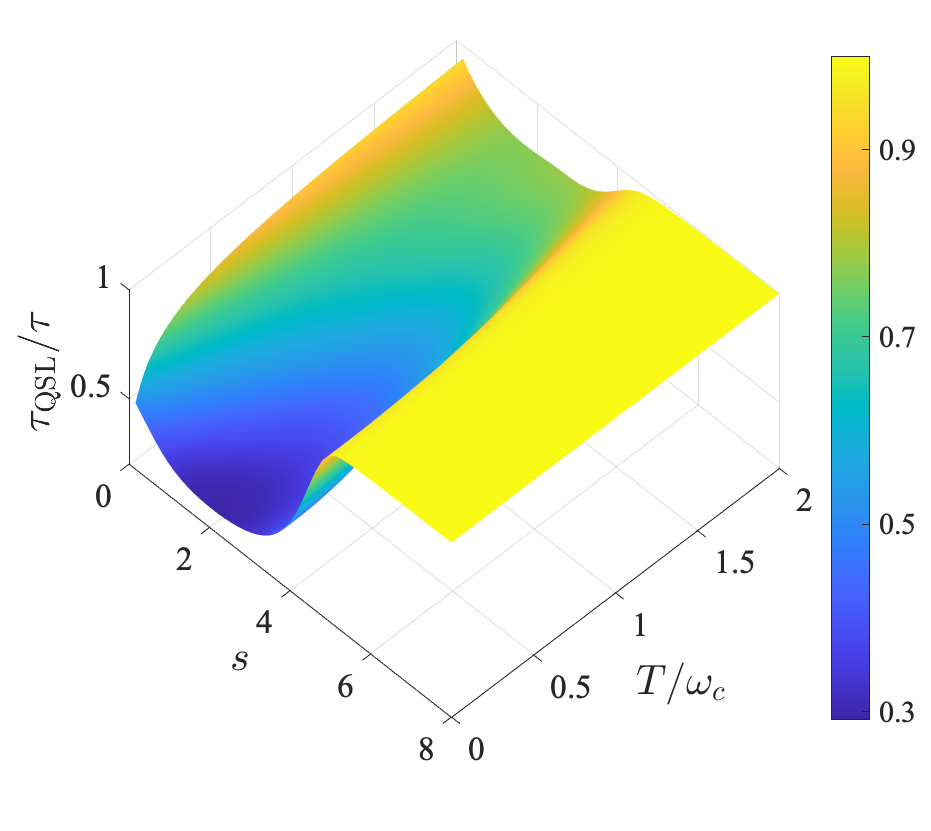}
		\caption{(Color online) The ratio $\tau_{\mathrm{QSL}}/\tau$ for different values of Ohmicity parameter $s$ and environmental temperature $T$. The transition frequency of the qubit system is chosen as $\omega_{0}=\omega_{c}$ and the evolution time is set as $\tau=10/\omega_{c}$.}
		\label{fig:interplay}
	\end{figure}

	 As depicted in Fig.~\ref{fig:interplay}, for a given environmental temperature $T$, as the Ohmicity parameter $s$ increases, 
	the ratio $\tau_{\mathrm{QSL}}/\tau$ first decreases, then increases and gradually tends to a steady value.
	In the region with Ohmicity parameter $0<s<4$, the ratio $\tau_{\mathrm{QSL}}/\tau$ increases with the increase of environmental temperature $T$.
	In the region with Ohmicity parameter $s>5$, the ratio $\tau_{\mathrm{QSL}}/\tau$ tends to saturation and the increase of environmental temperature $T$ has no significant impact on the ratio $\tau_{\mathrm{QSL}}/\tau$.
	This is mainly due to the fact that the dephasing factor decays very fast and its steady value is very small in the region with larger Ohmicity parameter $s$ as shown in Fig.~\ref{fig:dephasings}.
	Physically, it is the combined effect of the environmental temperature and the environmental spectral density that affects the QSLs time in the dephasing dynamics of the qubit system interacting with the thermal environments.
	Therefore, the QSLs time in dephasing qubit can be precisely modulated by engineering the spectral density of the environment.
	
	\section{Conclusions}
	\label{sec:conc}
	We have studied the QSLs of a dephasing qubit in a thermal environment with an engineerable spectral density. 
	By means of the geometric bound derived in terms of the trace distance metric, the QSLs time in the dephasing qubit does not depend on the initial state but is closely associated with the transition frequency of the system and the dynamical dephasing induced by the environment. 
	We explored the interplay between the QSLs time, environmental temperature and the spectral density of the environment.
	It is precisely the combined effect of the environmental temperature and the environmental spectral density that exerts a crucial influence on the QSLs time in the dephasing dynamics of the qubit system coupled to the thermal environments.
	It is shown that the increase of environmental temperature can enhance 
    the QSLs time.
	In addition, the increase in the Ohmicity parameter initially gives rise to a reduction in the QSLs time, and subsequently, it is followed by an upward trend in the QSLs time.
	Our investigation contributes to a deeper understanding of QSLs in the dynamics of open quantum systems and has potential application in the realization of optimal quantum evolution by engineering the spectral density of the environment.
	
	\begin{acknowledgments}
		This work was supported by the National Natural Science Foundation of China (Grant Nos. 12005121 and 12204277).
	\end{acknowledgments}
	
	\section*{Author declarations}
	\subsection*{Conflict of Interest}
	The authors have no conflicts to disclose.
	\subsection*{Author Contributions}
	\textbf{Xiangji Cai:} Conceptualization (lead); Formal analysis (equal); Investigation (equal); Writing – original draft (lead); Writing – review \& editing (equal). 
	\textbf{Yanyan Feng:} Formal analysis (equal); Investigation (equal); Writing – review \& editing (equal). 
	\textbf{Jin Ren:} Conceptualization (supporting); Formal analysis (equal); Investigation (equal); Writing – review \& editing (equal). 
	\textbf{Kang Lan:} Conceptualization (supporting); Formal analysis (equal); Investigation (equal); Writing – review \& editing (equal).
	\textbf{Shuning Sun:} Conceptualization (supporting); Formal analysis (equal); Investigation (equal); Writing – review \& editing (equal).
	\textbf{Xiangjia Meng:} Conceptualization (supporting); Formal analysis (equal); Investigation (equal); Writing – review \& editing (equal).
	\textbf{Artur Czerwinski:} Conceptualization (supporting); Formal analysis (equal); Investigation (equal); Writing – original draft (supporting); Writing – review \& editing (equal).
	
	\section*{Data Availability}
	The data that support the findings of this study are available from the corresponding author upon reasonable request.
	

\begin{thebibliography}{117}%
\makeatletter
\providecommand \@ifxundefined [1]{%
 \@ifx{#1\undefined}
}%
\providecommand \@ifnum [1]{%
 \ifnum #1\expandafter \@firstoftwo
 \else \expandafter \@secondoftwo
 \fi
}%
\providecommand \@ifx [1]{%
 \ifx #1\expandafter \@firstoftwo
 \else \expandafter \@secondoftwo
 \fi
}%
\providecommand \natexlab [1]{#1}%
\providecommand \enquote  [1]{``#1''}%
\providecommand \bibnamefont  [1]{#1}%
\providecommand \bibfnamefont [1]{#1}%
\providecommand \citenamefont [1]{#1}%
\providecommand \href@noop [0]{\@secondoftwo}%
\providecommand \href [0]{\begingroup \@sanitize@url \@href}%
\providecommand \@href[1]{\@@startlink{#1}\@@href}%
\providecommand \@@href[1]{\endgroup#1\@@endlink}%
\providecommand \@sanitize@url [0]{\catcode `\\12\catcode `\$12\catcode
  `\&12\catcode `\#12\catcode `\^12\catcode `\_12\catcode `\%12\relax}%
\providecommand \@@startlink[1]{}%
\providecommand \@@endlink[0]{}%
\providecommand \url  [0]{\begingroup\@sanitize@url \@url }%
\providecommand \@url [1]{\endgroup\@href {#1}{\urlprefix }}%
\providecommand \urlprefix  [0]{URL }%
\providecommand \Eprint [0]{\href }%
\providecommand \doibase [0]{https://doi.org/}%
\providecommand \selectlanguage [0]{\@gobble}%
\providecommand \bibinfo  [0]{\@secondoftwo}%
\providecommand \bibfield  [0]{\@secondoftwo}%
\providecommand \translation [1]{[#1]}%
\providecommand \BibitemOpen [0]{}%
\providecommand \bibitemStop [0]{}%
\providecommand \bibitemNoStop [0]{.\EOS\space}%
\providecommand \EOS [0]{\spacefactor3000\relax}%
\providecommand \BibitemShut  [1]{\csname bibitem#1\endcsname}%
\let\auto@bib@innerbib\@empty
\bibitem [{\citenamefont {Deffner}\ and\ \citenamefont
  {Campbell}(2017)}]{JPhysA50.453001}%
  \BibitemOpen
  \bibfield  {author} {\bibinfo {author} {\bibfnamefont {S.}~\bibnamefont
  {Deffner}}\ and\ \bibinfo {author} {\bibfnamefont {S.}~\bibnamefont
  {Campbell}},\ }\bibfield  {title} {\enquote {\bibinfo {title} {Quantum speed
  limits: from {H}eisenberg's uncertainty principle to optimal quantum
  control},}\ }\href {https://doi.org/10.1088/1751-8121/aa86c6} {\bibfield
  {journal} {\bibinfo  {journal} {J. Phys. A}\ }\textbf {\bibinfo {volume}
  {50}},\ \bibinfo {pages} {453001} (\bibinfo {year} {2017})}\BibitemShut
  {NoStop}%
\bibitem [{\citenamefont {Farhi}\ \emph {et~al.}(1998)\citenamefont {Farhi},
  \citenamefont {Goldstone}, \citenamefont {Gutmann},\ and\ \citenamefont
  {Sipser}}]{PhysRevLett.81.5442}%
  \BibitemOpen
  \bibfield  {author} {\bibinfo {author} {\bibfnamefont {E.}~\bibnamefont
  {Farhi}}, \bibinfo {author} {\bibfnamefont {J.}~\bibnamefont {Goldstone}},
  \bibinfo {author} {\bibfnamefont {S.}~\bibnamefont {Gutmann}},\ and\ \bibinfo
  {author} {\bibfnamefont {M.}~\bibnamefont {Sipser}},\ }\bibfield  {title}
  {\enquote {\bibinfo {title} {Limit on the speed of quantum computation in
  determining parity},}\ }\href {https://doi.org/10.1103/PhysRevLett.81.5442}
  {\bibfield  {journal} {\bibinfo  {journal} {Phys. Rev. Lett.}\ }\textbf
  {\bibinfo {volume} {81}},\ \bibinfo {pages} {5442} (\bibinfo {year}
  {1998})}\BibitemShut {NoStop}%
\bibitem [{\citenamefont {Lloyd}(2000)}]{Nature406.1047}%
  \BibitemOpen
  \bibfield  {author} {\bibinfo {author} {\bibfnamefont {S.}~\bibnamefont
  {Lloyd}},\ }\bibfield  {title} {\enquote {\bibinfo {title} {Ultimate physical
  limits to computation},}\ }\href {https://doi.org/10.1038/35023282}
  {\bibfield  {journal} {\bibinfo  {journal} {Nature}\ }\textbf {\bibinfo
  {volume} {406}},\ \bibinfo {pages} {1047} (\bibinfo {year}
  {2000})}\BibitemShut {NoStop}%
\bibitem [{\citenamefont {S{\o}rensen}\ \emph {et~al.}(2016)\citenamefont
  {S{\o}rensen}, \citenamefont {Pedersen}, \citenamefont {Munch}, \citenamefont
  {Haikka}, \citenamefont {Jensen}, \citenamefont {Planke}, \citenamefont
  {Andreasen}, \citenamefont {Gajdacz}, \citenamefont {M{\o}lmer},
  \citenamefont {Lieberoth},\ and\ \citenamefont {Sherson}}]{Nature532.210}%
  \BibitemOpen
  \bibfield  {author} {\bibinfo {author} {\bibfnamefont {J.~J. W.~H.}\
  \bibnamefont {S{\o}rensen}}, \bibinfo {author} {\bibfnamefont {M.~K.}\
  \bibnamefont {Pedersen}}, \bibinfo {author} {\bibfnamefont {M.}~\bibnamefont
  {Munch}}, \bibinfo {author} {\bibfnamefont {P.}~\bibnamefont {Haikka}},
  \bibinfo {author} {\bibfnamefont {J.~H.}\ \bibnamefont {Jensen}}, \bibinfo
  {author} {\bibfnamefont {T.}~\bibnamefont {Planke}}, \bibinfo {author}
  {\bibfnamefont {M.~G.}\ \bibnamefont {Andreasen}}, \bibinfo {author}
  {\bibfnamefont {M.}~\bibnamefont {Gajdacz}}, \bibinfo {author} {\bibfnamefont
  {K.}~\bibnamefont {M{\o}lmer}}, \bibinfo {author} {\bibfnamefont
  {A.}~\bibnamefont {Lieberoth}},\ and\ \bibinfo {author} {\bibfnamefont
  {J.~F.}\ \bibnamefont {Sherson}},\ }\bibfield  {title} {\enquote {\bibinfo
  {title} {Exploring the quantum speed limit with computer games},}\ }\href
  {https://doi.org/10.1038/nature17620} {\bibfield  {journal} {\bibinfo
  {journal} {Nature}\ }\textbf {\bibinfo {volume} {532}},\ \bibinfo {pages}
  {210} (\bibinfo {year} {2016})}\BibitemShut {NoStop}%
\bibitem [{\citenamefont {Suzuki}\ and\ \citenamefont
  {Takahashi}(2020)}]{PhysRevRes.2.032016}%
  \BibitemOpen
  \bibfield  {author} {\bibinfo {author} {\bibfnamefont {K.}~\bibnamefont
  {Suzuki}}\ and\ \bibinfo {author} {\bibfnamefont {K.}~\bibnamefont
  {Takahashi}},\ }\bibfield  {title} {\enquote {\bibinfo {title} {Performance
  evaluation of adiabatic quantum computation via quantum speed limits and
  possible applications to many-body systems},}\ }\href
  {https://doi.org/10.1103/PhysRevResearch.2.032016} {\bibfield  {journal}
  {\bibinfo  {journal} {Phys. Rev. Res.}\ }\textbf {\bibinfo {volume} {2}},\
  \bibinfo {pages} {032016} (\bibinfo {year} {2020})}\BibitemShut {NoStop}%
\bibitem [{\citenamefont {Mohan}, \citenamefont {Das},\ and\ \citenamefont
  {Pati}(2022)}]{NewJPhys.24.065003}%
  \BibitemOpen
  \bibfield  {author} {\bibinfo {author} {\bibfnamefont {B.}~\bibnamefont
  {Mohan}}, \bibinfo {author} {\bibfnamefont {S.}~\bibnamefont {Das}},\ and\
  \bibinfo {author} {\bibfnamefont {A.~K.}\ \bibnamefont {Pati}},\ }\bibfield
  {title} {\enquote {\bibinfo {title} {Quantum speed limits for information and
  coherence},}\ }\href {https://doi.org/10.1088/1367-2630/ac753c} {\bibfield
  {journal} {\bibinfo  {journal} {New J. Phys.}\ }\textbf {\bibinfo {volume}
  {24}},\ \bibinfo {pages} {065003} (\bibinfo {year} {2022})}\BibitemShut
  {NoStop}%
\bibitem [{\citenamefont {Murphy}\ \emph {et~al.}(2010)\citenamefont {Murphy},
  \citenamefont {Montangero}, \citenamefont {Giovannetti},\ and\ \citenamefont
  {Calarco}}]{PhysRevA82.022318}%
  \BibitemOpen
  \bibfield  {author} {\bibinfo {author} {\bibfnamefont {M.}~\bibnamefont
  {Murphy}}, \bibinfo {author} {\bibfnamefont {S.}~\bibnamefont {Montangero}},
  \bibinfo {author} {\bibfnamefont {V.}~\bibnamefont {Giovannetti}},\ and\
  \bibinfo {author} {\bibfnamefont {T.}~\bibnamefont {Calarco}},\ }\bibfield
  {title} {\enquote {\bibinfo {title} {Communication at the quantum speed limit
  along a spin chain},}\ }\href {https://doi.org/10.1103/PhysRevA.82.022318}
  {\bibfield  {journal} {\bibinfo  {journal} {Phys. Rev. A}\ }\textbf {\bibinfo
  {volume} {82}},\ \bibinfo {pages} {022318} (\bibinfo {year}
  {2010})}\BibitemShut {NoStop}%
\bibitem [{\citenamefont {Epstein}\ and\ \citenamefont
  {Whaley}(2017)}]{PhysRevA95.042314}%
  \BibitemOpen
  \bibfield  {author} {\bibinfo {author} {\bibfnamefont {J.~M.}\ \bibnamefont
  {Epstein}}\ and\ \bibinfo {author} {\bibfnamefont {K.~B.}\ \bibnamefont
  {Whaley}},\ }\bibfield  {title} {\enquote {\bibinfo {title} {Quantum speed
  limits for quantum-information-processing tasks},}\ }\href
  {https://doi.org/10.1103/PhysRevA.95.042314} {\bibfield  {journal} {\bibinfo
  {journal} {Phys. Rev. A}\ }\textbf {\bibinfo {volume} {95}},\ \bibinfo
  {pages} {042314} (\bibinfo {year} {2017})}\BibitemShut {NoStop}%
\bibitem [{\citenamefont {Caneva}\ \emph {et~al.}(2009)\citenamefont {Caneva},
  \citenamefont {Murphy}, \citenamefont {Calarco}, \citenamefont {Fazio},
  \citenamefont {Montangero}, \citenamefont {Giovannetti},\ and\ \citenamefont
  {Santoro}}]{PhysRevLett.103.240501}%
  \BibitemOpen
  \bibfield  {author} {\bibinfo {author} {\bibfnamefont {T.}~\bibnamefont
  {Caneva}}, \bibinfo {author} {\bibfnamefont {M.}~\bibnamefont {Murphy}},
  \bibinfo {author} {\bibfnamefont {T.}~\bibnamefont {Calarco}}, \bibinfo
  {author} {\bibfnamefont {R.}~\bibnamefont {Fazio}}, \bibinfo {author}
  {\bibfnamefont {S.}~\bibnamefont {Montangero}}, \bibinfo {author}
  {\bibfnamefont {V.}~\bibnamefont {Giovannetti}},\ and\ \bibinfo {author}
  {\bibfnamefont {G.~E.}\ \bibnamefont {Santoro}},\ }\bibfield  {title}
  {\enquote {\bibinfo {title} {Optimal control at the quantum speed limit},}\
  }\href {https://doi.org/10.1103/PhysRevLett.103.240501} {\bibfield  {journal}
  {\bibinfo  {journal} {Phys. Rev. Lett.}\ }\textbf {\bibinfo {volume} {103}},\
  \bibinfo {pages} {240501} (\bibinfo {year} {2009})}\BibitemShut {NoStop}%
\bibitem [{\citenamefont {Hegerfeldt}(2013)}]{PhysRevLett.111.260501}%
  \BibitemOpen
  \bibfield  {author} {\bibinfo {author} {\bibfnamefont {G.~C.}\ \bibnamefont
  {Hegerfeldt}},\ }\bibfield  {title} {\enquote {\bibinfo {title} {Driving at
  the quantum speed limit: Optimal control of a two-level system},}\ }\href
  {https://doi.org/10.1103/PhysRevLett.111.260501} {\bibfield  {journal}
  {\bibinfo  {journal} {Phys. Rev. Lett.}\ }\textbf {\bibinfo {volume} {111}},\
  \bibinfo {pages} {260501} (\bibinfo {year} {2013})}\BibitemShut {NoStop}%
\bibitem [{\citenamefont {Lloyd}\ and\ \citenamefont
  {Montangero}(2014)}]{PhysRevLett.113.010502}%
  \BibitemOpen
  \bibfield  {author} {\bibinfo {author} {\bibfnamefont {S.}~\bibnamefont
  {Lloyd}}\ and\ \bibinfo {author} {\bibfnamefont {S.}~\bibnamefont
  {Montangero}},\ }\bibfield  {title} {\enquote {\bibinfo {title} {Information
  theoretical analysis of quantum optimal control},}\ }\href
  {https://doi.org/10.1103/PhysRevLett.113.010502} {\bibfield  {journal}
  {\bibinfo  {journal} {Phys. Rev. Lett.}\ }\textbf {\bibinfo {volume} {113}},\
  \bibinfo {pages} {010502} (\bibinfo {year} {2014})}\BibitemShut {NoStop}%
\bibitem [{\citenamefont {Zhang}\ \emph {et~al.}(2021)\citenamefont {Zhang},
  \citenamefont {Kuang}, \citenamefont {Puri},\ and\ \citenamefont
  {Miller}}]{PhysRevLett.127.110506}%
  \BibitemOpen
  \bibfield  {author} {\bibinfo {author} {\bibfnamefont {H.}~\bibnamefont
  {Zhang}}, \bibinfo {author} {\bibfnamefont {Z.}~\bibnamefont {Kuang}},
  \bibinfo {author} {\bibfnamefont {S.}~\bibnamefont {Puri}},\ and\ \bibinfo
  {author} {\bibfnamefont {O.~D.}\ \bibnamefont {Miller}},\ }\bibfield  {title}
  {\enquote {\bibinfo {title} {Conservation-law-based global bounds to quantum
  optimal control},}\ }\href {https://doi.org/10.1103/PhysRevLett.127.110506}
  {\bibfield  {journal} {\bibinfo  {journal} {Phys. Rev. Lett.}\ }\textbf
  {\bibinfo {volume} {127}},\ \bibinfo {pages} {110506} (\bibinfo {year}
  {2021})}\BibitemShut {NoStop}%
\bibitem [{\citenamefont {Chu}, \citenamefont {Li},\ and\ \citenamefont
  {Cai}(2023)}]{PhysRevLett.130.170801}%
  \BibitemOpen
  \bibfield  {author} {\bibinfo {author} {\bibfnamefont {Y.}~\bibnamefont
  {Chu}}, \bibinfo {author} {\bibfnamefont {X.}~\bibnamefont {Li}},\ and\
  \bibinfo {author} {\bibfnamefont {J.}~\bibnamefont {Cai}},\ }\bibfield
  {title} {\enquote {\bibinfo {title} {Strong quantum metrological limit from
  many-body physics},}\ }\href {https://doi.org/10.1103/PhysRevLett.130.170801}
  {\bibfield  {journal} {\bibinfo  {journal} {Phys. Rev. Lett.}\ }\textbf
  {\bibinfo {volume} {130}},\ \bibinfo {pages} {170801} (\bibinfo {year}
  {2023})}\BibitemShut {NoStop}%
\bibitem [{\citenamefont {Herb}\ and\ \citenamefont
  {Degen}(2024)}]{PhysRevLett.133.210802}%
  \BibitemOpen
  \bibfield  {author} {\bibinfo {author} {\bibfnamefont {K.}~\bibnamefont
  {Herb}}\ and\ \bibinfo {author} {\bibfnamefont {C.~L.}\ \bibnamefont
  {Degen}},\ }\bibfield  {title} {\enquote {\bibinfo {title} {Quantum speed
  limit in quantum sensing},}\ }\href
  {https://doi.org/10.1103/PhysRevLett.133.210802} {\bibfield  {journal}
  {\bibinfo  {journal} {Phys. Rev. Lett.}\ }\textbf {\bibinfo {volume} {133}},\
  \bibinfo {pages} {210802} (\bibinfo {year} {2024})}\BibitemShut {NoStop}%
\bibitem [{\citenamefont {Mukhopadhyay}\ \emph {et~al.}(2018)\citenamefont
  {Mukhopadhyay}, \citenamefont {Misra}, \citenamefont {Bhattacharya},\ and\
  \citenamefont {Pati}}]{PhysRevE97.062116}%
  \BibitemOpen
  \bibfield  {author} {\bibinfo {author} {\bibfnamefont {C.}~\bibnamefont
  {Mukhopadhyay}}, \bibinfo {author} {\bibfnamefont {A.}~\bibnamefont {Misra}},
  \bibinfo {author} {\bibfnamefont {S.}~\bibnamefont {Bhattacharya}},\ and\
  \bibinfo {author} {\bibfnamefont {A.~K.}\ \bibnamefont {Pati}},\ }\bibfield
  {title} {\enquote {\bibinfo {title} {Quantum speed limit constraints on a
  nanoscale autonomous refrigerator},}\ }\href
  {https://doi.org/10.1103/PhysRevE.97.062116} {\bibfield  {journal} {\bibinfo
  {journal} {Phys. Rev. E}\ }\textbf {\bibinfo {volume} {97}},\ \bibinfo
  {pages} {062116} (\bibinfo {year} {2018})}\BibitemShut {NoStop}%
\bibitem [{\citenamefont {Mitchell}\ and\ \citenamefont
  {Palacios~Alvarez}(2020)}]{RevModPhys.92.021001}%
  \BibitemOpen
  \bibfield  {author} {\bibinfo {author} {\bibfnamefont {M.~W.}\ \bibnamefont
  {Mitchell}}\ and\ \bibinfo {author} {\bibfnamefont {S.}~\bibnamefont
  {Palacios~Alvarez}},\ }\bibfield  {title} {\enquote {\bibinfo {title}
  {Colloquium: Quantum limits to the energy resolution of magnetic field
  sensors},}\ }\href {https://doi.org/10.1103/RevModPhys.92.021001} {\bibfield
  {journal} {\bibinfo  {journal} {Rev. Mod. Phys.}\ }\textbf {\bibinfo {volume}
  {92}},\ \bibinfo {pages} {021001} (\bibinfo {year} {2020})}\BibitemShut
  {NoStop}%
\bibitem [{\citenamefont {Howard}\ \emph {et~al.}(2023)\citenamefont {Howard},
  \citenamefont {Lidiak}, \citenamefont {Jameson}, \citenamefont {Basyildiz},
  \citenamefont {Clark}, \citenamefont {Zhao}, \citenamefont {Bal},
  \citenamefont {Long}, \citenamefont {Pappas}, \citenamefont {Singh},\ and\
  \citenamefont {Gong}}]{PhysRevRes.5.043194}%
  \BibitemOpen
  \bibfield  {author} {\bibinfo {author} {\bibfnamefont {J.}~\bibnamefont
  {Howard}}, \bibinfo {author} {\bibfnamefont {A.}~\bibnamefont {Lidiak}},
  \bibinfo {author} {\bibfnamefont {C.}~\bibnamefont {Jameson}}, \bibinfo
  {author} {\bibfnamefont {B.}~\bibnamefont {Basyildiz}}, \bibinfo {author}
  {\bibfnamefont {K.}~\bibnamefont {Clark}}, \bibinfo {author} {\bibfnamefont
  {T.}~\bibnamefont {Zhao}}, \bibinfo {author} {\bibfnamefont {M.}~\bibnamefont
  {Bal}}, \bibinfo {author} {\bibfnamefont {J.}~\bibnamefont {Long}}, \bibinfo
  {author} {\bibfnamefont {D.~P.}\ \bibnamefont {Pappas}}, \bibinfo {author}
  {\bibfnamefont {M.}~\bibnamefont {Singh}},\ and\ \bibinfo {author}
  {\bibfnamefont {Z.}~\bibnamefont {Gong}},\ }\bibfield  {title} {\enquote
  {\bibinfo {title} {Implementing two-qubit gates at the quantum speed
  limit},}\ }\href {https://doi.org/10.1103/PhysRevResearch.5.043194}
  {\bibfield  {journal} {\bibinfo  {journal} {Phys. Rev. Res.}\ }\textbf
  {\bibinfo {volume} {5}},\ \bibinfo {pages} {043194} (\bibinfo {year}
  {2023})}\BibitemShut {NoStop}%
\bibitem [{\citenamefont {Mandelstam}\ and\ \citenamefont
  {Tamm}(1945)}]{JPhys.9.249}%
  \BibitemOpen
  \bibfield  {author} {\bibinfo {author} {\bibfnamefont {L.}~\bibnamefont
  {Mandelstam}}\ and\ \bibinfo {author} {\bibfnamefont {I.}~\bibnamefont
  {Tamm}},\ }\bibfield  {title} {\enquote {\bibinfo {title} {The uncertainty
  relation between energy and time in nonrelativistic quantum mechanics},}\
  }\href {https://doi.org/10.1007/978-3-642-74626-0_8} {\bibfield  {journal}
  {\bibinfo  {journal} {J. Phys. (USSR)}\ }\textbf {\bibinfo {volume} {9}},\
  \bibinfo {pages} {249} (\bibinfo {year} {1945})}\BibitemShut {NoStop}%
\bibitem [{\citenamefont {Margolus}\ and\ \citenamefont
  {Levitin}(1998)}]{PhysicaD120.188}%
  \BibitemOpen
  \bibfield  {author} {\bibinfo {author} {\bibfnamefont {N.}~\bibnamefont
  {Margolus}}\ and\ \bibinfo {author} {\bibfnamefont {L.~B.}\ \bibnamefont
  {Levitin}},\ }\bibfield  {title} {\enquote {\bibinfo {title} {The maximum
  speed of dynamical evolution},}\ }\href
  {https://doi.org/10.1016/S0167-2789(98)00054-2} {\bibfield  {journal}
  {\bibinfo  {journal} {Physica D}\ }\textbf {\bibinfo {volume} {120}},\
  \bibinfo {pages} {188} (\bibinfo {year} {1998})}\BibitemShut {NoStop}%
\bibitem [{\citenamefont {Giovannetti}, \citenamefont {Lloyd},\ and\
  \citenamefont {Maccone}(2003)}]{PhysRevA67.052109}%
  \BibitemOpen
  \bibfield  {author} {\bibinfo {author} {\bibfnamefont {V.}~\bibnamefont
  {Giovannetti}}, \bibinfo {author} {\bibfnamefont {S.}~\bibnamefont {Lloyd}},\
  and\ \bibinfo {author} {\bibfnamefont {L.}~\bibnamefont {Maccone}},\
  }\bibfield  {title} {\enquote {\bibinfo {title} {Quantum limits to dynamical
  evolution},}\ }\href {https://doi.org/10.1103/PhysRevA.67.052109} {\bibfield
  {journal} {\bibinfo  {journal} {Phys. Rev. A}\ }\textbf {\bibinfo {volume}
  {67}},\ \bibinfo {pages} {052109} (\bibinfo {year} {2003})}\BibitemShut
  {NoStop}%
\bibitem [{\citenamefont {Jones}\ and\ \citenamefont
  {Kok}(2010)}]{PhysRevA82.022107}%
  \BibitemOpen
  \bibfield  {author} {\bibinfo {author} {\bibfnamefont {P.~J.}\ \bibnamefont
  {Jones}}\ and\ \bibinfo {author} {\bibfnamefont {P.}~\bibnamefont {Kok}},\
  }\bibfield  {title} {\enquote {\bibinfo {title} {Geometric derivation of the
  quantum speed limit},}\ }\href {https://doi.org/10.1103/PhysRevA.82.022107}
  {\bibfield  {journal} {\bibinfo  {journal} {Phys. Rev. A}\ }\textbf {\bibinfo
  {volume} {82}},\ \bibinfo {pages} {022107} (\bibinfo {year}
  {2010})}\BibitemShut {NoStop}%
\bibitem [{\citenamefont {Deffner}\ and\ \citenamefont
  {Lutz}(2013{\natexlab{a}})}]{JPhysA46.335302}%
  \BibitemOpen
  \bibfield  {author} {\bibinfo {author} {\bibfnamefont {S.}~\bibnamefont
  {Deffner}}\ and\ \bibinfo {author} {\bibfnamefont {E.}~\bibnamefont {Lutz}},\
  }\bibfield  {title} {\enquote {\bibinfo {title} {Energy--time uncertainty
  relation for driven quantum systems},}\ }\href
  {https://doi.org/10.1088/1751-8113/46/33/335302} {\bibfield  {journal}
  {\bibinfo  {journal} {J. Phys. A}\ }\textbf {\bibinfo {volume} {46}},\
  \bibinfo {pages} {335302} (\bibinfo {year} {2013}{\natexlab{a}})}\BibitemShut
  {NoStop}%
\bibitem [{\citenamefont {Andersson}\ and\ \citenamefont
  {Heydari}(2014)}]{JPhysA47.215301}%
  \BibitemOpen
  \bibfield  {author} {\bibinfo {author} {\bibfnamefont {O.}~\bibnamefont
  {Andersson}}\ and\ \bibinfo {author} {\bibfnamefont {H.}~\bibnamefont
  {Heydari}},\ }\bibfield  {title} {\enquote {\bibinfo {title} {Quantum speed
  limits and optimal {H}amiltonians for driven systems in mixed states},}\
  }\href {https://doi.org/10.1088/1751-8113/47/21/215301} {\bibfield  {journal}
  {\bibinfo  {journal} {J. Phys. A}\ }\textbf {\bibinfo {volume} {47}},\
  \bibinfo {pages} {215301} (\bibinfo {year} {2014})}\BibitemShut {NoStop}%
\bibitem [{\citenamefont {Poggi}(2019)}]{PhysRevA99.042116}%
  \BibitemOpen
  \bibfield  {author} {\bibinfo {author} {\bibfnamefont {P.~M.}\ \bibnamefont
  {Poggi}},\ }\bibfield  {title} {\enquote {\bibinfo {title} {Geometric quantum
  speed limits and short-time accessibility to unitary operations},}\ }\href
  {https://doi.org/10.1103/PhysRevA.99.042116} {\bibfield  {journal} {\bibinfo
  {journal} {Phys. Rev. A}\ }\textbf {\bibinfo {volume} {99}},\ \bibinfo
  {pages} {042116} (\bibinfo {year} {2019})}\BibitemShut {NoStop}%
\bibitem [{\citenamefont {Deffner}(2017)}]{NewJPhys.19.103018}%
  \BibitemOpen
  \bibfield  {author} {\bibinfo {author} {\bibfnamefont {S.}~\bibnamefont
  {Deffner}},\ }\bibfield  {title} {\enquote {\bibinfo {title} {Geometric
  quantum speed limits: a case for {W}igner phase space},}\ }\href
  {https://doi.org/10.1088/1367-2630/aa83dc} {\bibfield  {journal} {\bibinfo
  {journal} {New J. Phys.}\ }\textbf {\bibinfo {volume} {19}},\ \bibinfo
  {pages} {103018} (\bibinfo {year} {2017})}\BibitemShut {NoStop}%
\bibitem [{\citenamefont {Hu}, \citenamefont {Sun},\ and\ \citenamefont
  {Zheng}(2020)}]{PhysRevA101.042107}%
  \BibitemOpen
  \bibfield  {author} {\bibinfo {author} {\bibfnamefont {X.}~\bibnamefont
  {Hu}}, \bibinfo {author} {\bibfnamefont {S.}~\bibnamefont {Sun}},\ and\
  \bibinfo {author} {\bibfnamefont {Y.}~\bibnamefont {Zheng}},\ }\bibfield
  {title} {\enquote {\bibinfo {title} {Quantum speed limit via the trajectory
  ensemble},}\ }\href {https://doi.org/10.1103/PhysRevA.101.042107} {\bibfield
  {journal} {\bibinfo  {journal} {Phys. Rev. A}\ }\textbf {\bibinfo {volume}
  {101}},\ \bibinfo {pages} {042107} (\bibinfo {year} {2020})}\BibitemShut
  {NoStop}%
\bibitem [{\citenamefont {Sun}\ and\ \citenamefont
  {Zheng}(2019)}]{PhysRevLett.123.180403}%
  \BibitemOpen
  \bibfield  {author} {\bibinfo {author} {\bibfnamefont {S.}~\bibnamefont
  {Sun}}\ and\ \bibinfo {author} {\bibfnamefont {Y.}~\bibnamefont {Zheng}},\
  }\bibfield  {title} {\enquote {\bibinfo {title} {Distinct bound of the
  quantum speed limit via the gauge invariant distance},}\ }\href
  {https://doi.org/10.1103/PhysRevLett.123.180403} {\bibfield  {journal}
  {\bibinfo  {journal} {Phys. Rev. Lett.}\ }\textbf {\bibinfo {volume} {123}},\
  \bibinfo {pages} {180403} (\bibinfo {year} {2019})}\BibitemShut {NoStop}%
\bibitem [{\citenamefont {Sun}\ \emph {et~al.}(2021)\citenamefont {Sun},
  \citenamefont {Peng}, \citenamefont {Hu},\ and\ \citenamefont
  {Zheng}}]{PhysRevLett.127.100404}%
  \BibitemOpen
  \bibfield  {author} {\bibinfo {author} {\bibfnamefont {S.}~\bibnamefont
  {Sun}}, \bibinfo {author} {\bibfnamefont {Y.}~\bibnamefont {Peng}}, \bibinfo
  {author} {\bibfnamefont {X.}~\bibnamefont {Hu}},\ and\ \bibinfo {author}
  {\bibfnamefont {Y.}~\bibnamefont {Zheng}},\ }\bibfield  {title} {\enquote
  {\bibinfo {title} {Quantum speed limit quantified by the changing rate of
  phase},}\ }\href {https://doi.org/10.1103/PhysRevLett.127.100404} {\bibfield
  {journal} {\bibinfo  {journal} {Phys. Rev. Lett.}\ }\textbf {\bibinfo
  {volume} {127}},\ \bibinfo {pages} {100404} (\bibinfo {year}
  {2021})}\BibitemShut {NoStop}%
\bibitem [{\citenamefont {Impens}\ \emph {et~al.}(2021)\citenamefont {Impens},
  \citenamefont {D'Angelis}, \citenamefont {Pinheiro},\ and\ \citenamefont
  {Gu\'ery-Odelin}}]{PhysRevA104.052620}%
  \BibitemOpen
  \bibfield  {author} {\bibinfo {author} {\bibfnamefont {F.}~\bibnamefont
  {Impens}}, \bibinfo {author} {\bibfnamefont {F.~M.}\ \bibnamefont
  {D'Angelis}}, \bibinfo {author} {\bibfnamefont {F.~A.}\ \bibnamefont
  {Pinheiro}},\ and\ \bibinfo {author} {\bibfnamefont {D.}~\bibnamefont
  {Gu\'ery-Odelin}},\ }\bibfield  {title} {\enquote {\bibinfo {title} {Time
  scaling and quantum speed limit in non-{H}ermitian {H}amiltonians},}\ }\href
  {https://doi.org/10.1103/PhysRevA.104.052620} {\bibfield  {journal} {\bibinfo
   {journal} {Phys. Rev. A}\ }\textbf {\bibinfo {volume} {104}},\ \bibinfo
  {pages} {052620} (\bibinfo {year} {2021})}\BibitemShut {NoStop}%
\bibitem [{\citenamefont {Shanahan}\ \emph {et~al.}(2018)\citenamefont
  {Shanahan}, \citenamefont {Chenu}, \citenamefont {Margolus},\ and\
  \citenamefont {{del Campo}}}]{PhysRevLett.120.070401}%
  \BibitemOpen
  \bibfield  {author} {\bibinfo {author} {\bibfnamefont {B.}~\bibnamefont
  {Shanahan}}, \bibinfo {author} {\bibfnamefont {A.}~\bibnamefont {Chenu}},
  \bibinfo {author} {\bibfnamefont {N.}~\bibnamefont {Margolus}},\ and\
  \bibinfo {author} {\bibfnamefont {A.}~\bibnamefont {{del Campo}}},\
  }\bibfield  {title} {\enquote {\bibinfo {title} {Quantum speed limits across
  the quantum-to-classical transition},}\ }\href
  {https://doi.org/10.1103/PhysRevLett.120.070401} {\bibfield  {journal}
  {\bibinfo  {journal} {Phys. Rev. Lett.}\ }\textbf {\bibinfo {volume} {120}},\
  \bibinfo {pages} {070401} (\bibinfo {year} {2018})}\BibitemShut {NoStop}%
\bibitem [{\citenamefont {Okuyama}\ and\ \citenamefont
  {Ohzeki}(2018)}]{PhysRevLett.120.070402}%
  \BibitemOpen
  \bibfield  {author} {\bibinfo {author} {\bibfnamefont {M.}~\bibnamefont
  {Okuyama}}\ and\ \bibinfo {author} {\bibfnamefont {M.}~\bibnamefont
  {Ohzeki}},\ }\bibfield  {title} {\enquote {\bibinfo {title} {Quantum speed
  limit is not quantum},}\ }\href
  {https://doi.org/10.1103/PhysRevLett.120.070402} {\bibfield  {journal}
  {\bibinfo  {journal} {Phys. Rev. Lett.}\ }\textbf {\bibinfo {volume} {120}},\
  \bibinfo {pages} {070402} (\bibinfo {year} {2018})}\BibitemShut {NoStop}%
\bibitem [{\citenamefont {Shiraishi}, \citenamefont {Funo},\ and\ \citenamefont
  {Saito}(2018)}]{PhysRevLett.121.070601}%
  \BibitemOpen
  \bibfield  {author} {\bibinfo {author} {\bibfnamefont {N.}~\bibnamefont
  {Shiraishi}}, \bibinfo {author} {\bibfnamefont {K.}~\bibnamefont {Funo}},\
  and\ \bibinfo {author} {\bibfnamefont {K.}~\bibnamefont {Saito}},\ }\bibfield
   {title} {\enquote {\bibinfo {title} {Speed limit for classical stochastic
  processes},}\ }\href {https://doi.org/10.1103/PhysRevLett.121.070601}
  {\bibfield  {journal} {\bibinfo  {journal} {Phys. Rev. Lett.}\ }\textbf
  {\bibinfo {volume} {121}},\ \bibinfo {pages} {070601} (\bibinfo {year}
  {2018})}\BibitemShut {NoStop}%
\bibitem [{\citenamefont {Poggi}, \citenamefont {Campbell},\ and\ \citenamefont
  {Deffner}(2021)}]{PRXQuantum2.040349}%
  \BibitemOpen
  \bibfield  {author} {\bibinfo {author} {\bibfnamefont {P.~M.}\ \bibnamefont
  {Poggi}}, \bibinfo {author} {\bibfnamefont {S.}~\bibnamefont {Campbell}},\
  and\ \bibinfo {author} {\bibfnamefont {S.}~\bibnamefont {Deffner}},\
  }\bibfield  {title} {\enquote {\bibinfo {title} {Diverging quantum speed
  limits: A herald of classicality},}\ }\href
  {https://doi.org/10.1103/PRXQuantum.2.040349} {\bibfield  {journal} {\bibinfo
   {journal} {PRX Quantum}\ }\textbf {\bibinfo {volume} {2}},\ \bibinfo {pages}
  {040349} (\bibinfo {year} {2021})}\BibitemShut {NoStop}%
\bibitem [{\citenamefont {Leggett}\ \emph {et~al.}(1987)\citenamefont
  {Leggett}, \citenamefont {Chakravarty}, \citenamefont {Dorsey}, \citenamefont
  {Fisher}, \citenamefont {Garg},\ and\ \citenamefont
  {Zwerger}}]{RevModPhys.59.1}%
  \BibitemOpen
  \bibfield  {author} {\bibinfo {author} {\bibfnamefont {A.~J.}\ \bibnamefont
  {Leggett}}, \bibinfo {author} {\bibfnamefont {S.}~\bibnamefont
  {Chakravarty}}, \bibinfo {author} {\bibfnamefont {A.~T.}\ \bibnamefont
  {Dorsey}}, \bibinfo {author} {\bibfnamefont {M.~P.~A.}\ \bibnamefont
  {Fisher}}, \bibinfo {author} {\bibfnamefont {A.}~\bibnamefont {Garg}},\ and\
  \bibinfo {author} {\bibfnamefont {W.}~\bibnamefont {Zwerger}},\ }\bibfield
  {title} {\enquote {\bibinfo {title} {Dynamics of the dissipative two-state
  system},}\ }\href {https://doi.org/10.1103/RevModPhys.59.1} {\bibfield
  {journal} {\bibinfo  {journal} {Rev. Mod. Phys.}\ }\textbf {\bibinfo {volume}
  {59}},\ \bibinfo {pages} {1} (\bibinfo {year} {1987})}\BibitemShut {NoStop}%
\bibitem [{\citenamefont {Breuer}\ and\ \citenamefont
  {Petruccione}(2002)}]{Breuerbook}%
  \BibitemOpen
  \bibfield  {author} {\bibinfo {author} {\bibfnamefont {H.~P.}\ \bibnamefont
  {Breuer}}\ and\ \bibinfo {author} {\bibfnamefont {F.}~\bibnamefont
  {Petruccione}},\ }\href@noop {} {\emph {\bibinfo {title} {The Theory of Open
  Quantum Systems}}}\ (\bibinfo  {publisher} {Oxford University Press},\
  \bibinfo {address} {New York},\ \bibinfo {year} {2002})\BibitemShut {NoStop}%
\bibitem [{\citenamefont {Schlosshauer}(2007)}]{Schlosshauerbook}%
  \BibitemOpen
  \bibfield  {author} {\bibinfo {author} {\bibfnamefont {M.}~\bibnamefont
  {Schlosshauer}},\ }\href@noop {} {\emph {\bibinfo {title} {Decoherence and
  the Quantum-to-Classical Transition}}}\ (\bibinfo  {publisher}
  {Springer-Verlag},\ \bibinfo {address} {Berlin, Heidelberg},\ \bibinfo {year}
  {2007})\BibitemShut {NoStop}%
\bibitem [{\citenamefont {Schlosshauer}(2019)}]{PhysRep.831.1}%
  \BibitemOpen
  \bibfield  {author} {\bibinfo {author} {\bibfnamefont {M.}~\bibnamefont
  {Schlosshauer}},\ }\bibfield  {title} {\enquote {\bibinfo {title} {Quantum
  decoherence},}\ }\href {https://doi.org/10.1016/j.physrep.2019.10.001}
  {\bibfield  {journal} {\bibinfo  {journal} {Phys. Rep.}\ }\textbf {\bibinfo
  {volume} {831}},\ \bibinfo {pages} {1} (\bibinfo {year} {2019})}\BibitemShut
  {NoStop}%
\bibitem [{\citenamefont {Tu}\ and\ \citenamefont
  {Zhang}(2008)}]{PhysRevB78.235311}%
  \BibitemOpen
  \bibfield  {author} {\bibinfo {author} {\bibfnamefont {M.~W.~Y.}\
  \bibnamefont {Tu}}\ and\ \bibinfo {author} {\bibfnamefont {W.-M.}\
  \bibnamefont {Zhang}},\ }\bibfield  {title} {\enquote {\bibinfo {title}
  {Non-{M}arkovian decoherence theory for a double-dot charge qubit},}\ }\href
  {https://doi.org/10.1103/PhysRevB.78.235311} {\bibfield  {journal} {\bibinfo
  {journal} {Phys. Rev. B}\ }\textbf {\bibinfo {volume} {78}},\ \bibinfo
  {pages} {235311} (\bibinfo {year} {2008})}\BibitemShut {NoStop}%
\bibitem [{\citenamefont {Jing}\ \emph {et~al.}(2018)\citenamefont {Jing},
  \citenamefont {Yu}, \citenamefont {Lam}, \citenamefont {You},\ and\
  \citenamefont {Wu}}]{PhysRevA97.012104}%
  \BibitemOpen
  \bibfield  {author} {\bibinfo {author} {\bibfnamefont {J.}~\bibnamefont
  {Jing}}, \bibinfo {author} {\bibfnamefont {T.}~\bibnamefont {Yu}}, \bibinfo
  {author} {\bibfnamefont {C.-H.}\ \bibnamefont {Lam}}, \bibinfo {author}
  {\bibfnamefont {J.~Q.}\ \bibnamefont {You}},\ and\ \bibinfo {author}
  {\bibfnamefont {L.-A.}\ \bibnamefont {Wu}},\ }\bibfield  {title} {\enquote
  {\bibinfo {title} {Control relaxation via dephasing: A
  quantum-state-diffusion study},}\ }\href
  {https://doi.org/10.1103/PhysRevA.97.012104} {\bibfield  {journal} {\bibinfo
  {journal} {Phys. Rev. A}\ }\textbf {\bibinfo {volume} {97}},\ \bibinfo
  {pages} {012104} (\bibinfo {year} {2018})}\BibitemShut {NoStop}%
\bibitem [{\citenamefont {Piilo}\ \emph {et~al.}(2008)\citenamefont {Piilo},
  \citenamefont {Maniscalco}, \citenamefont {H\"ark\"onen},\ and\ \citenamefont
  {Suominen}}]{PhysRevLett.100.180402}%
  \BibitemOpen
  \bibfield  {author} {\bibinfo {author} {\bibfnamefont {J.}~\bibnamefont
  {Piilo}}, \bibinfo {author} {\bibfnamefont {S.}~\bibnamefont {Maniscalco}},
  \bibinfo {author} {\bibfnamefont {K.}~\bibnamefont {H\"ark\"onen}},\ and\
  \bibinfo {author} {\bibfnamefont {K.-A.}\ \bibnamefont {Suominen}},\
  }\bibfield  {title} {\enquote {\bibinfo {title} {Non-{M}arkovian quantum
  jumps},}\ }\href {https://doi.org/10.1103/PhysRevLett.100.180402} {\bibfield
  {journal} {\bibinfo  {journal} {Phys. Rev. Lett.}\ }\textbf {\bibinfo
  {volume} {100}},\ \bibinfo {pages} {180402} (\bibinfo {year}
  {2008})}\BibitemShut {NoStop}%
\bibitem [{\citenamefont {Breuer}, \citenamefont {Laine},\ and\ \citenamefont
  {Piilo}(2009)}]{PhysRevLett.103.210401}%
  \BibitemOpen
  \bibfield  {author} {\bibinfo {author} {\bibfnamefont {H.}~\bibnamefont
  {Breuer}}, \bibinfo {author} {\bibfnamefont {E.}~\bibnamefont {Laine}},\ and\
  \bibinfo {author} {\bibfnamefont {J.}~\bibnamefont {Piilo}},\ }\bibfield
  {title} {\enquote {\bibinfo {title} {Measure for the degree of
  non-{M}arkovian behavior of quantum processes in open systems},}\ }\href
  {https://doi.org/10.1103/PhysRevLett.103.210401} {\bibfield  {journal}
  {\bibinfo  {journal} {Phys. Rev. Lett.}\ }\textbf {\bibinfo {volume} {103}},\
  \bibinfo {pages} {210401} (\bibinfo {year} {2009})}\BibitemShut {NoStop}%
\bibitem [{\citenamefont {Rivas}, \citenamefont {Huelga},\ and\ \citenamefont
  {Plenio}(2010)}]{PhysRevLett.105.050403}%
  \BibitemOpen
  \bibfield  {author} {\bibinfo {author} {\bibfnamefont {A.}~\bibnamefont
  {Rivas}}, \bibinfo {author} {\bibfnamefont {S.~F.}\ \bibnamefont {Huelga}},\
  and\ \bibinfo {author} {\bibfnamefont {M.~B.}\ \bibnamefont {Plenio}},\
  }\bibfield  {title} {\enquote {\bibinfo {title} {Entanglement and
  non-{M}arkovianity of quantum evolutions},}\ }\href
  {https://doi.org/10.1103/PhysRevLett.105.050403} {\bibfield  {journal}
  {\bibinfo  {journal} {Phys. Rev. Lett.}\ }\textbf {\bibinfo {volume} {105}},\
  \bibinfo {pages} {050403} (\bibinfo {year} {2010})}\BibitemShut {NoStop}%
\bibitem [{\citenamefont {Zhang}\ \emph {et~al.}(2012)\citenamefont {Zhang},
  \citenamefont {Lo}, \citenamefont {Xiong}, \citenamefont {Tu},\ and\
  \citenamefont {Nori}}]{PhysRevLett.109.170402}%
  \BibitemOpen
  \bibfield  {author} {\bibinfo {author} {\bibfnamefont {W.-M.}\ \bibnamefont
  {Zhang}}, \bibinfo {author} {\bibfnamefont {P.-Y.}\ \bibnamefont {Lo}},
  \bibinfo {author} {\bibfnamefont {H.-N.}\ \bibnamefont {Xiong}}, \bibinfo
  {author} {\bibfnamefont {M.~W.-Y.}\ \bibnamefont {Tu}},\ and\ \bibinfo
  {author} {\bibfnamefont {F.}~\bibnamefont {Nori}},\ }\bibfield  {title}
  {\enquote {\bibinfo {title} {General non-{M}arkovian dynamics of open quantum
  systems},}\ }\href {https://doi.org/10.1103/PhysRevLett.109.170402}
  {\bibfield  {journal} {\bibinfo  {journal} {Phys. Rev. Lett.}\ }\textbf
  {\bibinfo {volume} {109}},\ \bibinfo {pages} {170402} (\bibinfo {year}
  {2012})}\BibitemShut {NoStop}%
\bibitem [{\citenamefont {Rivas}, \citenamefont {Huelga},\ and\ \citenamefont
  {Plenio}(2014)}]{RepProgPhys.77.094001}%
  \BibitemOpen
  \bibfield  {author} {\bibinfo {author} {\bibfnamefont {A.}~\bibnamefont
  {Rivas}}, \bibinfo {author} {\bibfnamefont {S.~F.}\ \bibnamefont {Huelga}},\
  and\ \bibinfo {author} {\bibfnamefont {M.~B.}\ \bibnamefont {Plenio}},\
  }\bibfield  {title} {\enquote {\bibinfo {title} {Quantum non-{M}arkovianity:
  characterization, quantification and detection},}\ }\href
  {https://doi.org/10.1088/0034-4885/77/9/094001} {\bibfield  {journal}
  {\bibinfo  {journal} {Rep. Prog. Phys.}\ }\textbf {\bibinfo {volume} {77}},\
  \bibinfo {pages} {094001} (\bibinfo {year} {2014})}\BibitemShut {NoStop}%
\bibitem [{\citenamefont {Breuer}\ \emph {et~al.}(2016)\citenamefont {Breuer},
  \citenamefont {Laine}, \citenamefont {Piilo},\ and\ \citenamefont
  {Vacchini}}]{RevModPhys.88.021002}%
  \BibitemOpen
  \bibfield  {author} {\bibinfo {author} {\bibfnamefont {H.}~\bibnamefont
  {Breuer}}, \bibinfo {author} {\bibfnamefont {E.}~\bibnamefont {Laine}},
  \bibinfo {author} {\bibfnamefont {J.}~\bibnamefont {Piilo}},\ and\ \bibinfo
  {author} {\bibfnamefont {B.}~\bibnamefont {Vacchini}},\ }\bibfield  {title}
  {\enquote {\bibinfo {title} {\textit{Colloquium}: Non-{M}arkovian dynamics in
  open quantum systems},}\ }\href
  {https://doi.org/10.1103/RevModPhys.88.021002} {\bibfield  {journal}
  {\bibinfo  {journal} {Rev. Mod. Phys.}\ }\textbf {\bibinfo {volume} {88}},\
  \bibinfo {pages} {021002} (\bibinfo {year} {2016})}\BibitemShut {NoStop}%
\bibitem [{\citenamefont {de~Vega}\ and\ \citenamefont
  {Alonso}(2017)}]{RevModPhys.89.015001}%
  \BibitemOpen
  \bibfield  {author} {\bibinfo {author} {\bibfnamefont {I.}~\bibnamefont
  {de~Vega}}\ and\ \bibinfo {author} {\bibfnamefont {D.}~\bibnamefont
  {Alonso}},\ }\bibfield  {title} {\enquote {\bibinfo {title} {Dynamics of
  non-{M}arkovian open quantum systems},}\ }\href
  {https://doi.org/10.1103/RevModPhys.89.015001} {\bibfield  {journal}
  {\bibinfo  {journal} {Rev. Mod. Phys.}\ }\textbf {\bibinfo {volume} {89}},\
  \bibinfo {pages} {015001} (\bibinfo {year} {2017})}\BibitemShut {NoStop}%
\bibitem [{\citenamefont {Chru\'{s}ci\'{n}ski}\ and\ \citenamefont
  {Maniscalco}(2014)}]{PhysRevLett.112.120404}%
  \BibitemOpen
  \bibfield  {author} {\bibinfo {author} {\bibfnamefont {D.}~\bibnamefont
  {Chru\'{s}ci\'{n}ski}}\ and\ \bibinfo {author} {\bibfnamefont
  {S.}~\bibnamefont {Maniscalco}},\ }\bibfield  {title} {\enquote {\bibinfo
  {title} {Degree of non-{M}arkovianity of quantum evolution},}\ }\href
  {https://doi.org/10.1103/PhysRevLett.112.120404} {\bibfield  {journal}
  {\bibinfo  {journal} {Phys. Rev. Lett.}\ }\textbf {\bibinfo {volume} {112}},\
  \bibinfo {pages} {120404} (\bibinfo {year} {2014})}\BibitemShut {NoStop}%
\bibitem [{\citenamefont {Fanchini}\ \emph {et~al.}(2014)\citenamefont
  {Fanchini}, \citenamefont {Karpat}, \citenamefont {\c{C}akmak}, \citenamefont
  {Castelano}, \citenamefont {Aguilar}, \citenamefont {Far\'{\i}as},
  \citenamefont {Walborn}, \citenamefont {Ribeiro},\ and\ \citenamefont
  {de~Oliveira}}]{PhysRevLett.112.210402}%
  \BibitemOpen
  \bibfield  {author} {\bibinfo {author} {\bibfnamefont {F.~F.}\ \bibnamefont
  {Fanchini}}, \bibinfo {author} {\bibfnamefont {G.}~\bibnamefont {Karpat}},
  \bibinfo {author} {\bibfnamefont {B.}~\bibnamefont {\c{C}akmak}}, \bibinfo
  {author} {\bibfnamefont {L.~K.}\ \bibnamefont {Castelano}}, \bibinfo {author}
  {\bibfnamefont {G.~H.}\ \bibnamefont {Aguilar}}, \bibinfo {author}
  {\bibfnamefont {O.~J.}\ \bibnamefont {Far\'{\i}as}}, \bibinfo {author}
  {\bibfnamefont {S.~P.}\ \bibnamefont {Walborn}}, \bibinfo {author}
  {\bibfnamefont {P.~H.~S.}\ \bibnamefont {Ribeiro}},\ and\ \bibinfo {author}
  {\bibfnamefont {M.~C.}\ \bibnamefont {de~Oliveira}},\ }\bibfield  {title}
  {\enquote {\bibinfo {title} {Non-{M}arkovianity through accessible
  information},}\ }\href {https://doi.org/10.1103/PhysRevLett.112.210402}
  {\bibfield  {journal} {\bibinfo  {journal} {Phys. Rev. Lett.}\ }\textbf
  {\bibinfo {volume} {112}},\ \bibinfo {pages} {210402} (\bibinfo {year}
  {2014})}\BibitemShut {NoStop}%
\bibitem [{\citenamefont {Jing}\ \emph {et~al.}(2015)\citenamefont {Jing},
  \citenamefont {Wu}, \citenamefont {Byrd}, \citenamefont {You}, \citenamefont
  {Yu},\ and\ \citenamefont {Wang}}]{PhysRevLett.114.190502}%
  \BibitemOpen
  \bibfield  {author} {\bibinfo {author} {\bibfnamefont {J.}~\bibnamefont
  {Jing}}, \bibinfo {author} {\bibfnamefont {L.-A.}\ \bibnamefont {Wu}},
  \bibinfo {author} {\bibfnamefont {M.}~\bibnamefont {Byrd}}, \bibinfo {author}
  {\bibfnamefont {J.~Q.}\ \bibnamefont {You}}, \bibinfo {author} {\bibfnamefont
  {T.}~\bibnamefont {Yu}},\ and\ \bibinfo {author} {\bibfnamefont {Z.-M.}\
  \bibnamefont {Wang}},\ }\bibfield  {title} {\enquote {\bibinfo {title}
  {Nonperturbative leakage elimination operators and control of a three-level
  system},}\ }\href {https://doi.org/10.1103/PhysRevLett.114.190502} {\bibfield
   {journal} {\bibinfo  {journal} {Phys. Rev. Lett.}\ }\textbf {\bibinfo
  {volume} {114}},\ \bibinfo {pages} {190502} (\bibinfo {year}
  {2015})}\BibitemShut {NoStop}%
\bibitem [{\citenamefont {Luo}\ \emph {et~al.}(2019)\citenamefont {Luo},
  \citenamefont {Lin}, \citenamefont {You}, \citenamefont {Wu}, \citenamefont
  {Chatterjee},\ and\ \citenamefont {Yu}}]{PhysRevA100.062112}%
  \BibitemOpen
  \bibfield  {author} {\bibinfo {author} {\bibfnamefont {D.-W.}\ \bibnamefont
  {Luo}}, \bibinfo {author} {\bibfnamefont {H.-Q.}\ \bibnamefont {Lin}},
  \bibinfo {author} {\bibfnamefont {J.~Q.}\ \bibnamefont {You}}, \bibinfo
  {author} {\bibfnamefont {L.-A.}\ \bibnamefont {Wu}}, \bibinfo {author}
  {\bibfnamefont {R.}~\bibnamefont {Chatterjee}},\ and\ \bibinfo {author}
  {\bibfnamefont {T.}~\bibnamefont {Yu}},\ }\bibfield  {title} {\enquote
  {\bibinfo {title} {Geometric decoherence in diffusive open quantum
  systems},}\ }\href {https://doi.org/10.1103/PhysRevA.100.062112} {\bibfield
  {journal} {\bibinfo  {journal} {Phys. Rev. A}\ }\textbf {\bibinfo {volume}
  {100}},\ \bibinfo {pages} {062112} (\bibinfo {year} {2019})}\BibitemShut
  {NoStop}%
\bibitem [{\citenamefont {Chiang}\ and\ \citenamefont
  {Zhang}(2021)}]{PhysRevA103.013714}%
  \BibitemOpen
  \bibfield  {author} {\bibinfo {author} {\bibfnamefont {K.-T.}\ \bibnamefont
  {Chiang}}\ and\ \bibinfo {author} {\bibfnamefont {W.-M.}\ \bibnamefont
  {Zhang}},\ }\bibfield  {title} {\enquote {\bibinfo {title} {Non-{M}arkovian
  decoherence dynamics of strong-coupling hybrid quantum systems: A master
  equation approach},}\ }\href {https://doi.org/10.1103/PhysRevA.103.013714}
  {\bibfield  {journal} {\bibinfo  {journal} {Phys. Rev. A}\ }\textbf {\bibinfo
  {volume} {103}},\ \bibinfo {pages} {013714} (\bibinfo {year}
  {2021})}\BibitemShut {NoStop}%
\bibitem [{\citenamefont {Taddei}\ \emph {et~al.}(2013)\citenamefont {Taddei},
  \citenamefont {Escher}, \citenamefont {Davidovich},\ and\ \citenamefont {{de
  Matos Filho}}}]{PhysRevLett.110.050402}%
  \BibitemOpen
  \bibfield  {author} {\bibinfo {author} {\bibfnamefont {M.~M.}\ \bibnamefont
  {Taddei}}, \bibinfo {author} {\bibfnamefont {B.~M.}\ \bibnamefont {Escher}},
  \bibinfo {author} {\bibfnamefont {L.}~\bibnamefont {Davidovich}},\ and\
  \bibinfo {author} {\bibfnamefont {R.~L.}\ \bibnamefont {{de Matos Filho}}},\
  }\bibfield  {title} {\enquote {\bibinfo {title} {Quantum speed limit for
  physical processes},}\ }\href
  {https://doi.org/10.1103/PhysRevLett.110.050402} {\bibfield  {journal}
  {\bibinfo  {journal} {Phys. Rev. Lett.}\ }\textbf {\bibinfo {volume} {110}},\
  \bibinfo {pages} {050402} (\bibinfo {year} {2013})}\BibitemShut {NoStop}%
\bibitem [{\citenamefont {{del Campo}}\ \emph {et~al.}(2013)\citenamefont {{del
  Campo}}, \citenamefont {Egusquiza}, \citenamefont {Plenio},\ and\
  \citenamefont {Huelga}}]{PhysRevLett.110.050403}%
  \BibitemOpen
  \bibfield  {author} {\bibinfo {author} {\bibfnamefont {A.}~\bibnamefont {{del
  Campo}}}, \bibinfo {author} {\bibfnamefont {I.~L.}\ \bibnamefont
  {Egusquiza}}, \bibinfo {author} {\bibfnamefont {M.~B.}\ \bibnamefont
  {Plenio}},\ and\ \bibinfo {author} {\bibfnamefont {S.~F.}\ \bibnamefont
  {Huelga}},\ }\bibfield  {title} {\enquote {\bibinfo {title} {Quantum speed
  limits in open system dynamics},}\ }\href
  {https://doi.org/10.1103/PhysRevLett.110.050403} {\bibfield  {journal}
  {\bibinfo  {journal} {Phys. Rev. Lett.}\ }\textbf {\bibinfo {volume} {110}},\
  \bibinfo {pages} {050403} (\bibinfo {year} {2013})}\BibitemShut {NoStop}%
\bibitem [{\citenamefont {Deffner}\ and\ \citenamefont
  {Lutz}(2013{\natexlab{b}})}]{PhysRevLett.111.010402}%
  \BibitemOpen
  \bibfield  {author} {\bibinfo {author} {\bibfnamefont {S.}~\bibnamefont
  {Deffner}}\ and\ \bibinfo {author} {\bibfnamefont {E.}~\bibnamefont {Lutz}},\
  }\bibfield  {title} {\enquote {\bibinfo {title} {Quantum speed limit for
  non-{M}arkovian dynamics},}\ }\href
  {https://doi.org/10.1103/PhysRevLett.111.010402} {\bibfield  {journal}
  {\bibinfo  {journal} {Phys. Rev. Lett.}\ }\textbf {\bibinfo {volume} {111}},\
  \bibinfo {pages} {010402} (\bibinfo {year} {2013}{\natexlab{b}})}\BibitemShut
  {NoStop}%
\bibitem [{\citenamefont {Liu}, \citenamefont {Xu},\ and\ \citenamefont
  {Zhu}(2015)}]{PhysRevA91.022102}%
  \BibitemOpen
  \bibfield  {author} {\bibinfo {author} {\bibfnamefont {C.}~\bibnamefont
  {Liu}}, \bibinfo {author} {\bibfnamefont {Z.}~\bibnamefont {Xu}},\ and\
  \bibinfo {author} {\bibfnamefont {S.}~\bibnamefont {Zhu}},\ }\bibfield
  {title} {\enquote {\bibinfo {title} {Quantum-speed-limit time for multiqubit
  open systems},}\ }\href {https://doi.org/10.1103/PhysRevA.91.022102}
  {\bibfield  {journal} {\bibinfo  {journal} {Phys. Rev. A}\ }\textbf {\bibinfo
  {volume} {91}},\ \bibinfo {pages} {022102} (\bibinfo {year}
  {2015})}\BibitemShut {NoStop}%
\bibitem [{\citenamefont {Zhang}\ \emph
  {et~al.}(2015{\natexlab{a}})\citenamefont {Zhang}, \citenamefont {Han},
  \citenamefont {Xia}, \citenamefont {Cao},\ and\ \citenamefont
  {Fan}}]{PhysRevA91.032112}%
  \BibitemOpen
  \bibfield  {author} {\bibinfo {author} {\bibfnamefont {Y.}~\bibnamefont
  {Zhang}}, \bibinfo {author} {\bibfnamefont {W.}~\bibnamefont {Han}}, \bibinfo
  {author} {\bibfnamefont {Y.}~\bibnamefont {Xia}}, \bibinfo {author}
  {\bibfnamefont {J.}~\bibnamefont {Cao}},\ and\ \bibinfo {author}
  {\bibfnamefont {H.}~\bibnamefont {Fan}},\ }\bibfield  {title} {\enquote
  {\bibinfo {title} {Classical-driving-assisted quantum speed-up},}\ }\href
  {https://doi.org/10.1103/PhysRevA.91.032112} {\bibfield  {journal} {\bibinfo
  {journal} {Phys. Rev. A}\ }\textbf {\bibinfo {volume} {91}},\ \bibinfo
  {pages} {032112} (\bibinfo {year} {2015}{\natexlab{a}})}\BibitemShut
  {NoStop}%
\bibitem [{\citenamefont {Liu}\ \emph {et~al.}(2016)\citenamefont {Liu},
  \citenamefont {Yang}, \citenamefont {An},\ and\ \citenamefont
  {Xu}}]{PhysRevA93.020105}%
  \BibitemOpen
  \bibfield  {author} {\bibinfo {author} {\bibfnamefont {H.}~\bibnamefont
  {Liu}}, \bibinfo {author} {\bibfnamefont {W.~L.}\ \bibnamefont {Yang}},
  \bibinfo {author} {\bibfnamefont {J.}~\bibnamefont {An}},\ and\ \bibinfo
  {author} {\bibfnamefont {Z.}~\bibnamefont {Xu}},\ }\bibfield  {title}
  {\enquote {\bibinfo {title} {Mechanism for quantum speedup in open quantum
  systems},}\ }\href {https://doi.org/10.1103/PhysRevA.93.020105} {\bibfield
  {journal} {\bibinfo  {journal} {Phys. Rev. A}\ }\textbf {\bibinfo {volume}
  {93}},\ \bibinfo {pages} {020105} (\bibinfo {year} {2016})}\BibitemShut
  {NoStop}%
\bibitem [{\citenamefont {Mirkin}, \citenamefont {Toscano},\ and\ \citenamefont
  {Wisniacki}(2016)}]{PhysRevA94.052125}%
  \BibitemOpen
  \bibfield  {author} {\bibinfo {author} {\bibfnamefont {N.}~\bibnamefont
  {Mirkin}}, \bibinfo {author} {\bibfnamefont {F.}~\bibnamefont {Toscano}},\
  and\ \bibinfo {author} {\bibfnamefont {D.~A.}\ \bibnamefont {Wisniacki}},\
  }\bibfield  {title} {\enquote {\bibinfo {title} {Quantum-speed-limit bounds
  in an open quantum evolution},}\ }\href
  {https://doi.org/10.1103/PhysRevA.94.052125} {\bibfield  {journal} {\bibinfo
  {journal} {Phys. Rev. A}\ }\textbf {\bibinfo {volume} {94}},\ \bibinfo
  {pages} {052125} (\bibinfo {year} {2016})}\BibitemShut {NoStop}%
\bibitem [{\citenamefont {Cianciaruso}, \citenamefont {Maniscalco},\ and\
  \citenamefont {Adesso}(2017)}]{PhysRevA96.012105}%
  \BibitemOpen
  \bibfield  {author} {\bibinfo {author} {\bibfnamefont {M.}~\bibnamefont
  {Cianciaruso}}, \bibinfo {author} {\bibfnamefont {S.}~\bibnamefont
  {Maniscalco}},\ and\ \bibinfo {author} {\bibfnamefont {G.}~\bibnamefont
  {Adesso}},\ }\bibfield  {title} {\enquote {\bibinfo {title} {Role of
  non-{M}arkovianity and backflow of information in the speed of quantum
  evolution},}\ }\href {https://doi.org/10.1103/PhysRevA.96.012105} {\bibfield
  {journal} {\bibinfo  {journal} {Phys. Rev. A}\ }\textbf {\bibinfo {volume}
  {96}},\ \bibinfo {pages} {012105} (\bibinfo {year} {2017})}\BibitemShut
  {NoStop}%
\bibitem [{\citenamefont {Xu}, \citenamefont {Zhang},\ and\ \citenamefont
  {Liu}(2019)}]{PhysRevA100.052305}%
  \BibitemOpen
  \bibfield  {author} {\bibinfo {author} {\bibfnamefont {K.}~\bibnamefont
  {Xu}}, \bibinfo {author} {\bibfnamefont {G.-F.}\ \bibnamefont {Zhang}},\ and\
  \bibinfo {author} {\bibfnamefont {W.-M.}\ \bibnamefont {Liu}},\ }\bibfield
  {title} {\enquote {\bibinfo {title} {Quantum dynamical speedup in correlated
  noisy channels},}\ }\href {https://doi.org/10.1103/PhysRevA.100.052305}
  {\bibfield  {journal} {\bibinfo  {journal} {Phys. Rev. A}\ }\textbf {\bibinfo
  {volume} {100}},\ \bibinfo {pages} {052305} (\bibinfo {year}
  {2019})}\BibitemShut {NoStop}%
\bibitem [{\citenamefont {Pires}\ \emph {et~al.}(2016)\citenamefont {Pires},
  \citenamefont {Cianciaruso}, \citenamefont {C\'eleri}, \citenamefont
  {Adesso},\ and\ \citenamefont {Soares-Pinto}}]{PhysRevX6.021031}%
  \BibitemOpen
  \bibfield  {author} {\bibinfo {author} {\bibfnamefont {D.~P.}\ \bibnamefont
  {Pires}}, \bibinfo {author} {\bibfnamefont {M.}~\bibnamefont {Cianciaruso}},
  \bibinfo {author} {\bibfnamefont {L.~C.}\ \bibnamefont {C\'eleri}}, \bibinfo
  {author} {\bibfnamefont {G.}~\bibnamefont {Adesso}},\ and\ \bibinfo {author}
  {\bibfnamefont {D.~O.}\ \bibnamefont {Soares-Pinto}},\ }\bibfield  {title}
  {\enquote {\bibinfo {title} {Generalized geometric quantum speed limits},}\
  }\href {https://doi.org/10.1103/PhysRevX.6.021031} {\bibfield  {journal}
  {\bibinfo  {journal} {Phys. Rev. X}\ }\textbf {\bibinfo {volume} {6}},\
  \bibinfo {pages} {021031} (\bibinfo {year} {2016})}\BibitemShut {NoStop}%
\bibitem [{\citenamefont {Cai}\ and\ \citenamefont
  {Zheng}(2017)}]{PhysRevA95.052104}%
  \BibitemOpen
  \bibfield  {author} {\bibinfo {author} {\bibfnamefont {X.}~\bibnamefont
  {Cai}}\ and\ \bibinfo {author} {\bibfnamefont {Y.}~\bibnamefont {Zheng}},\
  }\bibfield  {title} {\enquote {\bibinfo {title} {Quantum dynamical speedup in
  a nonequilibrium environment},}\ }\href
  {https://doi.org/10.1103/PhysRevA.95.052104} {\bibfield  {journal} {\bibinfo
  {journal} {Phys. Rev. A}\ }\textbf {\bibinfo {volume} {95}},\ \bibinfo
  {pages} {052104} (\bibinfo {year} {2017})}\BibitemShut {NoStop}%
\bibitem [{\citenamefont {Teittinen}, \citenamefont {Lyyra},\ and\
  \citenamefont {Maniscalco}(2019)}]{NewJPhys.21.123041}%
  \BibitemOpen
  \bibfield  {author} {\bibinfo {author} {\bibfnamefont {J.}~\bibnamefont
  {Teittinen}}, \bibinfo {author} {\bibfnamefont {H.}~\bibnamefont {Lyyra}},\
  and\ \bibinfo {author} {\bibfnamefont {S.}~\bibnamefont {Maniscalco}},\
  }\bibfield  {title} {\enquote {\bibinfo {title} {There is no general
  connection between the quantum speed limit and non-{M}arkovianity},}\ }\href
  {https://doi.org/10.1088/1367-2630/ab59fe} {\bibfield  {journal} {\bibinfo
  {journal} {New J. Phys.}\ }\textbf {\bibinfo {volume} {21}},\ \bibinfo
  {pages} {123041} (\bibinfo {year} {2019})}\BibitemShut {NoStop}%
\bibitem [{\citenamefont {Marvian}\ and\ \citenamefont
  {Lidar}(2015)}]{PhysRevLett.115.210402}%
  \BibitemOpen
  \bibfield  {author} {\bibinfo {author} {\bibfnamefont {I.}~\bibnamefont
  {Marvian}}\ and\ \bibinfo {author} {\bibfnamefont {D.~A.}\ \bibnamefont
  {Lidar}},\ }\bibfield  {title} {\enquote {\bibinfo {title} {Quantum speed
  limits for leakage and decoherence},}\ }\href
  {https://doi.org/10.1103/PhysRevLett.115.210402} {\bibfield  {journal}
  {\bibinfo  {journal} {Phys. Rev. Lett.}\ }\textbf {\bibinfo {volume} {115}},\
  \bibinfo {pages} {210402} (\bibinfo {year} {2015})}\BibitemShut {NoStop}%
\bibitem [{\citenamefont {Zhang}\ \emph {et~al.}(2014)\citenamefont {Zhang},
  \citenamefont {Han}, \citenamefont {Xia}, \citenamefont {Cao},\ and\
  \citenamefont {Fan}}]{SciRep.4.4890}%
  \BibitemOpen
  \bibfield  {author} {\bibinfo {author} {\bibfnamefont {Y.}~\bibnamefont
  {Zhang}}, \bibinfo {author} {\bibfnamefont {W.}~\bibnamefont {Han}}, \bibinfo
  {author} {\bibfnamefont {Y.}~\bibnamefont {Xia}}, \bibinfo {author}
  {\bibfnamefont {J.}~\bibnamefont {Cao}},\ and\ \bibinfo {author}
  {\bibfnamefont {H.}~\bibnamefont {Fan}},\ }\bibfield  {title} {\enquote
  {\bibinfo {title} {Quantum speed limit for arbitrary initial states},}\
  }\href {https://doi.org/10.1038/srep04890} {\bibfield  {journal} {\bibinfo
  {journal} {Sci. Rep.}\ }\textbf {\bibinfo {volume} {4}},\ \bibinfo {pages}
  {4890} (\bibinfo {year} {2014})}\BibitemShut {NoStop}%
\bibitem [{\citenamefont {Xu}\ \emph {et~al.}(2018)\citenamefont {Xu},
  \citenamefont {Zhang}, \citenamefont {Xia}, \citenamefont {Wang},\ and\
  \citenamefont {Fan}}]{PhysRevA98.022114}%
  \BibitemOpen
  \bibfield  {author} {\bibinfo {author} {\bibfnamefont {K.}~\bibnamefont
  {Xu}}, \bibinfo {author} {\bibfnamefont {Y.-J.}\ \bibnamefont {Zhang}},
  \bibinfo {author} {\bibfnamefont {Y.-J.}\ \bibnamefont {Xia}}, \bibinfo
  {author} {\bibfnamefont {Z.~D.}\ \bibnamefont {Wang}},\ and\ \bibinfo
  {author} {\bibfnamefont {H.}~\bibnamefont {Fan}},\ }\bibfield  {title}
  {\enquote {\bibinfo {title} {Hierarchical-environment-assisted
  non-{M}arkovian speedup dynamics control},}\ }\href
  {https://doi.org/10.1103/PhysRevA.98.022114} {\bibfield  {journal} {\bibinfo
  {journal} {Phys. Rev. A}\ }\textbf {\bibinfo {volume} {98}},\ \bibinfo
  {pages} {022114} (\bibinfo {year} {2018})}\BibitemShut {NoStop}%
\bibitem [{\citenamefont {Wu}\ and\ \citenamefont
  {Yu}(2018)}]{PhysRevA98.042132}%
  \BibitemOpen
  \bibfield  {author} {\bibinfo {author} {\bibfnamefont {S.-x.}\ \bibnamefont
  {Wu}}\ and\ \bibinfo {author} {\bibfnamefont {C.-s.}\ \bibnamefont {Yu}},\
  }\bibfield  {title} {\enquote {\bibinfo {title} {Quantum speed limit for a
  mixed initial state},}\ }\href {https://doi.org/10.1103/PhysRevA.98.042132}
  {\bibfield  {journal} {\bibinfo  {journal} {Phys. Rev. A}\ }\textbf {\bibinfo
  {volume} {98}},\ \bibinfo {pages} {042132} (\bibinfo {year}
  {2018})}\BibitemShut {NoStop}%
\bibitem [{\citenamefont {Wang}\ and\ \citenamefont
  {P\'erez-Bernal}(2019)}]{PhysRevA100.022118}%
  \BibitemOpen
  \bibfield  {author} {\bibinfo {author} {\bibfnamefont {Q.}~\bibnamefont
  {Wang}}\ and\ \bibinfo {author} {\bibfnamefont {F.}~\bibnamefont
  {P\'erez-Bernal}},\ }\bibfield  {title} {\enquote {\bibinfo {title}
  {Excited-state quantum phase transition and the quantum-speed-limit time},}\
  }\href {https://doi.org/10.1103/PhysRevA.100.022118} {\bibfield  {journal}
  {\bibinfo  {journal} {Phys. Rev. A}\ }\textbf {\bibinfo {volume} {100}},\
  \bibinfo {pages} {022118} (\bibinfo {year} {2019})}\BibitemShut {NoStop}%
\bibitem [{\citenamefont {Funo}, \citenamefont {Shiraishi},\ and\ \citenamefont
  {Saito}(2019)}]{NewJPhys.21.013006}%
  \BibitemOpen
  \bibfield  {author} {\bibinfo {author} {\bibfnamefont {K.}~\bibnamefont
  {Funo}}, \bibinfo {author} {\bibfnamefont {N.}~\bibnamefont {Shiraishi}},\
  and\ \bibinfo {author} {\bibfnamefont {K.}~\bibnamefont {Saito}},\ }\bibfield
   {title} {\enquote {\bibinfo {title} {Speed limit for open quantum
  systems},}\ }\href {https://doi.org/10.1088/1367-2630/aaf9f5} {\bibfield
  {journal} {\bibinfo  {journal} {New J. Phys.}\ }\textbf {\bibinfo {volume}
  {21}},\ \bibinfo {pages} {013006} (\bibinfo {year} {2019})}\BibitemShut
  {NoStop}%
\bibitem [{\citenamefont {Shao}\ \emph {et~al.}(2020)\citenamefont {Shao},
  \citenamefont {Liu}, \citenamefont {Zhang}, \citenamefont {Yuan},\ and\
  \citenamefont {Liu}}]{PhysRevRes.2.023299}%
  \BibitemOpen
  \bibfield  {author} {\bibinfo {author} {\bibfnamefont {Y.}~\bibnamefont
  {Shao}}, \bibinfo {author} {\bibfnamefont {B.}~\bibnamefont {Liu}}, \bibinfo
  {author} {\bibfnamefont {M.}~\bibnamefont {Zhang}}, \bibinfo {author}
  {\bibfnamefont {H.}~\bibnamefont {Yuan}},\ and\ \bibinfo {author}
  {\bibfnamefont {J.}~\bibnamefont {Liu}},\ }\bibfield  {title} {\enquote
  {\bibinfo {title} {Operational definition of a quantum speed limit},}\ }\href
  {https://doi.org/10.1103/PhysRevResearch.2.023299} {\bibfield  {journal}
  {\bibinfo  {journal} {Phys. Rev. Res.}\ }\textbf {\bibinfo {volume} {2}},\
  \bibinfo {pages} {023299} (\bibinfo {year} {2020})}\BibitemShut {NoStop}%
\bibitem [{\citenamefont {O'Connor}, \citenamefont {Guarnieri},\ and\
  \citenamefont {Campbell}(2021)}]{PhysRevA103.022210}%
  \BibitemOpen
  \bibfield  {author} {\bibinfo {author} {\bibfnamefont {E.}~\bibnamefont
  {O'Connor}}, \bibinfo {author} {\bibfnamefont {G.}~\bibnamefont
  {Guarnieri}},\ and\ \bibinfo {author} {\bibfnamefont {S.}~\bibnamefont
  {Campbell}},\ }\bibfield  {title} {\enquote {\bibinfo {title} {Action quantum
  speed limits},}\ }\href {https://doi.org/10.1103/PhysRevA.103.022210}
  {\bibfield  {journal} {\bibinfo  {journal} {Phys. Rev. A}\ }\textbf {\bibinfo
  {volume} {103}},\ \bibinfo {pages} {022210} (\bibinfo {year}
  {2021})}\BibitemShut {NoStop}%
\bibitem [{\citenamefont {Nie}\ \emph {et~al.}(2021)\citenamefont {Nie},
  \citenamefont {Ren}, \citenamefont {He}, \citenamefont {Wu},\ and\
  \citenamefont {Wang}}]{PhysRevA104.052424}%
  \BibitemOpen
  \bibfield  {author} {\bibinfo {author} {\bibfnamefont {S.-S.}\ \bibnamefont
  {Nie}}, \bibinfo {author} {\bibfnamefont {F.-H.}\ \bibnamefont {Ren}},
  \bibinfo {author} {\bibfnamefont {R.-H.}\ \bibnamefont {He}}, \bibinfo
  {author} {\bibfnamefont {J.}~\bibnamefont {Wu}},\ and\ \bibinfo {author}
  {\bibfnamefont {Z.-M.}\ \bibnamefont {Wang}},\ }\bibfield  {title} {\enquote
  {\bibinfo {title} {Control cost and quantum speed limit time in controlled
  almost-exact state transmission in open systems},}\ }\href
  {https://doi.org/10.1103/PhysRevA.104.052424} {\bibfield  {journal} {\bibinfo
   {journal} {Phys. Rev. A}\ }\textbf {\bibinfo {volume} {104}},\ \bibinfo
  {pages} {052424} (\bibinfo {year} {2021})}\BibitemShut {NoStop}%
\bibitem [{\citenamefont {Fogarty}\ \emph {et~al.}(2020)\citenamefont
  {Fogarty}, \citenamefont {Deffner}, \citenamefont {Busch},\ and\
  \citenamefont {Campbell}}]{PhysRevLett.124.110601}%
  \BibitemOpen
  \bibfield  {author} {\bibinfo {author} {\bibfnamefont {T.}~\bibnamefont
  {Fogarty}}, \bibinfo {author} {\bibfnamefont {S.}~\bibnamefont {Deffner}},
  \bibinfo {author} {\bibfnamefont {T.}~\bibnamefont {Busch}},\ and\ \bibinfo
  {author} {\bibfnamefont {S.}~\bibnamefont {Campbell}},\ }\bibfield  {title}
  {\enquote {\bibinfo {title} {Orthogonality catastrophe as a consequence of
  the quantum speed limit},}\ }\href
  {https://doi.org/10.1103/PhysRevLett.124.110601} {\bibfield  {journal}
  {\bibinfo  {journal} {Phys. Rev. Lett.}\ }\textbf {\bibinfo {volume} {124}},\
  \bibinfo {pages} {110601} (\bibinfo {year} {2020})}\BibitemShut {NoStop}%
\bibitem [{\citenamefont {Yadin}, \citenamefont {Imai},\ and\ \citenamefont
  {G\"uhne}(2024)}]{PhysRevLett.132.230404}%
  \BibitemOpen
  \bibfield  {author} {\bibinfo {author} {\bibfnamefont {B.}~\bibnamefont
  {Yadin}}, \bibinfo {author} {\bibfnamefont {S.}~\bibnamefont {Imai}},\ and\
  \bibinfo {author} {\bibfnamefont {O.}~\bibnamefont {G\"uhne}},\ }\bibfield
  {title} {\enquote {\bibinfo {title} {Quantum speed limit for states and
  observables of perturbed open systems},}\ }\href
  {https://doi.org/10.1103/PhysRevLett.132.230404} {\bibfield  {journal}
  {\bibinfo  {journal} {Phys. Rev. Lett.}\ }\textbf {\bibinfo {volume} {132}},\
  \bibinfo {pages} {230404} (\bibinfo {year} {2024})}\BibitemShut {NoStop}%
\bibitem [{\citenamefont {Peng}\ \emph {et~al.}(2024)\citenamefont {Peng},
  \citenamefont {C\'eleri}, \citenamefont {Basit},\ and\ \citenamefont
  {Xianlong}}]{PhysRevA110.052433}%
  \BibitemOpen
  \bibfield  {author} {\bibinfo {author} {\bibfnamefont {Z.}~\bibnamefont
  {Peng}}, \bibinfo {author} {\bibfnamefont {L.~C.}\ \bibnamefont {C\'eleri}},
  \bibinfo {author} {\bibfnamefont {A.}~\bibnamefont {Basit}},\ and\ \bibinfo
  {author} {\bibfnamefont {G.}~\bibnamefont {Xianlong}},\ }\bibfield  {title}
  {\enquote {\bibinfo {title} {Effects of reservoir squeezing on the
  amplification of quantum correlation and the quantum speed limit},}\ }\href
  {https://doi.org/10.1103/PhysRevA.110.052433} {\bibfield  {journal} {\bibinfo
   {journal} {Phys. Rev. A}\ }\textbf {\bibinfo {volume} {110}},\ \bibinfo
  {pages} {052433} (\bibinfo {year} {2024})}\BibitemShut {NoStop}%
\bibitem [{\citenamefont {Campaioli}\ \emph {et~al.}(2018)\citenamefont
  {Campaioli}, \citenamefont {Pollock}, \citenamefont {Binder},\ and\
  \citenamefont {Modi}}]{PhysRevLett.120.060409}%
  \BibitemOpen
  \bibfield  {author} {\bibinfo {author} {\bibfnamefont {F.}~\bibnamefont
  {Campaioli}}, \bibinfo {author} {\bibfnamefont {F.~A.}\ \bibnamefont
  {Pollock}}, \bibinfo {author} {\bibfnamefont {F.~C.}\ \bibnamefont
  {Binder}},\ and\ \bibinfo {author} {\bibfnamefont {K.}~\bibnamefont {Modi}},\
  }\bibfield  {title} {\enquote {\bibinfo {title} {Tightening quantum speed
  limits for almost all states},}\ }\href
  {https://doi.org/10.1103/PhysRevLett.120.060409} {\bibfield  {journal}
  {\bibinfo  {journal} {Phys. Rev. Lett.}\ }\textbf {\bibinfo {volume} {120}},\
  \bibinfo {pages} {060409} (\bibinfo {year} {2018})}\BibitemShut {NoStop}%
\bibitem [{\citenamefont {Campaioli}, \citenamefont {Pollock},\ and\
  \citenamefont {Modi}(2019)}]{Quantum3.168}%
  \BibitemOpen
  \bibfield  {author} {\bibinfo {author} {\bibfnamefont {F.}~\bibnamefont
  {Campaioli}}, \bibinfo {author} {\bibfnamefont {F.~A.}\ \bibnamefont
  {Pollock}},\ and\ \bibinfo {author} {\bibfnamefont {K.}~\bibnamefont
  {Modi}},\ }\bibfield  {title} {\enquote {\bibinfo {title} {Tight, robust, and
  feasible quantum speed limits for open dynamics},}\ }\href
  {https://doi.org/10.22331/q-2019-08-05-168} {\bibfield  {journal} {\bibinfo
  {journal} {Quantum}\ }\textbf {\bibinfo {volume} {3}},\ \bibinfo {pages}
  {168} (\bibinfo {year} {2019})}\BibitemShut {NoStop}%
\bibitem [{\citenamefont {Mai}\ and\ \citenamefont
  {Yu}(2023)}]{PhysRevA108.052207}%
  \BibitemOpen
  \bibfield  {author} {\bibinfo {author} {\bibfnamefont {Z.-y.}\ \bibnamefont
  {Mai}}\ and\ \bibinfo {author} {\bibfnamefont {C.-s.}\ \bibnamefont {Yu}},\
  }\bibfield  {title} {\enquote {\bibinfo {title} {Tight and attainable quantum
  speed limit for open systems},}\ }\href
  {https://doi.org/10.1103/PhysRevA.108.052207} {\bibfield  {journal} {\bibinfo
   {journal} {Phys. Rev. A}\ }\textbf {\bibinfo {volume} {108}},\ \bibinfo
  {pages} {052207} (\bibinfo {year} {2023})}\BibitemShut {NoStop}%
\bibitem [{\citenamefont {Pires}(2022)}]{PhysRevA106.012403}%
  \BibitemOpen
  \bibfield  {author} {\bibinfo {author} {\bibfnamefont {D.~P.}\ \bibnamefont
  {Pires}},\ }\bibfield  {title} {\enquote {\bibinfo {title} {Unified entropies
  and quantum speed limits for nonunitary dynamics},}\ }\href
  {https://doi.org/10.1103/PhysRevA.106.012403} {\bibfield  {journal} {\bibinfo
   {journal} {Phys. Rev. A}\ }\textbf {\bibinfo {volume} {106}},\ \bibinfo
  {pages} {012403} (\bibinfo {year} {2022})}\BibitemShut {NoStop}%
\bibitem [{\citenamefont {Garc\'{\i}a-Pintos}\ \emph
  {et~al.}(2022)\citenamefont {Garc\'{\i}a-Pintos}, \citenamefont {Nicholson},
  \citenamefont {Green}, \citenamefont {del Campo},\ and\ \citenamefont
  {Gorshkov}}]{PhysRevX12.011038}%
  \BibitemOpen
  \bibfield  {author} {\bibinfo {author} {\bibfnamefont {L.~P.}\ \bibnamefont
  {Garc\'{\i}a-Pintos}}, \bibinfo {author} {\bibfnamefont {S.~B.}\ \bibnamefont
  {Nicholson}}, \bibinfo {author} {\bibfnamefont {J.~R.}\ \bibnamefont
  {Green}}, \bibinfo {author} {\bibfnamefont {A.}~\bibnamefont {del Campo}},\
  and\ \bibinfo {author} {\bibfnamefont {A.~V.}\ \bibnamefont {Gorshkov}},\
  }\bibfield  {title} {\enquote {\bibinfo {title} {Unifying quantum and
  classical speed limits on observables},}\ }\href
  {https://doi.org/10.1103/PhysRevX.12.011038} {\bibfield  {journal} {\bibinfo
  {journal} {Phys. Rev. X}\ }\textbf {\bibinfo {volume} {12}},\ \bibinfo
  {pages} {011038} (\bibinfo {year} {2022})}\BibitemShut {NoStop}%
\bibitem [{\citenamefont {Van~Vu}\ and\ \citenamefont
  {Saito}(2023)}]{PhysRevX13.011013}%
  \BibitemOpen
  \bibfield  {author} {\bibinfo {author} {\bibfnamefont {T.}~\bibnamefont
  {Van~Vu}}\ and\ \bibinfo {author} {\bibfnamefont {K.}~\bibnamefont {Saito}},\
  }\bibfield  {title} {\enquote {\bibinfo {title} {Thermodynamic unification of
  optimal transport: Thermodynamic uncertainty relation, minimum dissipation,
  and thermodynamic speed limits},}\ }\href
  {https://doi.org/10.1103/PhysRevX.13.011013} {\bibfield  {journal} {\bibinfo
  {journal} {Phys. Rev. X}\ }\textbf {\bibinfo {volume} {13}},\ \bibinfo
  {pages} {011013} (\bibinfo {year} {2023})}\BibitemShut {NoStop}%
\bibitem [{\citenamefont {Hadipour}\ \emph {et~al.}(2022)\citenamefont
  {Hadipour}, \citenamefont {Haseli}, \citenamefont {Dolatkhah}, \citenamefont
  {Haddadi},\ and\ \citenamefont {Czerwinski}}]{Photonics9.875}%
  \BibitemOpen
  \bibfield  {author} {\bibinfo {author} {\bibfnamefont {M.}~\bibnamefont
  {Hadipour}}, \bibinfo {author} {\bibfnamefont {S.}~\bibnamefont {Haseli}},
  \bibinfo {author} {\bibfnamefont {H.}~\bibnamefont {Dolatkhah}}, \bibinfo
  {author} {\bibfnamefont {S.}~\bibnamefont {Haddadi}},\ and\ \bibinfo {author}
  {\bibfnamefont {A.}~\bibnamefont {Czerwinski}},\ }\bibfield  {title}
  {\enquote {\bibinfo {title} {Quantum speed limit for a moving qubit inside a
  leaky cavity},}\ }\href {https://doi.org/10.3390/photonics9110875} {\bibfield
   {journal} {\bibinfo  {journal} {Photonics}\ }\textbf {\bibinfo {volume}
  {9}},\ \bibinfo {pages} {875} (\bibinfo {year} {2022})}\BibitemShut {NoStop}%
\bibitem [{\citenamefont {Wei}\ \emph {et~al.}(2023)\citenamefont {Wei},
  \citenamefont {Han}, \citenamefont {Zhang}, \citenamefont {Du}, \citenamefont
  {Xia},\ and\ \citenamefont {Fan}}]{PhysRevD108.126011}%
  \BibitemOpen
  \bibfield  {author} {\bibinfo {author} {\bibfnamefont {Z.-D.}\ \bibnamefont
  {Wei}}, \bibinfo {author} {\bibfnamefont {W.}~\bibnamefont {Han}}, \bibinfo
  {author} {\bibfnamefont {Y.-J.}\ \bibnamefont {Zhang}}, \bibinfo {author}
  {\bibfnamefont {S.-J.}\ \bibnamefont {Du}}, \bibinfo {author} {\bibfnamefont
  {Y.-J.}\ \bibnamefont {Xia}},\ and\ \bibinfo {author} {\bibfnamefont
  {H.}~\bibnamefont {Fan}},\ }\bibfield  {title} {\enquote {\bibinfo {title}
  {Non-{M}arkovian speedup dynamics of a photon induced by gravitational
  redshift},}\ }\href {https://doi.org/10.1103/PhysRevD.108.126011} {\bibfield
  {journal} {\bibinfo  {journal} {Phys. Rev. D}\ }\textbf {\bibinfo {volume}
  {108}},\ \bibinfo {pages} {126011} (\bibinfo {year} {2023})}\BibitemShut
  {NoStop}%
\bibitem [{\citenamefont {Shrimali}, \citenamefont {Bhowmick},\ and\
  \citenamefont {Pati}(2025)}]{PhysRevA111.022445}%
  \BibitemOpen
  \bibfield  {author} {\bibinfo {author} {\bibfnamefont {D.}~\bibnamefont
  {Shrimali}}, \bibinfo {author} {\bibfnamefont {S.}~\bibnamefont {Bhowmick}},\
  and\ \bibinfo {author} {\bibfnamefont {A.~K.}\ \bibnamefont {Pati}},\
  }\bibfield  {title} {\enquote {\bibinfo {title} {Quantum speed limit on the
  production of quantumness of observables},}\ }\href
  {https://doi.org/10.1103/PhysRevA.111.022445} {\bibfield  {journal} {\bibinfo
   {journal} {Phys. Rev. A}\ }\textbf {\bibinfo {volume} {111}},\ \bibinfo
  {pages} {022445} (\bibinfo {year} {2025})}\BibitemShut {NoStop}%
\bibitem [{\citenamefont {Marvian}, \citenamefont {Spekkens},\ and\
  \citenamefont {Zanardi}(2016)}]{PhysRevA93.052331}%
  \BibitemOpen
  \bibfield  {author} {\bibinfo {author} {\bibfnamefont {I.}~\bibnamefont
  {Marvian}}, \bibinfo {author} {\bibfnamefont {R.~W.}\ \bibnamefont
  {Spekkens}},\ and\ \bibinfo {author} {\bibfnamefont {P.}~\bibnamefont
  {Zanardi}},\ }\bibfield  {title} {\enquote {\bibinfo {title} {Quantum speed
  limits, coherence, and asymmetry},}\ }\href
  {https://doi.org/10.1103/PhysRevA.93.052331} {\bibfield  {journal} {\bibinfo
  {journal} {Phys. Rev. A}\ }\textbf {\bibinfo {volume} {93}},\ \bibinfo
  {pages} {052331} (\bibinfo {year} {2016})}\BibitemShut {NoStop}%
\bibitem [{\citenamefont {Lan}, \citenamefont {Xie},\ and\ \citenamefont
  {Cai}(2022)}]{NewJPhys.24.055003}%
  \BibitemOpen
  \bibfield  {author} {\bibinfo {author} {\bibfnamefont {K.}~\bibnamefont
  {Lan}}, \bibinfo {author} {\bibfnamefont {S.}~\bibnamefont {Xie}},\ and\
  \bibinfo {author} {\bibfnamefont {X.}~\bibnamefont {Cai}},\ }\bibfield
  {title} {\enquote {\bibinfo {title} {Geometric quantum speed limits for
  {M}arkovian dynamics in open quantum systems},}\ }\href
  {https://doi.org/10.1088/1367-2630/ac696b} {\bibfield  {journal} {\bibinfo
  {journal} {New J. Phys.}\ }\textbf {\bibinfo {volume} {24}},\ \bibinfo
  {pages} {055003} (\bibinfo {year} {2022})}\BibitemShut {NoStop}%
\bibitem [{\citenamefont {Wu}\ and\ \citenamefont
  {An}(2023)}]{PhysRevA108.012204}%
  \BibitemOpen
  \bibfield  {author} {\bibinfo {author} {\bibfnamefont {W.}~\bibnamefont
  {Wu}}\ and\ \bibinfo {author} {\bibfnamefont {J.-H.}\ \bibnamefont {An}},\
  }\bibfield  {title} {\enquote {\bibinfo {title} {Quantum speed limit from a
  quantum-state-diffusion method},}\ }\href
  {https://doi.org/10.1103/PhysRevA.108.012204} {\bibfield  {journal} {\bibinfo
   {journal} {Phys. Rev. A}\ }\textbf {\bibinfo {volume} {108}},\ \bibinfo
  {pages} {012204} (\bibinfo {year} {2023})}\BibitemShut {NoStop}%
\bibitem [{\citenamefont {Zheng}\ and\ \citenamefont
  {Peng}(2024)}]{ResultsPhys.57.107315}%
  \BibitemOpen
  \bibfield  {author} {\bibinfo {author} {\bibfnamefont {L.}~\bibnamefont
  {Zheng}}\ and\ \bibinfo {author} {\bibfnamefont {Y.}~\bibnamefont {Peng}},\
  }\bibfield  {title} {\enquote {\bibinfo {title} {Quantum speed limits of
  quantumsystem in colored environments},}\ }\href
  {https://doi.org/10.1016/j.rinp.2023.107315} {\bibfield  {journal} {\bibinfo
  {journal} {Results Phys.}\ }\textbf {\bibinfo {volume} {57}},\ \bibinfo
  {pages} {107315} (\bibinfo {year} {2024})}\BibitemShut {NoStop}%
\bibitem [{\citenamefont {Mai}\ and\ \citenamefont
  {Yu}(2024)}]{PhysRevA110.042425}%
  \BibitemOpen
  \bibfield  {author} {\bibinfo {author} {\bibfnamefont {Z.-y.}\ \bibnamefont
  {Mai}}\ and\ \bibinfo {author} {\bibfnamefont {C.-s.}\ \bibnamefont {Yu}},\
  }\bibfield  {title} {\enquote {\bibinfo {title} {Quantum speed limit in terms
  of coherence variations},}\ }\href
  {https://doi.org/10.1103/PhysRevA.110.042425} {\bibfield  {journal} {\bibinfo
   {journal} {Phys. Rev. A}\ }\textbf {\bibinfo {volume} {110}},\ \bibinfo
  {pages} {042425} (\bibinfo {year} {2024})}\BibitemShut {NoStop}%
\bibitem [{\citenamefont {Hu}, \citenamefont {Sun},\ and\ \citenamefont
  {Zheng}(2022)}]{JChemPhys.156.134113}%
  \BibitemOpen
  \bibfield  {author} {\bibinfo {author} {\bibfnamefont {X.}~\bibnamefont
  {Hu}}, \bibinfo {author} {\bibfnamefont {S.}~\bibnamefont {Sun}},\ and\
  \bibinfo {author} {\bibfnamefont {Y.}~\bibnamefont {Zheng}},\ }\bibfield
  {title} {\enquote {\bibinfo {title} {Witnessing localization of a quantum
  state via quantum speed limits in a driven avoided-level crossing system},}\
  }\href {https://doi.org/10.1063/5.0078207} {\bibfield  {journal} {\bibinfo
  {journal} {J. Chem. Phys.}\ }\textbf {\bibinfo {volume} {156}},\ \bibinfo
  {pages} {134113} (\bibinfo {year} {2022})}\BibitemShut {NoStop}%
\bibitem [{\citenamefont {Cimmarusti}\ \emph {et~al.}(2015)\citenamefont
  {Cimmarusti}, \citenamefont {Yan}, \citenamefont {Patterson}, \citenamefont
  {Corcos}, \citenamefont {Orozco},\ and\ \citenamefont
  {Deffner}}]{PhysRevLett.114.233602}%
  \BibitemOpen
  \bibfield  {author} {\bibinfo {author} {\bibfnamefont {A.~D.}\ \bibnamefont
  {Cimmarusti}}, \bibinfo {author} {\bibfnamefont {Z.}~\bibnamefont {Yan}},
  \bibinfo {author} {\bibfnamefont {B.~D.}\ \bibnamefont {Patterson}}, \bibinfo
  {author} {\bibfnamefont {L.~P.}\ \bibnamefont {Corcos}}, \bibinfo {author}
  {\bibfnamefont {L.~A.}\ \bibnamefont {Orozco}},\ and\ \bibinfo {author}
  {\bibfnamefont {S.}~\bibnamefont {Deffner}},\ }\bibfield  {title} {\enquote
  {\bibinfo {title} {Environment-assisted speed-up of the field evolution in
  cavity quantum electrodynamics},}\ }\href
  {https://doi.org/10.1103/PhysRevLett.114.233602} {\bibfield  {journal}
  {\bibinfo  {journal} {Phys. Rev. Lett.}\ }\textbf {\bibinfo {volume} {114}},\
  \bibinfo {pages} {233602} (\bibinfo {year} {2015})}\BibitemShut {NoStop}%
\bibitem [{\citenamefont {{del Campo}}(2021)}]{PhysRevLett.126.180603}%
  \BibitemOpen
  \bibfield  {author} {\bibinfo {author} {\bibfnamefont {A.}~\bibnamefont {{del
  Campo}}},\ }\bibfield  {title} {\enquote {\bibinfo {title} {Probing quantum
  speed limits with ultracold gases},}\ }\href
  {https://doi.org/10.1103/PhysRevLett.126.180603} {\bibfield  {journal}
  {\bibinfo  {journal} {Phys. Rev. Lett.}\ }\textbf {\bibinfo {volume} {126}},\
  \bibinfo {pages} {180603} (\bibinfo {year} {2021})}\BibitemShut {NoStop}%
\bibitem [{\citenamefont {Lu}\ \emph {et~al.}(2024)\citenamefont {Lu},
  \citenamefont {Liu}, \citenamefont {Liu}, \citenamefont {Rao}, \citenamefont
  {Lao}, \citenamefont {Wu}, \citenamefont {Zhu},\ and\ \citenamefont
  {Luo}}]{NewJPhys.26.013043}%
  \BibitemOpen
  \bibfield  {author} {\bibinfo {author} {\bibfnamefont {P.}~\bibnamefont
  {Lu}}, \bibinfo {author} {\bibfnamefont {T.}~\bibnamefont {Liu}}, \bibinfo
  {author} {\bibfnamefont {Y.}~\bibnamefont {Liu}}, \bibinfo {author}
  {\bibfnamefont {X.}~\bibnamefont {Rao}}, \bibinfo {author} {\bibfnamefont
  {Q.}~\bibnamefont {Lao}}, \bibinfo {author} {\bibfnamefont {H.}~\bibnamefont
  {Wu}}, \bibinfo {author} {\bibfnamefont {F.}~\bibnamefont {Zhu}},\ and\
  \bibinfo {author} {\bibfnamefont {L.}~\bibnamefont {Luo}},\ }\bibfield
  {title} {\enquote {\bibinfo {title} {Realizing quantum speed limit in open
  system with a $\mathcal{P}\mathcal{T}$-symmetric trapped-ion qubit},}\ }\href
  {https://doi.org/10.1088/1367-2630/ad1a28} {\bibfield  {journal} {\bibinfo
  {journal} {New J. Phys.}\ }\textbf {\bibinfo {volume} {26}},\ \bibinfo
  {pages} {013043} (\bibinfo {year} {2024})}\BibitemShut {NoStop}%
\bibitem [{\citenamefont {Pires}\ \emph {et~al.}(2024)\citenamefont {Pires},
  \citenamefont {deAzevedo}, \citenamefont {Diogo O. Soares-Pinto},\ and\
  \citenamefont {Filgueiras}}]{CommunPhys.7.142}%
  \BibitemOpen
  \bibfield  {author} {\bibinfo {author} {\bibfnamefont {D.~P.}\ \bibnamefont
  {Pires}}, \bibinfo {author} {\bibfnamefont {E.~R.}\ \bibnamefont
  {deAzevedo}}, \bibinfo {author} {\bibfnamefont {F.~B.}\ \bibnamefont {Diogo
  O. Soares-Pinto}},\ and\ \bibinfo {author} {\bibfnamefont {J.~G.}\
  \bibnamefont {Filgueiras}},\ }\bibfield  {title} {\enquote {\bibinfo {title}
  {Experimental investigation of geometric quantum speed limits in an open
  quantum system},}\ }\href {https://doi.org/10.1038/s42005-024-01634-5}
  {\bibfield  {journal} {\bibinfo  {journal} {Commun. Phys.}\ }\textbf
  {\bibinfo {volume} {7}},\ \bibinfo {pages} {142} (\bibinfo {year}
  {2024})}\BibitemShut {NoStop}%
\bibitem [{\citenamefont {Wu}\ \emph {et~al.}(2024)\citenamefont {Wu},
  \citenamefont {Yuan}, \citenamefont {Zhang}, \citenamefont {Zhu},
  \citenamefont {Deng}, \citenamefont {Zhang}, \citenamefont {Zhang},
  \citenamefont {Guo}, \citenamefont {Wang}, \citenamefont {Huang},
  \citenamefont {Song}, \citenamefont {Li}, \citenamefont {Wang}, \citenamefont
  {Wang},\ and\ \citenamefont {Agarwal}}]{PhysRevA110.042215}%
  \BibitemOpen
  \bibfield  {author} {\bibinfo {author} {\bibfnamefont {Y.}~\bibnamefont
  {Wu}}, \bibinfo {author} {\bibfnamefont {J.}~\bibnamefont {Yuan}}, \bibinfo
  {author} {\bibfnamefont {C.}~\bibnamefont {Zhang}}, \bibinfo {author}
  {\bibfnamefont {Z.}~\bibnamefont {Zhu}}, \bibinfo {author} {\bibfnamefont
  {J.}~\bibnamefont {Deng}}, \bibinfo {author} {\bibfnamefont {X.}~\bibnamefont
  {Zhang}}, \bibinfo {author} {\bibfnamefont {P.}~\bibnamefont {Zhang}},
  \bibinfo {author} {\bibfnamefont {Q.}~\bibnamefont {Guo}}, \bibinfo {author}
  {\bibfnamefont {Z.}~\bibnamefont {Wang}}, \bibinfo {author} {\bibfnamefont
  {J.}~\bibnamefont {Huang}}, \bibinfo {author} {\bibfnamefont
  {C.}~\bibnamefont {Song}}, \bibinfo {author} {\bibfnamefont {H.}~\bibnamefont
  {Li}}, \bibinfo {author} {\bibfnamefont {D.-W.}\ \bibnamefont {Wang}},
  \bibinfo {author} {\bibfnamefont {H.}~\bibnamefont {Wang}},\ and\ \bibinfo
  {author} {\bibfnamefont {G.~S.}\ \bibnamefont {Agarwal}},\ }\bibfield
  {title} {\enquote {\bibinfo {title} {Testing the unified bounds of the
  quantum speed limit},}\ }\href {https://doi.org/10.1103/PhysRevA.110.042215}
  {\bibfield  {journal} {\bibinfo  {journal} {Phys. Rev. A}\ }\textbf {\bibinfo
  {volume} {110}},\ \bibinfo {pages} {042215} (\bibinfo {year}
  {2024})}\BibitemShut {NoStop}%
\bibitem [{\citenamefont {Palma}, \citenamefont {antti Suominen},\ and\
  \citenamefont {Ekert}(1996)}]{ProcRSocA452.567}%
  \BibitemOpen
  \bibfield  {author} {\bibinfo {author} {\bibfnamefont {G.~M.}\ \bibnamefont
  {Palma}}, \bibinfo {author} {\bibfnamefont {K.}~\bibnamefont {antti
  Suominen}},\ and\ \bibinfo {author} {\bibfnamefont {A.}~\bibnamefont
  {Ekert}},\ }\bibfield  {title} {\enquote {\bibinfo {title} {Quantum computers
  and dissipation},}\ }\href {https://doi.org/10.1098/rspa.1996.0029}
  {\bibfield  {journal} {\bibinfo  {journal} {Proc. R. Soc. A}\ }\textbf
  {\bibinfo {volume} {452}},\ \bibinfo {pages} {567} (\bibinfo {year}
  {1996})}\BibitemShut {NoStop}%
\bibitem [{\citenamefont {Reina}, \citenamefont {Quiroga},\ and\ \citenamefont
  {Johnson}(2002)}]{PhysRevA65.032326}%
  \BibitemOpen
  \bibfield  {author} {\bibinfo {author} {\bibfnamefont {J.~H.}\ \bibnamefont
  {Reina}}, \bibinfo {author} {\bibfnamefont {L.}~\bibnamefont {Quiroga}},\
  and\ \bibinfo {author} {\bibfnamefont {N.~F.}\ \bibnamefont {Johnson}},\
  }\bibfield  {title} {\enquote {\bibinfo {title} {Decoherence of quantum
  registers},}\ }\href {https://doi.org/10.1103/PhysRevA.65.032326} {\bibfield
  {journal} {\bibinfo  {journal} {Phys. Rev. A}\ }\textbf {\bibinfo {volume}
  {65}},\ \bibinfo {pages} {032326} (\bibinfo {year} {2002})}\BibitemShut
  {NoStop}%
\bibitem [{\citenamefont {Haikka}\ \emph {et~al.}(2011)\citenamefont {Haikka},
  \citenamefont {McEndoo}, \citenamefont {De~Chiara}, \citenamefont {Palma},\
  and\ \citenamefont {Maniscalco}}]{PhysRevA84.031602}%
  \BibitemOpen
  \bibfield  {author} {\bibinfo {author} {\bibfnamefont {P.}~\bibnamefont
  {Haikka}}, \bibinfo {author} {\bibfnamefont {S.}~\bibnamefont {McEndoo}},
  \bibinfo {author} {\bibfnamefont {G.}~\bibnamefont {De~Chiara}}, \bibinfo
  {author} {\bibfnamefont {G.~M.}\ \bibnamefont {Palma}},\ and\ \bibinfo
  {author} {\bibfnamefont {S.}~\bibnamefont {Maniscalco}},\ }\bibfield  {title}
  {\enquote {\bibinfo {title} {Quantifying, characterizing, and controlling
  information flow in ultracold atomic gases},}\ }\href
  {https://doi.org/10.1103/PhysRevA.84.031602} {\bibfield  {journal} {\bibinfo
  {journal} {Phys. Rev. A}\ }\textbf {\bibinfo {volume} {84}},\ \bibinfo
  {pages} {031602(R)} (\bibinfo {year} {2011})}\BibitemShut {NoStop}%
\bibitem [{\citenamefont {Haikka}, \citenamefont {Johnson},\ and\ \citenamefont
  {Maniscalco}(2013)}]{PhysRevA87.010103}%
  \BibitemOpen
  \bibfield  {author} {\bibinfo {author} {\bibfnamefont {P.}~\bibnamefont
  {Haikka}}, \bibinfo {author} {\bibfnamefont {T.~H.}\ \bibnamefont
  {Johnson}},\ and\ \bibinfo {author} {\bibfnamefont {S.}~\bibnamefont
  {Maniscalco}},\ }\bibfield  {title} {\enquote {\bibinfo {title}
  {Non-{M}arkovianity of local dephasing channels and time-invariant
  discord},}\ }\href {https://doi.org/10.1103/PhysRevA.87.010103} {\bibfield
  {journal} {\bibinfo  {journal} {Phys. Rev. A}\ }\textbf {\bibinfo {volume}
  {87}},\ \bibinfo {pages} {010103(R)} (\bibinfo {year} {2013})}\BibitemShut
  {NoStop}%
\bibitem [{\citenamefont {Addis}\ \emph
  {et~al.}(2014{\natexlab{a}})\citenamefont {Addis}, \citenamefont {Brebner},
  \citenamefont {Haikka},\ and\ \citenamefont
  {Maniscalco}}]{PhysRevA89.024101}%
  \BibitemOpen
  \bibfield  {author} {\bibinfo {author} {\bibfnamefont {C.}~\bibnamefont
  {Addis}}, \bibinfo {author} {\bibfnamefont {G.}~\bibnamefont {Brebner}},
  \bibinfo {author} {\bibfnamefont {P.}~\bibnamefont {Haikka}},\ and\ \bibinfo
  {author} {\bibfnamefont {S.}~\bibnamefont {Maniscalco}},\ }\bibfield  {title}
  {\enquote {\bibinfo {title} {Coherence trapping and information backflow in
  dephasing qubits},}\ }\href {https://doi.org/10.1103/PhysRevA.89.024101}
  {\bibfield  {journal} {\bibinfo  {journal} {Phys. Rev. A}\ }\textbf {\bibinfo
  {volume} {89}},\ \bibinfo {pages} {024101} (\bibinfo {year}
  {2014}{\natexlab{a}})}\BibitemShut {NoStop}%
\bibitem [{\citenamefont {Addis}\ \emph
  {et~al.}(2014{\natexlab{b}})\citenamefont {Addis}, \citenamefont {Bylicka},
  \citenamefont {Chru\'{s}ci\'{n}ski},\ and\ \citenamefont
  {Maniscalco}}]{PhysRevA90.052103}%
  \BibitemOpen
  \bibfield  {author} {\bibinfo {author} {\bibfnamefont {C.}~\bibnamefont
  {Addis}}, \bibinfo {author} {\bibfnamefont {B.}~\bibnamefont {Bylicka}},
  \bibinfo {author} {\bibfnamefont {D.}~\bibnamefont {Chru\'{s}ci\'{n}ski}},\
  and\ \bibinfo {author} {\bibfnamefont {S.}~\bibnamefont {Maniscalco}},\
  }\bibfield  {title} {\enquote {\bibinfo {title} {Comparative study of
  non-{M}arkovianity measures in exactly solvable one- and two-qubit models},}\
  }\href {https://doi.org/10.1103/PhysRevA.90.052103} {\bibfield  {journal}
  {\bibinfo  {journal} {Phys. Rev. A}\ }\textbf {\bibinfo {volume} {90}},\
  \bibinfo {pages} {052103} (\bibinfo {year} {2014}{\natexlab{b}})}\BibitemShut
  {NoStop}%
\bibitem [{\citenamefont {Zhang}\ \emph
  {et~al.}(2015{\natexlab{b}})\citenamefont {Zhang}, \citenamefont {Han},
  \citenamefont {Xia}, \citenamefont {Yu},\ and\ \citenamefont
  {Fan}}]{SciRep.5.13359}%
  \BibitemOpen
  \bibfield  {author} {\bibinfo {author} {\bibfnamefont {Y.-J.}\ \bibnamefont
  {Zhang}}, \bibinfo {author} {\bibfnamefont {W.}~\bibnamefont {Han}}, \bibinfo
  {author} {\bibfnamefont {Y.-J.}\ \bibnamefont {Xia}}, \bibinfo {author}
  {\bibfnamefont {Y.-M.}\ \bibnamefont {Yu}},\ and\ \bibinfo {author}
  {\bibfnamefont {H.}~\bibnamefont {Fan}},\ }\bibfield  {title} {\enquote
  {\bibinfo {title} {Role of initial system-bath correlation on coherence
  trapping},}\ }\href {https://doi.org/10.1038/srep13359} {\bibfield  {journal}
  {\bibinfo  {journal} {Sci. Rep.}\ }\textbf {\bibinfo {volume} {5}},\ \bibinfo
  {pages} {13359} (\bibinfo {year} {2015}{\natexlab{b}})}\BibitemShut {NoStop}%
\bibitem [{\citenamefont {Cai}\ and\ \citenamefont
  {Zheng}(2018)}]{JChemPhys.149.094107}%
  \BibitemOpen
  \bibfield  {author} {\bibinfo {author} {\bibfnamefont {X.}~\bibnamefont
  {Cai}}\ and\ \bibinfo {author} {\bibfnamefont {Y.}~\bibnamefont {Zheng}},\
  }\bibfield  {title} {\enquote {\bibinfo {title} {Non-{M}arkovian decoherence
  dynamics in nonequilibrium environments},}\ }\href
  {https://doi.org/10.1063/1.5039891} {\bibfield  {journal} {\bibinfo
  {journal} {J. Chem. Phys.}\ }\textbf {\bibinfo {volume} {149}},\ \bibinfo
  {pages} {094107} (\bibinfo {year} {2018})}\BibitemShut {NoStop}%
\bibitem [{\citenamefont {Benedetti}\ \emph {et~al.}(2018)\citenamefont
  {Benedetti}, \citenamefont {Salari~Sehdaran}, \citenamefont {Zandi},\ and\
  \citenamefont {Paris}}]{PhysRevA97.012126}%
  \BibitemOpen
  \bibfield  {author} {\bibinfo {author} {\bibfnamefont {C.}~\bibnamefont
  {Benedetti}}, \bibinfo {author} {\bibfnamefont {F.}~\bibnamefont
  {Salari~Sehdaran}}, \bibinfo {author} {\bibfnamefont {M.~H.}\ \bibnamefont
  {Zandi}},\ and\ \bibinfo {author} {\bibfnamefont {M.~G.~A.}\ \bibnamefont
  {Paris}},\ }\bibfield  {title} {\enquote {\bibinfo {title} {Quantum probes
  for the cutoff frequency of {O}hmic environments},}\ }\href
  {https://doi.org/10.1103/PhysRevA.97.012126} {\bibfield  {journal} {\bibinfo
  {journal} {Phys. Rev. A}\ }\textbf {\bibinfo {volume} {97}},\ \bibinfo
  {pages} {012126} (\bibinfo {year} {2018})}\BibitemShut {NoStop}%
\bibitem [{\citenamefont {Gebbia}\ \emph {et~al.}(2020)\citenamefont {Gebbia},
  \citenamefont {Benedetti}, \citenamefont {Benatti}, \citenamefont
  {Floreanini}, \citenamefont {Bina},\ and\ \citenamefont
  {Paris}}]{PhysRevA101.032112}%
  \BibitemOpen
  \bibfield  {author} {\bibinfo {author} {\bibfnamefont {F.}~\bibnamefont
  {Gebbia}}, \bibinfo {author} {\bibfnamefont {C.}~\bibnamefont {Benedetti}},
  \bibinfo {author} {\bibfnamefont {F.}~\bibnamefont {Benatti}}, \bibinfo
  {author} {\bibfnamefont {R.}~\bibnamefont {Floreanini}}, \bibinfo {author}
  {\bibfnamefont {M.}~\bibnamefont {Bina}},\ and\ \bibinfo {author}
  {\bibfnamefont {M.~G.~A.}\ \bibnamefont {Paris}},\ }\bibfield  {title}
  {\enquote {\bibinfo {title} {Two-qubit quantum probes for the temperature of
  an {O}hmic environment},}\ }\href
  {https://doi.org/10.1103/PhysRevA.101.032112} {\bibfield  {journal} {\bibinfo
   {journal} {Phys. Rev. A}\ }\textbf {\bibinfo {volume} {101}},\ \bibinfo
  {pages} {032112} (\bibinfo {year} {2020})}\BibitemShut {NoStop}%
\bibitem [{\citenamefont {Meng}\ \emph {et~al.}(2023)\citenamefont {Meng},
  \citenamefont {Sun}, \citenamefont {Wang}, \citenamefont {Ren}, \citenamefont
  {Cai},\ and\ \citenamefont {Czerwinski}}]{Entropy25.634}%
  \BibitemOpen
  \bibfield  {author} {\bibinfo {author} {\bibfnamefont {X.}~\bibnamefont
  {Meng}}, \bibinfo {author} {\bibfnamefont {Y.}~\bibnamefont {Sun}}, \bibinfo
  {author} {\bibfnamefont {Q.}~\bibnamefont {Wang}}, \bibinfo {author}
  {\bibfnamefont {J.}~\bibnamefont {Ren}}, \bibinfo {author} {\bibfnamefont
  {X.}~\bibnamefont {Cai}},\ and\ \bibinfo {author} {\bibfnamefont
  {A.}~\bibnamefont {Czerwinski}},\ }\bibfield  {title} {\enquote {\bibinfo
  {title} {Dephasing dynamics in a non-equilibrium fluctuating environment},}\
  }\href {https://doi.org/10.3390/e25040634} {\bibfield  {journal} {\bibinfo
  {journal} {Entropy}\ }\textbf {\bibinfo {volume} {25}},\ \bibinfo {pages}
  {634} (\bibinfo {year} {2023})}\BibitemShut {NoStop}%
\bibitem [{\citenamefont {Breuer}(2004)}]{PhysRevA70.012106}%
  \BibitemOpen
  \bibfield  {author} {\bibinfo {author} {\bibfnamefont {H.-P.}\ \bibnamefont
  {Breuer}},\ }\bibfield  {title} {\enquote {\bibinfo {title} {Genuine quantum
  trajectories for non-{M}arkovian processes},}\ }\href
  {https://doi.org/10.1103/PhysRevA.70.012106} {\bibfield  {journal} {\bibinfo
  {journal} {Phys. Rev. A}\ }\textbf {\bibinfo {volume} {70}},\ \bibinfo
  {pages} {012106} (\bibinfo {year} {2004})}\BibitemShut {NoStop}%
\bibitem [{\citenamefont {Laine}, \citenamefont {Piilo},\ and\ \citenamefont
  {Breuer}(2010)}]{PhysRevA81.062115}%
  \BibitemOpen
  \bibfield  {author} {\bibinfo {author} {\bibfnamefont {E.~M.}\ \bibnamefont
  {Laine}}, \bibinfo {author} {\bibfnamefont {J.}~\bibnamefont {Piilo}},\ and\
  \bibinfo {author} {\bibfnamefont {H.~P.}\ \bibnamefont {Breuer}},\ }\bibfield
   {title} {\enquote {\bibinfo {title} {Measure for the non-{M}arkovianity of
  quantum processes},}\ }\href {https://doi.org/10.1103/PhysRevA.81.062115}
  {\bibfield  {journal} {\bibinfo  {journal} {Phys. Rev. A}\ }\textbf {\bibinfo
  {volume} {81}},\ \bibinfo {pages} {062115} (\bibinfo {year}
  {2010})}\BibitemShut {NoStop}%
\bibitem [{\citenamefont {Uhlmann}(1995)}]{RepMathPhys.36.461}%
  \BibitemOpen
  \bibfield  {author} {\bibinfo {author} {\bibfnamefont {A.}~\bibnamefont
  {Uhlmann}},\ }\bibfield  {title} {\enquote {\bibinfo {title} {Geometric
  phases and related structures},}\ }\href
  {https://doi.org/10.1016/0034-4877(96)83640-8} {\bibfield  {journal}
  {\bibinfo  {journal} {Rep. Prog. Phys.}\ }\textbf {\bibinfo {volume} {36}},\
  \bibinfo {pages} {461} (\bibinfo {year} {1995})}\BibitemShut {NoStop}%
\bibitem [{\citenamefont {Gibilisco}\ and\ \citenamefont
  {Isola}(2003)}]{JMathPhys.44.3752}%
  \BibitemOpen
  \bibfield  {author} {\bibinfo {author} {\bibfnamefont {P.}~\bibnamefont
  {Gibilisco}}\ and\ \bibinfo {author} {\bibfnamefont {T.}~\bibnamefont
  {Isola}},\ }\bibfield  {title} {\enquote {\bibinfo {title} {{W}igner-{Y}anase
  information on quantum state space: {T}he geometric approach},}\ }\href
  {https://doi.org/10.1063/1.1598279} {\bibfield  {journal} {\bibinfo
  {journal} {J. Math. Phys.}\ }\textbf {\bibinfo {volume} {44}},\ \bibinfo
  {pages} {3752} (\bibinfo {year} {2003})}\BibitemShut {NoStop}%
\bibitem [{\citenamefont {Marvian}\ and\ \citenamefont
  {Spekkens}(2014)}]{NatCommun.5.3821}%
  \BibitemOpen
  \bibfield  {author} {\bibinfo {author} {\bibfnamefont {I.}~\bibnamefont
  {Marvian}}\ and\ \bibinfo {author} {\bibfnamefont {R.~W.}\ \bibnamefont
  {Spekkens}},\ }\bibfield  {title} {\enquote {\bibinfo {title} {Extending
  {N}oether’s theorem by quantifying the asymmetry of quantum states},}\
  }\href {https://doi.org/10.1038/ncomms4821} {\bibfield  {journal} {\bibinfo
  {journal} {Nat. Commun.}\ }\textbf {\bibinfo {volume} {5}},\ \bibinfo {pages}
  {3821} (\bibinfo {year} {2014})}\BibitemShut {NoStop}%
\bibitem [{\citenamefont {DiVincenzo}(1995)}]{PhysRevA51.1015}%
  \BibitemOpen
  \bibfield  {author} {\bibinfo {author} {\bibfnamefont {D.~P.}\ \bibnamefont
  {DiVincenzo}},\ }\bibfield  {title} {\enquote {\bibinfo {title} {Two-bit
  gates are universal for quantum computation},}\ }\href
  {https://doi.org/10.1103/PhysRevA.51.1015} {\bibfield  {journal} {\bibinfo
  {journal} {Phys. Rev. A}\ }\textbf {\bibinfo {volume} {51}},\ \bibinfo
  {pages} {1015} (\bibinfo {year} {1995})}\BibitemShut {NoStop}%
\bibitem [{\citenamefont {Weiss}(1999)}]{Weissbook}%
  \BibitemOpen
  \bibfield  {author} {\bibinfo {author} {\bibfnamefont {U.}~\bibnamefont
  {Weiss}},\ }\href@noop {} {\emph {\bibinfo {title} {Quantum Dissipative
  Systems}}}\ (\bibinfo  {publisher} {World Scientific},\ \bibinfo {address}
  {Singapore},\ \bibinfo {year} {1999})\BibitemShut {NoStop}%
\bibitem [{\citenamefont {Cirone}\ \emph {et~al.}(2009)\citenamefont {Cirone},
  \citenamefont {Chiara}, \citenamefont {Palma},\ and\ \citenamefont
  {Recati}}]{NewJPhys.11.103055}%
  \BibitemOpen
  \bibfield  {author} {\bibinfo {author} {\bibfnamefont {M.~A.}\ \bibnamefont
  {Cirone}}, \bibinfo {author} {\bibfnamefont {G.~D.}\ \bibnamefont {Chiara}},
  \bibinfo {author} {\bibfnamefont {G.~M.}\ \bibnamefont {Palma}},\ and\
  \bibinfo {author} {\bibfnamefont {A.}~\bibnamefont {Recati}},\ }\bibfield
  {title} {\enquote {\bibinfo {title} {Collective decoherence of cold atoms
  coupled to a {B}ose–{E}instein condensate},}\ }\href
  {https://doi.org/10.1088/1367-2630/11/10/103055} {\bibfield  {journal}
  {\bibinfo  {journal} {New J. Phys.}\ }\textbf {\bibinfo {volume} {11}},\
  \bibinfo {pages} {103055} (\bibinfo {year} {2009})}\BibitemShut {NoStop}%
\bibitem [{\citenamefont {Haikka}, \citenamefont {McEndoo},\ and\ \citenamefont
  {Maniscalco}(2013)}]{PhysRevA87.012127}%
  \BibitemOpen
  \bibfield  {author} {\bibinfo {author} {\bibfnamefont {P.}~\bibnamefont
  {Haikka}}, \bibinfo {author} {\bibfnamefont {S.}~\bibnamefont {McEndoo}},\
  and\ \bibinfo {author} {\bibfnamefont {S.}~\bibnamefont {Maniscalco}},\
  }\bibfield  {title} {\enquote {\bibinfo {title} {Non-{M}arkovian probes in
  ultracold gases},}\ }\href {https://doi.org/10.1103/PhysRevA.87.012127}
  {\bibfield  {journal} {\bibinfo  {journal} {Phys. Rev. A}\ }\textbf {\bibinfo
  {volume} {87}},\ \bibinfo {pages} {012127} (\bibinfo {year}
  {2013})}\BibitemShut {NoStop}%
\bibitem [{\citenamefont {Cai}\ and\ \citenamefont
  {Zheng}(2016)}]{PhysRevA94.042110}%
  \BibitemOpen
  \bibfield  {author} {\bibinfo {author} {\bibfnamefont {X.}~\bibnamefont
  {Cai}}\ and\ \bibinfo {author} {\bibfnamefont {Y.}~\bibnamefont {Zheng}},\
  }\bibfield  {title} {\enquote {\bibinfo {title} {Decoherence induced by
  non-{M}arkovian noise in a nonequilibrium environment},}\ }\href
  {https://doi.org/10.1103/PhysRevA.94.042110} {\bibfield  {journal} {\bibinfo
  {journal} {Phys. Rev. A}\ }\textbf {\bibinfo {volume} {94}},\ \bibinfo
  {pages} {042110} (\bibinfo {year} {2016})}\BibitemShut {NoStop}%
\bibitem [{\citenamefont {Cai}(2020)}]{SciRep.10.88}%
  \BibitemOpen
  \bibfield  {author} {\bibinfo {author} {\bibfnamefont {X.}~\bibnamefont
  {Cai}},\ }\bibfield  {title} {\enquote {\bibinfo {title} {Quantum dephasing
  induced by non-{M}arkovian random telegraph noise},}\ }\href
  {https://doi.org/10.1038/s41598-019-57081-8} {\bibfield  {journal} {\bibinfo
  {journal} {Sci. Rep.}\ }\textbf {\bibinfo {volume} {10}},\ \bibinfo {pages}
  {88} (\bibinfo {year} {2020})}\BibitemShut {NoStop}%
\end{thebibliography}%


%
	%
	
\end{document}